\documentclass[twoside,twocolumn,english,prc,amsmath,a4paper,myheadings]{revtex4}
\usepackage[T1]{fontenc}
\usepackage{geometry}
\usepackage{verbatim}
\usepackage{amsmath}
\usepackage{color}
\usepackage{graphicx}
\usepackage{amssymb}

\makeatletter

\providecommand{\tabularnewline}{\\}

\def\@oddhead{\rightmark \hfill  Analysing radial flow features in pPb  \hfill \thepage}
\def\@evenhead{\thepage \hfill K. Werner et al.\hfill}
\def\fnum@table{\tablename~{\bf\thetable}}
\def\fnum@figure{\figurename~{\bf\thefigure}}
\def\tablename{\footnotesize{\bf Table}}
\def\figurename{\footnotesize{\bf Figure}}

\usepackage{dcolumn}

\def\citet{\cite}
\newcommand{\dd}{\partial}

\makeatother

\usepackage{babel}

\begin{document}

\title{Analysing radial flow features in p-Pb and p-p collisions at several
TeV by studying identified particle production in EPOS3 }

\author{{\normalsize K.$\,$Werner$^{(a)}$, B. Guiot$^{(a)}$, Iu.$\,$Karpenko$^{(b,c)}$,
T.$\,$Pierog$^{(d)}$}}

\address{$^{(a)}$ SUBATECH, University of Nantes -- IN2P3/CNRS-- EMN, Nantes,
France}

\address{$^{(b)}$ Bogolyubov Institute for Theoretical Physics, Kiev 143,
03680, Ukraine}

\address{$^{(c)}$ FIAS, Johann Wolfgang Goethe Universitaet, Frankfurt am
Main, Germany}

\address{$^{(d)}$Karlsruhe Inst. of Technology, KIT, Campus North, Inst.
f. Kernphysik, Germany}

\begin{abstract}
Experimental transverse momentum spectra of identified particles in
p-Pb collisions at 5.02 TeV show many similarities to the corresponding
Pb-Pb results, the latter ones usually being interpreted in term of
hydrodynamic flow. We analyse these data using EPOS3, an event generator
based on a 3D+1 viscous hydrodynamical evolution starting from flux
tube initial conditions, which are generated in the Gribov-Regge multiple
scattering framework. An individual scattering is referred to as Pomeron,
identified with a parton ladder, eventually showing up as flux tubes
(or strings). Each parton ladder is composed of a pQCD hard process,
plus initial and final state linear parton emission. Nonlinear effects
are considered by using saturation scales $Q_{s}$, depending on the
energy and the number of participants connected to the Pomeron in
question. We compute transverse momentum ($p_{t}$) spectra of pions,
kaons, protons, lambdas, and $\Xi$ baryons in p-Pb and p-p scattering,
compared to experimental data and many other models. In this way we
show in a quantitative fashion that p-Pb data (and even p-p ones)
show the typical {}``flow effect'' of enhanced particle production
at intermediate $p_{t}$ values, more and more visible with increasing
hadron mass.
\end{abstract}
\maketitle

\section{Introduction}

Collective hydrodynamic flow seems to be well established in heavy
ion (HI) collisions at energies between 200 and 2760 AGeV, whereas
p-p and p-nucleus (p-A) collisions are often considered to be simple
reference systems, showing {}``normal'' behavior, such that deviations
in HI collisions with respect to p-p or p-A reveal {}``new physics''.
Surprisingly, the first results from p-Pb at 5.02 TeV on the transverse
momentum dependence of azimuthal anisotropies and particle yields
are very similar to the observations in HI scattering. In this paper
we will focus on transverse momentum spectra of identified particles.
The CMS collaboration showed recently \citet{cms} that the shapes
of transverse momentum spectra of pions, kaons, and protons change
in a characteristic way with multiplicity, which looks like an increasing
contribution from radial flow with multiplicity. A similar conclusion
can be drawn from recent measurements from ALICE \citet{alice}concerning
transverse momentum spectra of pions, kaons, protons, and lambdas.
In particular, the ratio lambda over kaon shows a peak structure,
similar as in HI, more and more pronounced with increasing multiplicity. 

Do we see radial flow in p-Pb collisions? In order to answer this
question, we will employ the EPOS3 approach, well suited for this
problem, since it provides within a unique theoretical scheme the
initial conditions for a hydrodynamical evolution in p-p, p-A, and
HI collisions. The initial conditions are generated in the Gribov-Regge
multiple scattering framework. An individual scattering is referred
to as Pomeron, identified with a parton ladder, eventually showing
up as flux tubes (also called strings). Each parton ladder is composed
of a pQCD hard process, plus initial and final state linear parton
emission. Our formalism is referred to as {}``Parton based Gribov
Regge Theory'' and described in very detail in \citet{hajo}. Based
on these initial conditions, we performed already ideal hydrodynamical
calculations \citet{epos2,jetbulk,kw1,kw2} to analyse HI and p-p
scattering at RHIC and LHC. In this paper we discuss two major improvements:
a more sophisticated treatment of nonlinear effects in the parton
evolution by considering individual (per Pomeron) saturation scales,
and a 3D+1 viscous hydrodynamical evolution. There are also changes
in our core-corona procedure, which %
\begin{figure}[b]
\begin{centering}
\includegraphics[angle=270,scale=0.24]{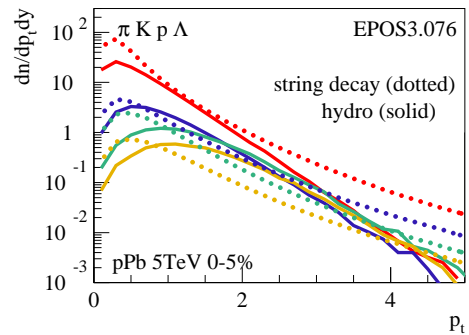}
\par\end{centering}

\caption{(Color online) Identified particle spectra as a function of $p_{t}$,
for central (0-5\%) p-Pb collisions at 5.02 TeV. We show results for
particle production from string decay, i.e. EPOS without hydro (dotted
curves), and particle production from pure hydro, without corona (solid
lines). In both cases, we show (from top to bottom) pions, kaons,
protons, and lambdas.\label{fig:flow}}

\end{figure}
amounts to separate the initial energy of the flux tubes into a part
which constitutes the initial conditions for hydro (core) and the
particles which leave the {}``matter''. This is crucial as well
in proton-nucleus collisions (as in all other collision types).

To understand the results discussed later in this paper, we show in
fig. \ref{fig:flow} the effect of flow on identified particle spectra,
by comparing $p_{t}$ distributions from pure string decay to spectra
from a pure hydrodynamic evolution. In case of string fragmentation,
heavier particles are strongly suppressed compared to lighter ones,
but the shapes are not so different. This picture changes completely
in the fluid case: The heavier the particle, the more it gets shifted
from low to intermediate $p_{t}$. This is a direct consequence of
the fact that the particles are produced from fluid cells characterized
by radial flow velocities, which gives more transverse momentum to
heavier particles.

There are few other studies of hydrodynamic expansion in proton-nucleus
systems. In \citet{hydro-bozek}, fluctuating initial conditions based
on the so-called Monte Carlo Glauber model (which is actually a wounded
nucleon model) are employed, followed by a viscous hydrodynamical
evolution. Also \citet{hydro-schenke} uses fluctuating initial conditions,
here based on both Glauber Monte Carlo and Glasma initial conditions.
Finally in \citet{hydro-qin}, ideal hydrodynamical calculations are
performed, starting from smooth Glauber model initial conditions. 

In the following chapters II to VII, we discuss the different elements
of the EPOS3 model. In chapters VIII to X, we report results on p-Pb
and p-p scattering, comparing EPOS3 with data and other models, which
leads to conclusions concerning hydrodynamical flow. Data points are
systematically shown with statistical errors only, unless mentioned
otherwise (often the error bars are too small to be visible). When
comparing simulations to data, we always adopt the same multiplicity
definition as in experiment.

\section{Multiple Pomeron exchange and saturation}

The starting point is a multiple scattering approach corresponding
to a marriage of Gribov-Regge theory \citet{gri-bla62,gri-che61,gri-reg59,gri62,gri65,gri68,gri69,gri69b,gri69c,gri70,gri73}
and perturbative QCD (pQCD), which has the advantage of being applicable
to deep inelastic lepton-proton scattering, as well as proton-proton
(p-p) , proton-nucleus (p-A), and nucleus-nucleus (A-A) collisions
(see \citet{hajo}). A very important aspect of this formalism is
its ability to provide \textbf{exclusive} cross sections, a necessary
requirement for Monte Carlo applications, the latter ones becoming
more and more popular after the discovery of the importance of event-by-event
fluctuations even in A-A collisions.

\begin{figure}[tb]
\begin{minipage}[t][1\totalheight]{1\columnwidth}%
\noindent \textcolor{blue}{\Large \hspace*{-0.5cm}}%
\begin{minipage}[c][1\totalheight]{0.45\columnwidth}%
\noindent \begin{center}
\textcolor{blue}{\Large \[
\sigma^{\mathrm{tot}}=\!\sum_{\mathrm{cut\, P}}\int\!\!\!\sum_{\mathrm{uncut\, P}}\int\]
}
\par\end{center}%
\end{minipage}%
\begin{minipage}[c][1\totalheight]{0.5\columnwidth}%
\noindent \begin{center}
\includegraphics[scale=0.2]{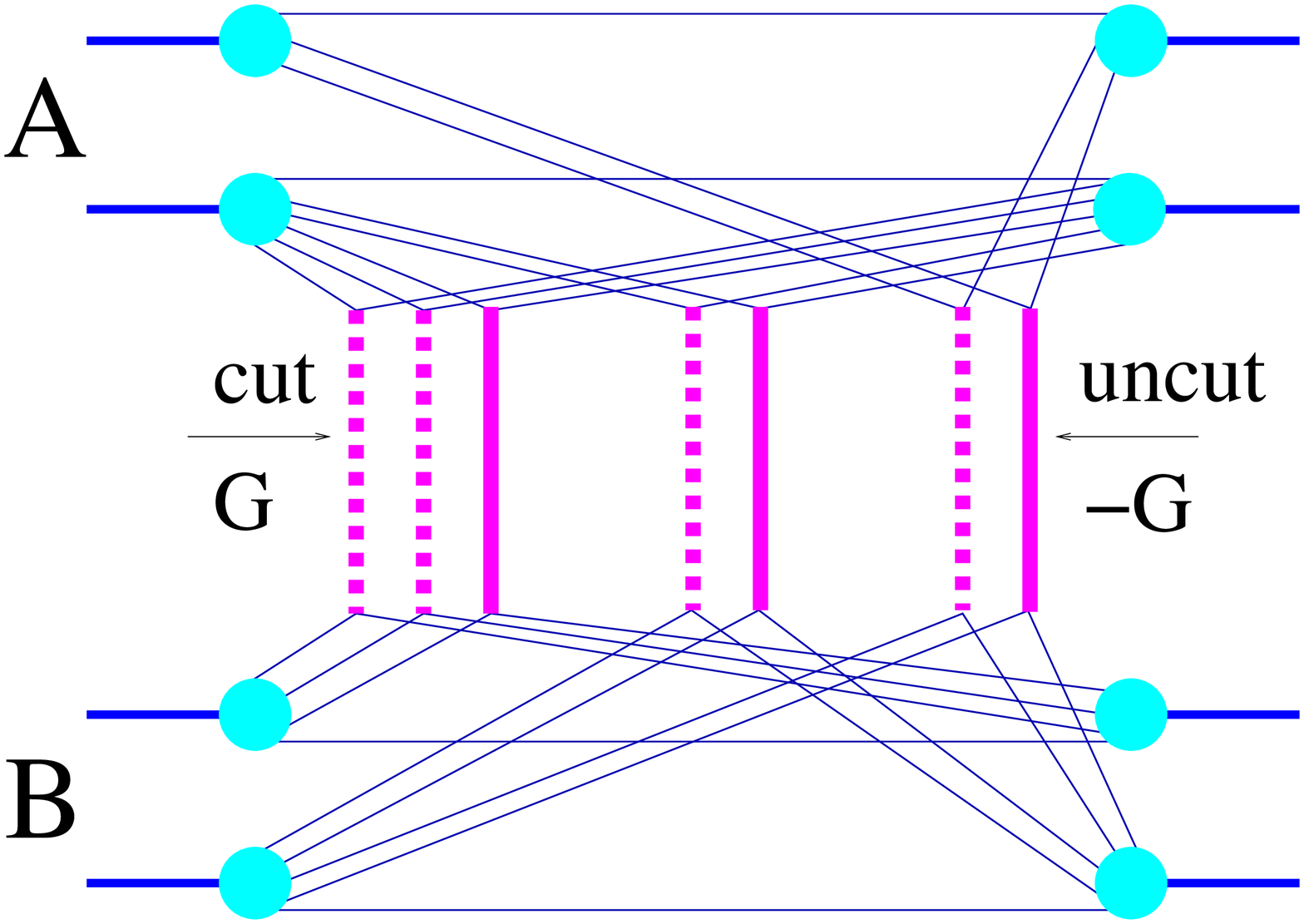}
\par\end{center}%
\end{minipage}\vspace{-0.5cm}
\begin{minipage}[t][1\totalheight]{0.95\columnwidth}%
\textcolor{blue}{\LARGE \[
\qquad\qquad\underbrace{\qquad\qquad\quad\qquad\qquad\qquad}\]
}%
\end{minipage}

\vspace{-0.7cm}
\begin{minipage}[t][1\totalheight]{0.95\columnwidth}%
\textcolor{blue}{\Large \[
\qquad\qquad\qquad d\sigma_{\mathrm{exclusive}}\]
}%
\end{minipage}%
\end{minipage}

\caption{(Color online) The total cross section expressed in terms of cut (dashed
lines) and uncut (solid lines) Pomerons, for nucleus-nucleus, proton-nucleus,
and proton-proton collisions. Partial summations allow to obtain exclusive
cross sections. \label{muscatt}}

\end{figure}

Gribov-Regge theory starts from the hypothesis that the T-matrix of
the scattering process can be written as a product of elementary objects
(later) referred to as Pomerons ($I\!\! P$). In \citet{hajo}, we
generalize the approach by adding energy-momentum conservation into
the expression for the T-matrix. Squaring the T-matrix can be done
conveniently using the Cutcosky cutting rule technique, and one obtains
for the total cross section an expression as illustrated in fig. \ref{muscatt},
expressed in terms of cut and uncut Pomerons. The mathematical expressions
corresponding to the cut Pomeron is referred to as $G$ (to be discussed
later), for the uncut one we have $-G$ (see \citet{hajo}). %

It is of course very useful to have an explicit formula for the total
cross section, but the real power of the expression in fig. \ref{muscatt}
is the fact that partial summations (as indicated in the figure) provide
exclusive cross sections for subprocesses, as for example the cross
section for triple scattering (triple $I\!\! P$ exchange) in p-p
collisions, or the cross section for a given number of $I\!\! P$
exchanges in p-A. This discussion is in particular important for p-A
scattering, because finally the number of Pomeron exchanges characterizes
the geometry of a collision, not the number of participants or the
number of collisions.

The following formulas are somewhat simplified, not showing explicitly
summations over parton flavors, the precise formulas being given in
\citet{hajo}. The expression corresponding to fig. \ref{muscatt}
is\begin{align*}
 & \sigma^{\mathrm{tot}}=\int d^{2}b\int\prod_{i=1}^{A}d^{2}b_{i}^{A}\, dz_{i}^{A}\,\rho_{A}(\sqrt{(b_{i}^{A})^{2}+(z_{i}^{A})^{2}})\qquad\qquad\qquad\\
 & \qquad\qquad\prod_{j=1}^{B}d^{2}b_{j}^{B}\, dz_{j}^{B}\,\rho_{B}(\sqrt{(b_{j}^{B})^{2}+(z_{j}^{B})^{2}})\\
 & \qquad\qquad\sum_{m_{1}l_{1}}\ldots\sum_{m_{AB}l_{AB}}(1-\delta_{0\Sigma m_{k}})\end{align*}
\vspace{-0.6cm}
\[
\qquad\quad\int\,\prod_{k=1}^{AB}\bigg(\prod_{\mu=1}^{m_{k}}dx_{k,\mu}^{+}dx_{k,\mu}^{-}\,\prod_{\lambda=1}^{l_{k}}d\tilde{x}_{k,\lambda}^{+}d\tilde{x}_{k,\lambda}^{-}\bigg)\Bigg{\{}\]
\vspace{-0.4cm}
\begin{align*}
 & \quad\prod_{k=1}^{AB}\bigg{(}\frac{1}{m_{k}!}\frac{1}{l_{k}!}\prod_{\mu=1}^{m_{k}}G(x_{k,\mu}^{+},x_{k,\mu}^{-},s,|\vec{b}+\vec{b}_{\pi(k)}^{A}-\vec{b}_{\tau(k)}^{B}|)\\
 & \qquad\qquad\prod_{\lambda=1}^{l_{k}}-G(\tilde{x}_{k,\lambda}^{+},\tilde{x}_{k,\lambda}^{-},s,|\vec{b}+\vec{b}_{\pi(k)}^{A}-\vec{b}_{\tau(k)}^{B}|)\bigg)\end{align*}

\vspace{-0.4cm}
\begin{align}
 & \quad\prod_{i=1}^{A}F_{\mathrm{remn}}\bigg(1-\sum_{\pi(k)=i}x_{k,\mu,}^{+}-\sum_{\pi(k)=i}\tilde{x}_{k,\lambda}^{+}\bigg)\nonumber \\
 & \quad\prod_{j=1}^{B}F_{\mathrm{remn}}\bigg(1-\sum_{\tau(k)=j}x_{k,\mu}^{-}-\sum_{\tau(k)=j}\tilde{x}_{k,\lambda}^{-}\bigg)\quad\Bigg\},\label{eq:master}\end{align}
where $A$ and $B$ are the number of nucleons of the two nuclei,
$(\vec{b}_{i}^{A/B},z_{i}^{A/B})$ the nucleon coordinates, $\rho^{A/B}$
the nuclear densities, $x_{k,\mu}^{+/-}$ and $\tilde{x}_{k,\mu}^{+/-}$
the light cone momentum fractions of respectively the cut and uncut
Pomerons. The functions $\pi(k)$ and $\tau(k)$ refer to the projectile
and target nucleon linked to nucleon-nucleon pair (or {}``collision
number'') $k$, and we use $F_{\mathrm{remn}}(x)=[x\theta(x)\theta(1-x)]^{\alpha}$
with $\theta$ being the Heaviside function, which ensures energy
conservation. This is the master formula of our approach, because
it allows to compute (doing partial summation) exclusive cross section
calculations for particular sub-processes. The formula is also valid
for p-p scattering, here we have simply $A=B=1$, and $\rho_{A/B}(\vec{x})=\delta(\vec{x})$.

The single Pomeron contribution $G$ is the imaginary part of the
transverse Fourier transform of the single Pomeron exchange amplitude
$T$ divided by the cms energy $\hat{s}$. The amplitude $T$ is given
as a sum over several terms, see \citet{hajo}. One contributions
is the soft one, $T_{\mathrm{soft}}$, corresponding to a soft Pomeron
exchange, parametrized in Regge pole fashion. The most important contribution
at high energies is the semihard contribution $T_{\mathrm{sea}-\mathrm{sea}}$,
with \begin{eqnarray}
iT_{\mathrm{sea}-\mathrm{sea}}(\hat{s},t) & = & \int_{0}^{1}\!\frac{dz^{+}}{z^{+}}\frac{dz^{-}}{z^{-}}\,\mathrm{Im}\, T_{\mathrm{soft}}\!\!\left(\frac{s_{0}}{z^{+}},t\right)\label{eq:t-sea-sea}\\
 &  & \,\mathrm{Im}\, T_{\mathrm{soft}}\!\!\left(\frac{s_{0}}{z^{-}},t\right)\, iT_{\mathrm{hard}}(z^{+}z^{-}\hat{s},t),\nonumber \end{eqnarray}
with the hard scattering amplitude $T_{\mathrm{hard}}$ given as

\begin{equation}
T_{\mathrm{hard}}=i\hat{s}\,\sigma_{hard}(\hat{s})\,\exp(R_{\mathrm{hard}}^{2}t),\label{eq:t-hard}\end{equation}
 and with \begin{eqnarray}
\sigma_{\mathrm{hard}}(\hat{s},Q_{0}^{2}) & = & \frac{1}{2\hat{s}}2\mathrm{Im}\, T_{\mathrm{hard}}\!\!\left(\hat{s},t=0\right)\nonumber \\
 & = & K\,\int dx_{B}^{+}dx_{B}^{-}dp_{t}^{2}\frac{d\sigma_{\mathrm{Born}}^{ml}}{dp_{t}^{2}}(x_{B}^{+}x_{B}^{-}\hat{s},p_{t}^{2})\nonumber \\
 &  & \: E_{\mathrm{QCD}}(x_{B}^{+},Q_{0}^{2},M_{F}^{2})\, E_{\mathrm{QCD}}(x_{B}^{-},Q_{0}^{2},M_{F}^{2})\nonumber \\
\nonumber \\ &  & \qquad\qquad\:\theta\!\left(M_{F}^{2}-Q_{0}^{2}\right),\label{eq:sigh}\end{eqnarray}
based on the fact that the real part of $T_{\mathrm{hard}}$ can be
neglected and its slope $R_{\mathrm{hard}}^{2}$ is very small \citet{gri-lip86,gri-rys92}
(and finally taken to be zero). The functions $E_{\mathrm{QCD}}$
represent the linear parton evolution, following the same evolution
equations as the usual parton distribution functions, but here the
initial condition is $E_{\mathrm{QCD}}\left(z,Q_{0}^{2},Q_{0}^{2}\right)=\delta(1-z).$
We use $M_{F}^{2}=p_{t}^{2}/4$. So far, $Q_{0}$ has been a constant,
but this will change as discussed below. 

In addition to the {}``sea-sea'' contribution as discussed above,
we have {}``val-val'', {}``sea-val'', and {}``val-sea'' (see
\citet{hajo}), where {}``sea'' and {}``val'' refer to sea or
valence quarks on respectively the projectile and target side, initiating
the parton ladder.

The formula eq. \ref{eq:sigh} looks very similar to the usual factorization
formula used to compute inclusive cross sections, but here we use
it to compute the single Pomeron T-matrix.

It is known \citet{hajo} that our formalism as described so far is
incomplete, for example total cross sections will grow power like
at high energies, violating the famous Froissart bound. The missing
element is an explicit treatment of non-linear effects concerning
the parton evolutions. It is known that parton saturation effects
play an important role \citet{sat1,sat2,sat3,sat4,sat5,sat6}, and
can be summarized by the so-called saturation scale $Q_{s}$, representing
the virtuality scale below which non-linear affects (like gluon-gluon
fusion) become important. Popular expressions for the $A$ and $x$
dependence (respectively mass number and longitudinal momentum fraction)
are \begin{equation}
Q_{s}^{2}\sim\frac{A^{1/3}}{x^{\lambda}},\end{equation}
or (for the centrality dependence)\begin{equation}
Q_{s}^{2}\sim\frac{N_{\mathrm{part}}}{x^{\lambda}},\end{equation}
with $N_{\mathrm{part}}$ being the number of participating nucleons. 

We adapt the above formulas to our formalism and use for each Pomeron
\begin{equation}
Q_{s}^{2}=B_{\mathrm{sat}}\frac{N_{\mathrm{part}}}{(1/\hat{s})^{\lambda}},\end{equation}
where $\hat{s}$ is the cms energy of the Pomeron and $N_{part}$
the number of participants. We use $\lambda=0.25$. These individual
scales $Q_{s}$ replace the constant values $Q_{0}$ in the above
formulas. The proportionality constant $B_{\mathrm{sat}}$ is chosen
to assure binary scaling in p-A and A-A at high $p_{t}$.

How to compute $N_{part}$? First one might think of estimating simply
the number of participating nucleons. For example for a given Pomeron
exchanged between projectile nucleon $i$ and target nucleon $j$,
one counts the projectile nucleons being closer to nucleon $j$ than
some transverse distance $b_{0}$, \begin{equation}
N_{\mathrm{part}}^{\mathrm{proj}}=\sum_{\mathrm{proj\, nucleons\, i'}}\Theta(b_{0}-|\vec{b}+\vec{b_{i'}}-\vec{b_{j}}|),\end{equation}
with $\vec{b}$ being the impact parameter, and with $\vec{b_{i'}}$
and $\vec{b_{j}}$ referring to the transverse positions of the nucleons
in the nuclei. A corresponding formula applies for the target participants.
We want to go further and estimate the number of participating partons,
since we expect already saturation effects in proton-proton scattering.
So we use actually\begin{equation}
N_{\mathrm{part}}^{\mathrm{proj}}=\sum_{\mathrm{proj\, nucleons\, i'}}\!\!\! f_{\mathrm{part}}\left(|\vec{b}+\vec{b_{i'}}-\vec{b_{j}}|\right),\end{equation}
with\begin{equation}
f_{\mathrm{part}}(b)=\Theta(b_{0}-b)\: g\left(A_{\mathrm{sat}}\exp(-b^{2}/4\pi\lambda_{\mathrm{soft}})\right),\end{equation}

\noindent where $\exp(-b^{2}/4\pi\lambda_{\mathrm{soft}})$ is our
{}``usual'' $b$--dependence of the single Pomeron amplitudes, with
$\lambda_{\mathrm{soft}}=2R_{\mathrm{part}}^{2}+\alpha'\!_{\mathrm{soft}}\ln\!(s/s_{0}).$
Here, $R_{\mathrm{part}}$ and $\alpha'_{\mathrm{soft}}$ are soft
Pomeron parameters (see \citet{hajo}). The phenomenological function
$g(x)=x/(1-\exp(-x))$ is the average of a Poisson distribution with
at least one scattering. We compute correspondingly the number of
target participants, and then we define $N_{\mathrm{part}}$ to be
the maximum of the two numbers $N_{\mathrm{part}}^{\mathrm{proj}}$
and $N_{\mathrm{part}}^{\mathrm{targ}}$. %
\begin{figure}[tb]
\begin{centering}
\includegraphics[angle=270,scale=0.24]{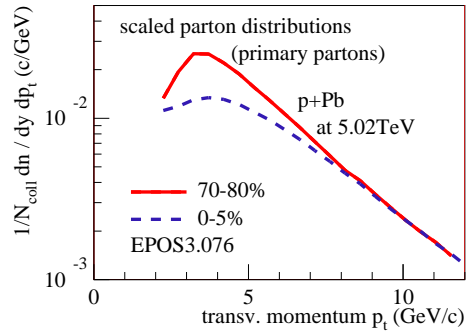}
\par\end{centering}

\caption{(Color online) Scaled parton distributions as a function of $p_{t}$,
for central (0-5\%) and peripheral (70-80\%) p-Pb collisions at 5TeV.\label{fig:scal}}

\end{figure}

The effect of the saturation scale can be seen clearly when comparing
transverse momentum distributions of primary partons (originating
from the hard scattering process) for central and peripheral p-Pb
collisions, see fig. \ref{fig:scal}. We plot scaled distributions,
i.e. the p-A results divided by the number $N_{\mathrm{coll}}$ of
binary collisions. Whereas the two curves coincide for large $p_{t}$,
the low $p_{t}$ region in the central case is significantly suppressed
compared to the peripheral one. In other words: We obtain a different
centrality dependence of the parton multiplicity $n$ at low and high
$p_{t}$:

\begin{itemize}
\item \noindent $n\{\mathrm{high}\, p_{t}\}$ grows with $N_{\mathrm{coll}}$
(= binary scaling)
\item \noindent $n\{\mathrm{low}\, p_{t}\}$ grows less than $N_{\mathrm{coll}}$
(the statement also holds for the integrated multiplicity).
\end{itemize}
\noindent There is considerable confusion about these different scaling
behaviors at low and high $p_{t}.$ It is often referred to as soft
and hard contributions, with the soft one scaling as the number of
participants (or wounded nucleons). But this is an old concept from
low energy scattering, where projectile and target fragmentation play
a role ad mid-rapidity, which is not at all the case in the TeV energy
domain. To be clear: here there is nothing soft, we just have more
or less screening at different $p_{t}$, governed by the saturation
scale.

Introducing a saturation scale to account for nonlinear effects is
new in EPOS3, and it replaces the procedures introduced in \citet{kwladsplit}
and used in EPOS2. Whereas the new procedure is very simple and clear
concerning its definition, the numerical implementation turned out
to be very difficult, due to the fact that we use for many quantities
pre-fabricated tables, allowing to do fast interpolations during the
Monte Carlo iterations (as a reminder: as explained in \citet{hajo},
we use Metropolis techniques to deal with the multidimensional phase
space). But only the new method gives a consistent picture, and provides
what is expected from common sense, like binary scaling at high $p_{t}$,
which is not the case in the old method. The latter one is in particular
unable (for whatever parameter choice) to reproduce experimental p-Pb
results at the LHC, showing a nuclear modification factor (rescaled
p-Pb / p-p) to be unity at large $p_{t}$, whereas the new method
perfectly reproduced these data.

The saturation scale procedure is therefore a substantial improvement
of our scheme, not only compatible with new theoretical developments
during the past two decades \citet{sat1,sat2,sat3,sat4,sat5,sat6},
but also allowing a self-consistent treatment of soft and hard physics
in a unique approach.

\section{Flux tubes}

\noindent Our master formula eq. (\ref{eq:master}) allows to compute
total cross sections, and (even more importantly) partial cross sections
for particular multiple scattering configurations in p-p, p-A, and
AA scatterings. The corresponding integrands can be interpreted as
probability distributions of such configurations, and serve as basis
of Monte Carlo applications (see \citet{hajo}). Here, contrary to
many other Monte Carlo calculations, our events are real physical
events, there is no need to introduce {}``test particles'', all
kinds of fluctuations can be treated based on event-by-event fluctuations. 

\begin{figure}[b]
\begin{centering}
\includegraphics[scale=0.2]{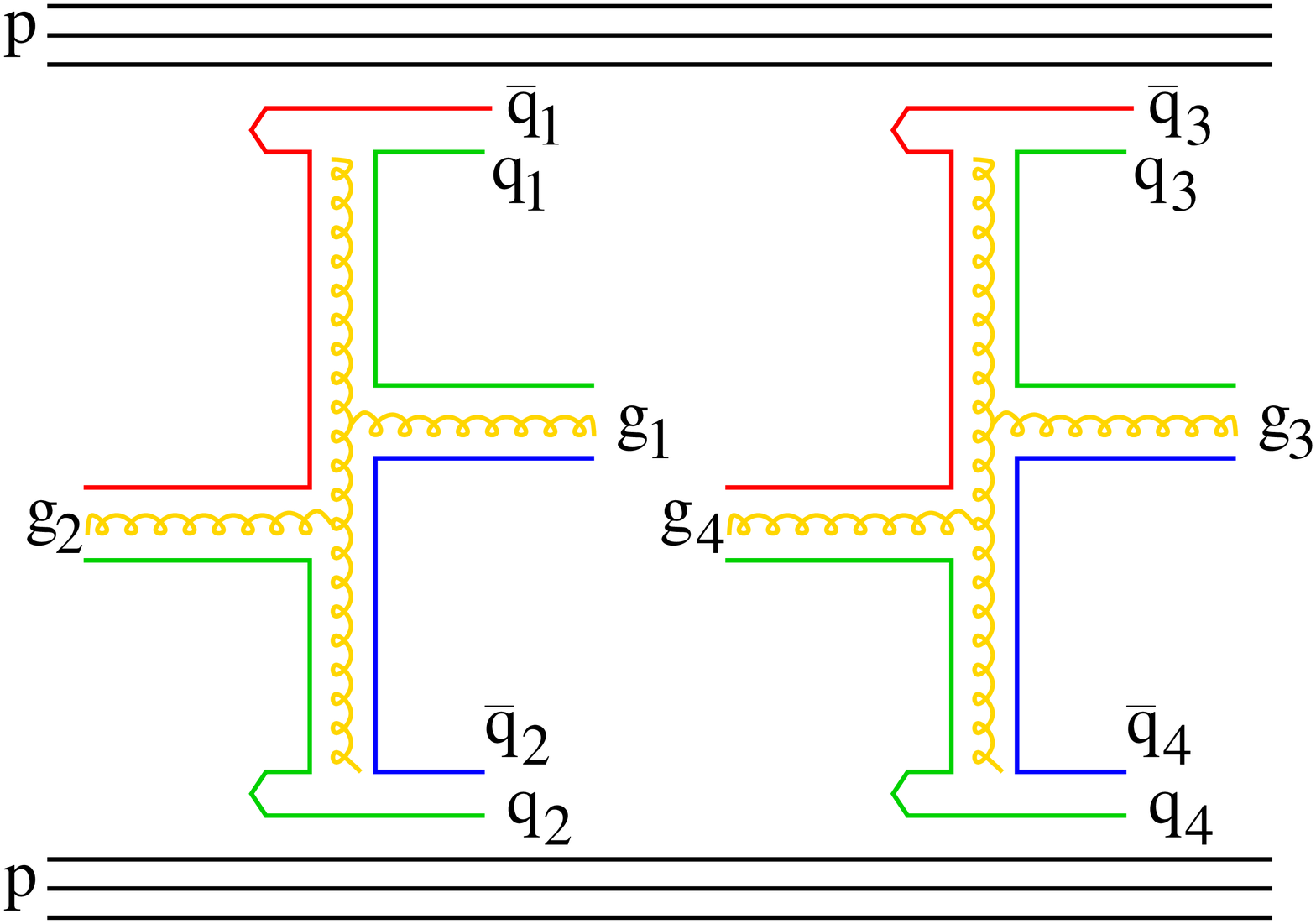}
\par\end{centering}

\caption{(Color online) Color flow, for double Pomeron exchange, for a simplified
scattering, without initial and final state cascade.\label{ldfmultcol.eps}}

\end{figure}

Generating an event is done in several steps: 

\begin{itemize}
\item Step 1 amounts to generate the multiple scattering configuration according
to eq. (\ref{eq:master}), characterized by the number of cut Pomerons
per possible nucleon-nucleon pair, and the light cone momentum fractions
$x^{\pm}$ of the Pomeron ends. For example for Au-Au or Pb-Pb collisions,
with around 40000 nucleon-nucleon pairs, one has up to $10^{6}$variables
to generate, which requires sophisticated Monte Carlo methods \citet{hajo}.
\item Step 2 amounts to generate, for a given configuration, the partons
associated to each Pomeron, based on the expressions representing
a cut Pomeron, eqs. (\ref{eq:t-sea-sea},\ref{eq:t-hard},\ref{eq:sigh}).
This time we are not using the integrals in these equations (needed
in step 1), but their integrands, which serve as probability distributions.
\end{itemize}
\noindent The chain of partons corresponding to a given Pomeron is
referred to as parton ladder.  These ladders are identified with flux
tubes, as explained in \citet{hajo}. As a first step, for a given
scattering, one considers the color flow. In Fig. \ref{ldfmultcol.eps}),
we show as an example two cut Pomerons of the {}``sea-sea'' contribution,
with a simple $gg\to gg$ elementary scattering, without initial and
final state cascade. The projectile and target remnants stay always
color-neutral (they simply become excited). The actual interactions
concern sea quarks, in the example of Fig. \ref{ldfmultcol.eps} the
(anti)quarks $q_{1}$,$\bar{q}_{1}$ $q_{2}$,$\bar{q}_{2}$ for the
first Pomeron, and $q_{3}$,$\bar{q}_{3}$ $q_{4}$,$\bar{q}_{4}$
for the second one. As a first step, one considers the color flow,
shown in Fig. \ref{ldfmultcol.eps} by the red, blue, and green lines.
Once the color flow is identified, one follows the line from a quark,
via intermediate gluons, till an antiquark is found. In the example,
we have $q_{1}-g_{1}-\bar{q}_{2}$ and $q_{2}-g_{2}-\bar{q}_{1}$
for the first Pomeron, and $q_{3}-g_{3}-\bar{q}_{4}$ and $q_{4}-g_{4}-\bar{q}_{3}$
for the second one. Each of these four parton sequences is identified
with a so-called {}``kinky string''.

\noindent The relativistic string picture \citet{string1,string2,string3}
is very attractive, because its dynamics is essentially derived from
general principles as covariance and gauge invariance. The simplest
possible string is a surface $X(\alpha,\beta)$ in 3+1 dimensional
space-time, with piecewise constant initial velocities $\partial X/\partial\beta$.
These velocities are identified with parton velocities,%
\begin{figure}[b]
\begin{raggedright}
{\large \hspace*{1.2cm}(a)}
\par\end{raggedright}{\large \par}

\vspace*{-0.5cm}

\begin{centering}
\includegraphics[scale=0.1]{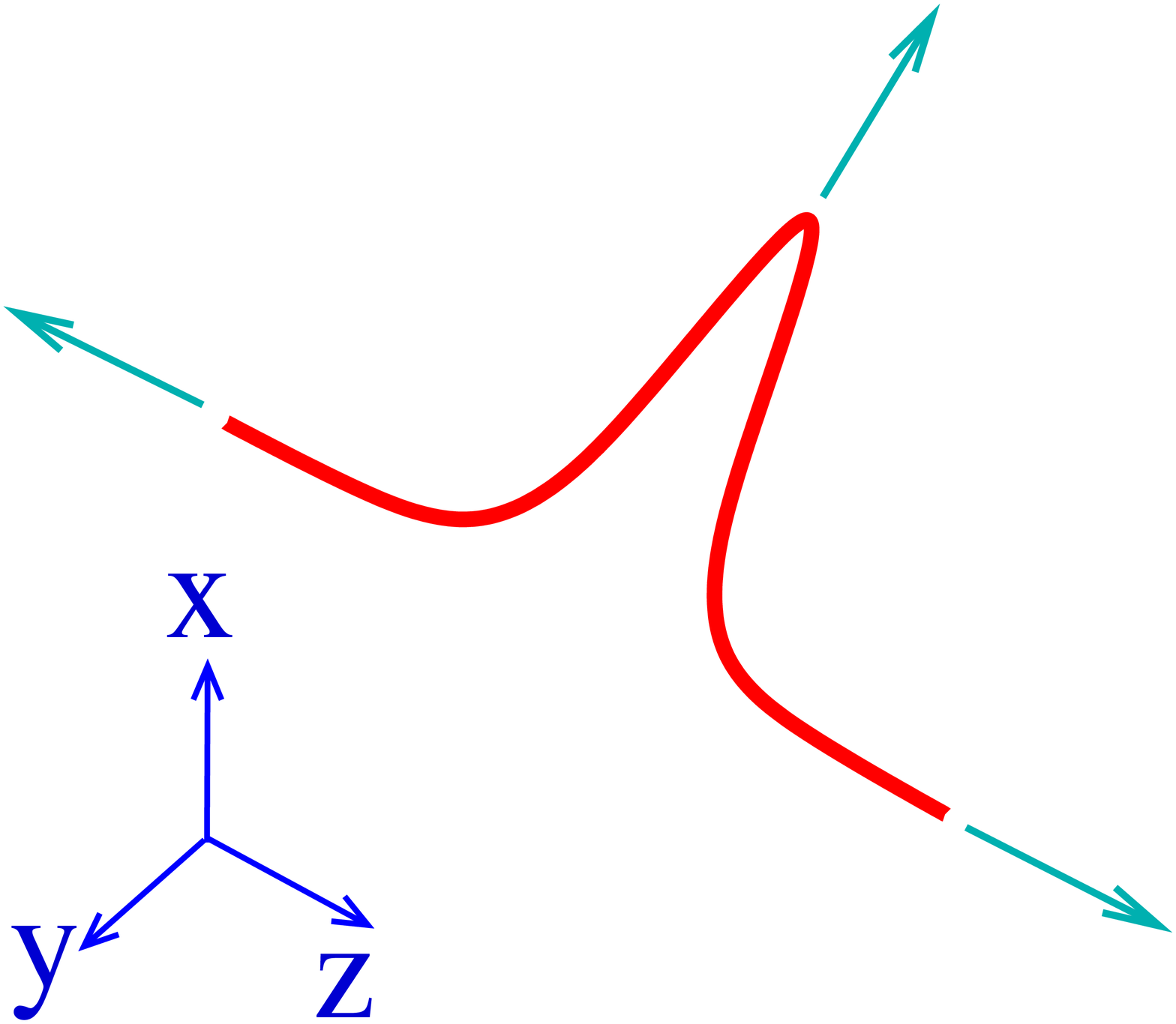}
\par\end{centering}

\begin{raggedright}
{\large \hspace*{1.2cm}(b)}
\par\end{raggedright}{\large \par}

\vspace*{-0.5cm}

\begin{centering}
\includegraphics[scale=0.15]{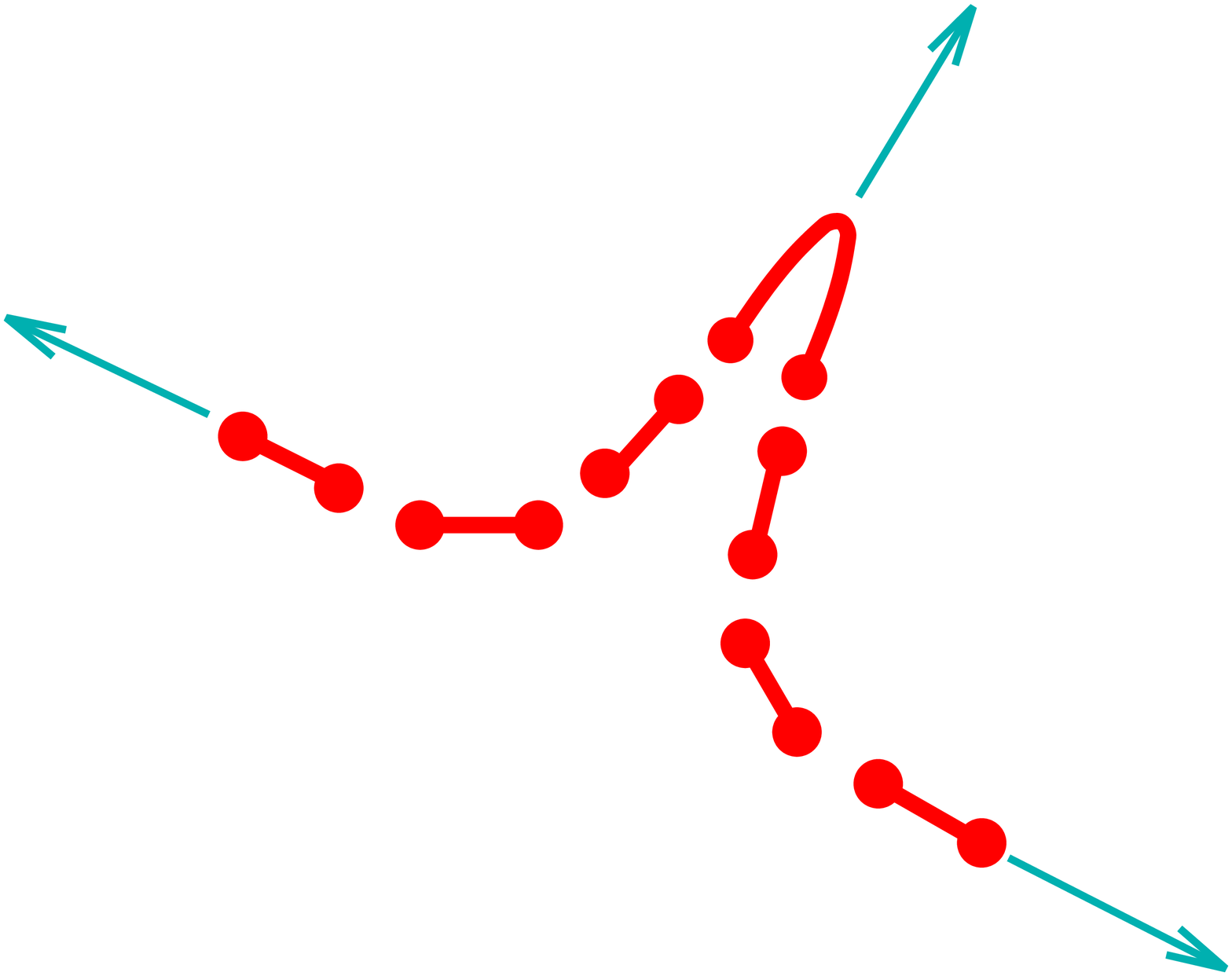}
\par\end{centering}

\caption{(Color online) (a) Flux tube with transversely moving part (kinky
string) in space, at given proper time. \protect \\
(b) Flux tube breaking via $q-\bar{q}$ production, which screens
the color field (Schwinger mechanism).\label{stgfrag1.eps}}

\end{figure}
which provides a one to one mapping from partons to strings. For details
see \citet{hajo,epos2}. In the above example, we have four strings
with a single kink each. The physical picture behind the {}``kinky
string'' is an essentially one-dimensional {}``color flux tube''
(with eventually a finite but very small transverse dimension). 

\noindent The high transverse momentum ($p_{t})$ partons will show
up as transversely moving string pieces, see Fig. \ref{stgfrag1.eps}(a).
Despite the fact that in the TeV energy range most processes are hard,
and despite the theoretical importance of very high $p_{t}$ partons,
it should not be forgotten that the latter processes are rare, most
kinks carry only few GeV of transverse momentum, and the energy is
nevertheless essentially longitudinal. In case of elementary reactions,
the strings will break (see Fig. \ref{stgfrag1.eps}(b) via the production
of quark-antiquark pairs according to the so-called area law \citet{artru2,mor87,hajo,epos2}.
The string segments are identified with final hadrons and resonances. 

\begin{figure}[tb]
\begin{centering}
\includegraphics[angle=270,scale=0.24]{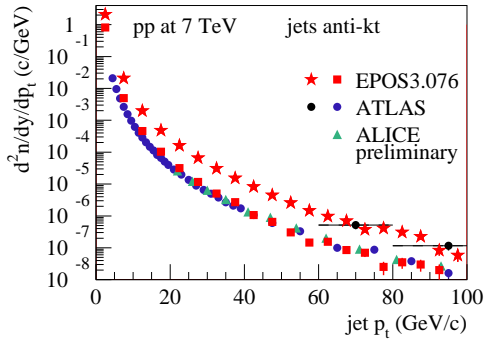}
\par\end{centering}

\caption{(Color online) Inclusive $p_{t}$ distribution of jets. We show the
calculation (red stars) compared to ATLAS data \citet{atlasjet} (black
circles). We also show the calculated $p_{t}$ distribution of charged
particle jets (red squares) compared to data from ATLAS (blue circles)
\citet{atlasjet2} and ALICE (green triangles) \citet{alicejet}.
\label{fig:jets}}

\end{figure}
This picture has been very successful to describe particle production
in electron-positron annihilation or in proton-proton scattering at
very high energies. In the latter case, not only low $p_{t}$ particles
are described correctly, for example for p-p scattering at 7 TeV \citet{kw1,kw2},
also jet production is covered. As discussed earlier, the high transverse
momenta of the hard partons show up as kinks, transversely moving
string regions. After string breaking, the string pieces from these
transversely moving areas represent the jets of particles associated
with the hard partons. To demonstrate that this picture also works
quantitatively, we compute the inclusive $p_{t}$ distribution of
jets, reconstructed with the anti-kt algorithm \citet{antikt} and
compare with data from ATLAS \citet{atlasjet,atlasjet2} and ALICE
\citet{alicejet}, see Fig. \ref{fig:jets}.

\section{Core-corona procedure for proton-nucleus scattering}

In heavy ion collisions and also in high multiplicity events in proton-proton
and proton-nucleus scattering at very high energies, the density of
strings will be so high that the strings cannot decay independently
as described above. Here we have to modify the procedure as discussed
in the following. The starting point are still the flux tubes (kinky
strings) discussed earlier. Some of these flux tubes will constitute
bulk matter which thermalizes and expands collectively -- this is
the so-called {}``core''. Other segments, being close to the surface
or having a large transverse momentum, will leave the {}``bulk matter''
and show up as hadrons (including jet-hadrons), this is the so-called
{}``corona''. 

In principle the core--corona separation is a dynamical process. However,
the knowledge of the initial transverse momenta $p_{t}$ of string
segments and their density $\rho(x,y)$ allows already an estimate
about the fate of these string segments. By {}``initial'' we mean
some early proper time $\tau_{0}$ which is a parameter of the model.
In a first version of this {}``core-corona'' approach \citet{core},
the core was simply defined by the string segment density (being bigger
than some critical density $\rho_{0}$). More recently \citet{jetbulk},
we also considered the transverse momentum of the segments, to allow
high $p_{t}$ segment to leave the bulk part. This procedure was able
to describe flow features and jet production at the same time. 

Whereas our core-corona procedures (old and new ones) are always based
on flux-tubes (coming from Gribov-Regge multiple scattering), there
are also core-corona models \citet{core-bec,core-aich1,core-aich2,core-aich3}
based on the {}``wounded nucleon approach'' , where the core multiplicity
is proportional to the number of participating nucleons having suffered
at least two collisions, whereas the nucleons colliding only once
contribute to the corona.

%
{}

In our new core-corona procedure, for the moment optimized for p-p
and p-A scattering, string segments constitute bulk matter or escape,
depending on their transverse momenta $p_{t}$ and the local string
density $\rho$. We compute for each string segment \begin{equation}
p_{t}^{\mathrm{new}}=p_{t}-f_{\mathrm{Eloss}}\int_{\gamma}\rho\, dL,\end{equation}
where $\gamma$ is the trajectory of the segment, and $f_{\mathrm{Eloss}}$
a nonzero constant for $p_{t}<p_{t,1}$, zero for $p_{t}>p_{t,2}$,
and with a linear interpolation between $p_{t,1}$ and $p_{t,2}$.
If a segment has a positive $p_{t}^{\mathrm{new}}$, it is allowed
to escape -- it is a corona particle. Otherwise, the segment contributes
to the core. 

\begin{figure}[tb]
\begin{centering}
\includegraphics[angle=270,scale=0.24]{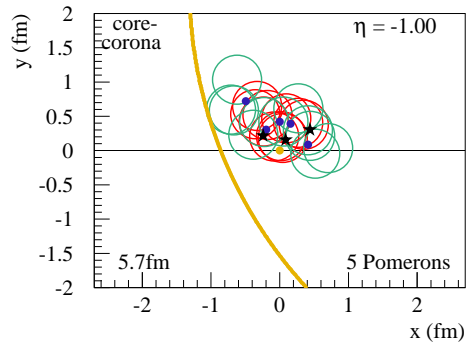}
\par\end{centering}

\caption{(Color online) Core-corona separation in a randomly chosen p-Pb event
(with an impact parameter of 5.7 fm), in the transverse plane at space-time
rapidity $\eta_{s}=-1$. We show the positions of the projectile nucleon
(yellow dot), the Pb surface (yellow line), the hit target nucleons
(stars), Pomerons (blue dots), as well as the core (red circles) and
the corona (green circles) string segments. \label{fig:core-corona}}

\end{figure}

In fig. \ref{fig:core-corona},%
we show as an example the core-corona separation in a randomly chosen
p-Pb event, by plotting the transverse plane at space-time rapidity
$\eta_{s}=-1$. %
\begin{figure}[tb]
\includegraphics[angle=270,scale=0.58]{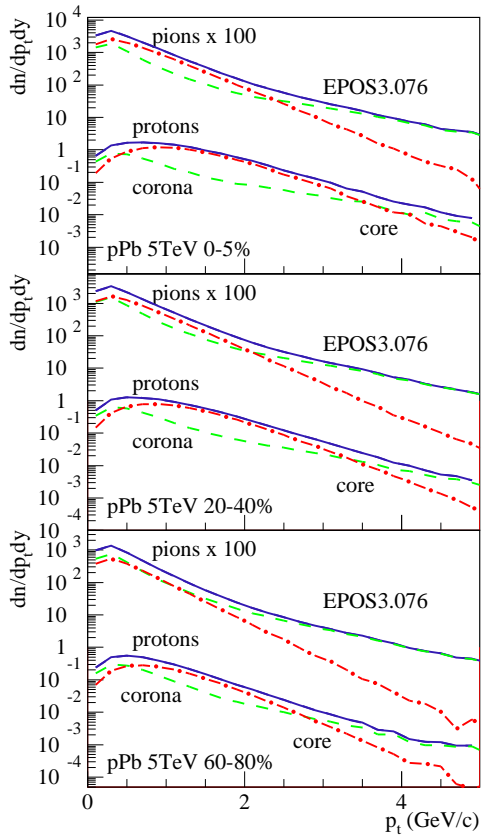}

\caption{(Color online) Core (red dashed-dotted lines) and corona contributions
(green dashed lines) to the production of pions (upper curves, multiplied
by 100) and protons (lower curves), for different centralities in
p-Pb collision at 5TeV. The blue solid lines are the sum of core and
corona. The calculations are done based on the hydrodynamical evolution
as described in the next chapter, without employing a hadronic cascade.
\label{z-selb-10.ps}}

\end{figure}
The yellow dot at $x=y=0$ is the position of the projectile proton,
the yellow line represents the target Pb surface (considering a hard
sphere for the plot, whereas all calculations are done with a realistic
Wood-Saxon distribution). The black stars mark the nucleons of the
Pb nucleus, hit by the proton. The blue dots mark the transverse positions
of the Pomerons, the flux-tubes are scattered around these Pomeron
points. Flux-tube segments contributing to the core are shown as red
circles, the green ones represent the corona. The latter ones will
show up as hadrons, whereas the core provides the initial condition
of a hydrodynamical evolution (discussed in the next chapter), where
the particles will be produces later at {}``freeze-out'' from the
flowing medium, which occurs at some {}``hadronization temperature''
$T_{H}$ \citet{epos2}. After this {}``hadronization'' the hadrons
still interact among each other, realized via a hadronic cascade procedure
\citet{urqmd}, already discussed in \citet{epos2}.

In fig. \ref{z-selb-10.ps}, we show how core and corona contribute
to the production of pions and protons, for different centralities
(based on impact parameter). The corona contributions dominate completely
the high $p_{t}$ regions, for all centralities. For central collisions
(0-5\%), the core dominates for both pion and protons at low $p_{t}$,
but the dominance (core over corona) is much more pronounced for protons,
and the crossing (core=corona) happens at larger $p_{t}$ (3.5 GeV/c)
for the protons compared to pions (2-2.5 GeV/c). The fact that the
core is much more visible in protons compared to pions is a consequence
of radial flow: when particles are produced in a radially flowing
medium, the heavier particles acquire more transverse momentum than
the light ones. It is a mass effect (lambdas look similar to protons,
kaons are in between pions and protons). Going to more peripheral
collisions, the flow effects get smaller, but even for peripheral
events (60-80\%), we still have flow (actually even in p-p!).

\section{Viscous Hydrodynamics }

The core extracted as described above provides the initial condition
for a hydrodynamic evolution. As explained in \citet{epos2}, we compute
the energy momentum tensor and the flavor flow vector at some position
$x$ (at $\tau=\tau_{0}$) from the four-momenta of the bulk string
segments. The time $\tau=\tau_{0}$ is as well taken to be the initial
time for the hydrodynamic evolution. This seems to be a drastic simplification,
the justification being as follows: we imagine to have a purely longitudinal
scenario (described by flux tubes) till some proper time $\tau_{\mathrm{flux}}<\tau_{0}$.
During this stage there is practically no transverse expansion, and
the energy per unit of space-time rapidity does not change. This property
should not change drastically beyond $\tau_{\mathrm{flux}}$, so we
assume it will continue to hold during thermalization between $\tau_{\mathrm{flux}}$
and $\tau_{0}$. So although we cannot say anything about the precise
mechanism which leads to thermalization, and therefore we cannot compute
the real $T^{\mu\nu}$, we expect at least the elements $T^{00}$
and $T^{0i}$ to stay close to the flux tube values, and we can use
the flux tube results to compute the energy density, as explained
in the following.. 

Based on the four-momenta of string segments, we compute the energy
momentum tensor and the flavor flow vector at some position $x$ (at
$\tau=\tau_{0}$) as \citet{epos2} \textcolor{black}{\begin{eqnarray}
T^{\mu\nu}(x) & = & \sum_{i}\frac{\delta p_{i}^{\mu}\delta p_{i}^{\nu}}{\delta p_{i}^{0}}g(x-x_{i}),\\
N_{q}^{\mu}(x) & = & \sum_{i}\frac{\delta p_{i}^{\mu}}{\delta p_{i}^{0}}\, q_{i}\, g(x-x_{i}),\end{eqnarray}
 where $q\in{u,d,s}$ represents the net flavor content of the string
segments, and\begin{equation}
\delta p=\left\{ \frac{\partial X(\alpha,\beta)}{\partial\beta}\delta\alpha+\frac{\partial X(\alpha,\beta)}{\partial\alpha}\delta\beta\right\} \quad\end{equation}
are the four-momenta} of the segments. The function $g$ is a Gaussian
smoothing kernel with a transverse width $\sigma_{\bot}$= 0.25 fm.
The Lorentz transformation into the comoving frame gives\begin{equation}
\Lambda^{\alpha}\,_{\mu}\Lambda^{\beta}\,_{\nu}T^{\mu\nu}=T_{\mathrm{com}}^{\mu\nu},\end{equation}
where we define the comoving frame such that the first column of $T_{\mathrm{com}}$
is of the form $(\varepsilon,0,0,0)^{T}$.  This provides four equations
for the energy density $\varepsilon$ in the comoving frame, and the
flow velocity components $v^{i}$ , which may be solved iteratively
\citet{epos2}. The flavor density is then calculated as\begin{equation}
f_{q}=N_{q}u,\end{equation}
with $u$ being the flow four-velocity. 

For the hydrodynamic calculations, we construct the equation of state
as\begin{equation}
p=p_{Q}+\lambda\,(p_{H}-p_{Q}),\end{equation}
where $p_{H}$ is the pressure of a resonance gas, and $p_{Q}$ the
pressure of an ideal quark gluon plasma, including bag pressure. We
use\begin{equation}
\lambda=\exp\left(-z-3z^{2}\right)\Theta(T-T_{c})+\Theta(T_{c}-T),\end{equation}
with\begin{equation}
z=x/(1+x/0.77),\quad x=(T-T_{c})/\delta,\end{equation}
using $\delta=0.24\exp(-\mu_{b}^{2}/0.4^{2})$.%
\begin{figure}[tb]
\begin{centering}
\includegraphics[angle=270,scale=0.24]{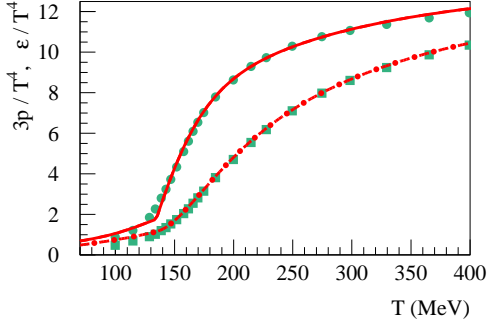}
\par\end{centering}

\caption{(Color online) Energy density and pressure versus temperature, for
our equation-of-state (lines) compared to lattice data \citet{lattice}
(points). \label{cap:eos1}}

\end{figure}
This $\lambda$ provides an equation of state in agreement with recent
lattice data \citet{lattice}, see Fig. \ref{cap:eos1}. 

In the following, we describe the 3+1D viscous hydrodynamic approach
and the corresponding hydrodynamic component of the EPOS3 model, which
we call vHLLE (viscous HLLE-based algorithm) \citet{yuri}. For the
hydrodynamic evolution we choose Milne coordinates (and its natural
frame) for the $t-z$ plane in space-time ($z$ being the collision
axis). The new coordinates are expressed in terms of Minkowski coordinates
$\{t,x,y,z\}$ as $\tau=\sqrt{t^{2}-z^{2}}$, $\eta=\frac{1}{2}\mathrm{ln}(t+z)/(t-z)$,
while the definitions of $x$ and $y$ coordinates are unchanged.
The transformation tensor $M$ is\begin{equation}
\left\{ M_{\,\nu}^{\mu}\right\} =\left(\begin{array}{cccc}
\cosh\eta & \quad0\quad & \quad0\quad & \quad-\sinh\eta\\
0 & \quad1\quad & \quad0\quad & \quad0\\
0 & \quad0\quad & \quad1\quad & \quad0\\
-\frac{1}{\tau}\sinh\eta & \quad0\quad & \quad0\quad & \quad\frac{1}{\tau}\cosh\eta\end{array}\right).\end{equation}
We choose $(+,-,-,-)$ signature of $g_{\mu\nu}$ in Minkowski space-time,
so in Milne coordinates the invariant interval is $ds^{2}=dt^{2}-dx^{2}-dy^{2}-\tau^{2}d\eta^{2}$
and the metric tensor is \begin{equation}
g^{\mu\nu}=\mathrm{diag}(1,-1,-1,-1/\tau^{2}).\end{equation}
Although space-time is still flat, there are nontrivial Christoffel
symbols, the nonzero components being \begin{equation}
\Gamma_{\tau\eta}^{\eta}=\Gamma_{\eta\tau}^{\eta}=1/\tau,\quad\Gamma_{\eta\eta}^{\tau}=\tau.\end{equation}

The hydrodynamic equations are given as: \begin{equation}
\dd_{;\nu}T^{\mu\nu}=\dd_{\nu}T^{\mu\nu}+\Gamma_{\nu\lambda}^{\mu}T^{\nu\lambda}+\Gamma_{\nu\lambda}^{\nu}T^{\mu\lambda}=0\label{hydro-eqns}\end{equation}
All source terms in (\ref{hydro-eqns}) coming from $\Gamma_{\nu\lambda}^{\nu}T^{\mu\lambda}$
are proportional to $1/\tau$, which makes them dominant for the hydrodynamic
evolution at small enough $\tau$ and would eventually require to
apply a higher order numerical time integration scheme. We circumvent
this by defining $\tilde{T}^{\mu\nu}$ via \begin{equation}
\begin{cases}
T^{\mu\nu}=\tilde{T}^{\mu\nu},\ \quad\,\mu,\nu\neq\eta\,,\\
T^{\mu\eta}=\tilde{T}^{\mu\eta}/\tau,\ \;\mu\neq\eta\,\,,\\
T^{\eta\eta}=\tilde{T}^{\eta\eta}/\tau^{2}\quad,\end{cases}\end{equation}
and we obtain from eq. (\ref{hydro-eqns}) the following equations
for $\tau\tilde{T}^{\mu\nu}$: \begin{align}
 & \tilde{\dd}_{\nu}(\tau\tilde{T}^{\tau\nu})+\frac{1}{\tau}(\tau\tilde{T}^{\eta\eta})=0\,,\nonumber \\
 & \tilde{\dd}_{\nu}(\tau\tilde{T}^{x\nu})=0\:.,\nonumber \\
 & \tilde{\dd}_{\nu}(\tau\tilde{T}^{y\nu})=0\:,\label{eqns1}\\
 & \tilde{\dd}_{\nu}(\tau\tilde{T}^{\eta\nu})+\frac{1}{\tau}\tau\tilde{T}^{\eta\tau}=0\,,\nonumber \end{align}
with\begin{equation}
\tilde{\dd}\equiv\bigg(\dd/\dd\tau,\dd/\dd x,\dd/\dd y,(1/\tau)\dd/\dd\eta\bigg),\end{equation}
and thus all the $\tilde{T}^{\mu\nu}$ have the same units, as well
as $\tilde{\dd}_{\mu}$-- namely {[}1/length]. The actual conservative
variables used in the code are therefore $Q^{\mu}=\tau\cdot\tilde{T}^{\mu\tau}$,
fluxes are $\tau\cdot\tilde{T}^{ij}$. Then (\ref{eqns1}) are the
explicit form of energy-momentum conservation equations which are
solved numerically.

The energy-momentum tensor can be expanded in the case of a viscous
fluid as \begin{equation}
T^{\mu\nu}=\epsilon u^{\mu}u^{\nu}-(p+\Pi)\Delta^{\mu\nu}+\pi^{\mu\nu},\end{equation}
where $\Delta^{\mu\nu}=g^{\mu\nu}-u^{\mu}u^{\nu}$ is the projector
orthogonal to $u^{\mu}$ ($u^{\mu}$ being the flow field), and finally
$\pi^{\mu\nu}$ and $\Pi$ are the shear stress tensor and bulk pressure,
respectively. Expressing the four-velocities $u_{MS}$ in Minkowski
space-time in terms of the longitudinal / transverse rapidities as\begin{align}
u_{MS}= & \big(\cosh\eta_{f}\cosh\eta_{T}\,,\,\sinh\eta_{T}\cos\phi\,,\nonumber \\
 & \quad\sinh\eta_{T}\sin\phi\,,\,\sinh\eta_{f}\cosh\eta_{T}\big),\end{align}
we get in Milne coordinates $u^{\mu}=M_{\,\nu}^{\mu}u_{MS}^{\nu},$
which gives\begin{align}
u= & \big(\,\cosh(\eta_{f}-\eta)\cosh\eta_{T}\,,\,\sinh\eta_{T}\cos\phi\,,\nonumber \\
 & \quad\sinh\eta_{T}\sin\phi\,,\,\tau^{-1}\sinh(\eta_{f}-\eta)\cosh\eta_{T}\,\big).\label{umu}\end{align}
Defining $\tilde{u}=(u^{\tau},u^{x},u^{y},\tau u^{\eta})$, we have
$\tilde{T}^{\eta\eta}=(\epsilon+p+\Pi)\tilde{u}^{\eta}\tilde{u}^{\eta}+(p+\Pi)+\pi^{\eta\eta}$,
so both $\tilde{u}$ and $\tilde{T}^{\eta\eta}$ do not include the
factor $1/\tau$ any more.

The hydrodynamic code used solves the equations of relativistic viscous
hydrodynamics in Israel-Stewart framework \citet{IsraelStewart}.
In particular we solve the following equations for the shear stress
tensor and bulk pressure, neglecting vorticity terms: \begin{align}
<u^{\gamma}\dd_{;\gamma}\pi^{\mu\nu}> & =-\frac{\pi^{\mu\nu}-\pi_{\text{NS}}^{\mu\nu}}{\tau_{\pi}}-\frac{4}{3}\pi^{\mu\nu}\dd_{;\gamma}u^{\gamma},\label{evolutionShear}\\
u^{\gamma}\dd_{;\gamma}\Pi & =-\frac{\Pi-\Pi_{\text{NS}}}{\tau_{\Pi}}-\frac{4}{3}\Pi\dd_{;\gamma}u^{\gamma},\label{evolutionBulk}\end{align}
\vspace{-0.6cm}

\noindent \vspace{-0.3cm}
where \[
<A^{\mu\nu}>=(\frac{1}{2}\Delta_{\alpha}^{\mu}\Delta_{\beta}^{\nu}+\frac{1}{2}\Delta_{\alpha}^{\nu}\Delta_{\beta}^{\mu}-\frac{1}{3}\Delta^{\mu\nu}\Delta_{\alpha\beta})A^{\alpha\beta}\]
 denotes the symmetric and traceless part of $A^{\mu\nu}$ being orthogonal
to $u^{\mu}$, and \begin{align}
\pi_{\text{NS}}^{\mu\nu} & =\eta(\Delta^{\mu\lambda}\dd_{;\lambda}u^{\nu}+\Delta{\nu\lambda}\dd_{;\lambda}u^{\mu})-\frac{2}{3}\eta\Delta^{\mu\nu}\dd_{;\lambda}u^{\lambda},\nonumber \\
\Pi_{\text{NS}} & =-\zeta\dd_{;\lambda}u^{\lambda}\end{align}
 are the values of the shear stress tensor and bulk pressure in the
limiting Navier-Stokes case.

For the purpose of our current study, we do not include the baryon/electric
charge diffusion. The same choice for the evolutionary equations (\ref{evolutionShear})
was used in the recent studies of p-A collisions employing relativistic
viscous hydrodynamics \citet{hydro-schenke,hydro-bozek}.

In the same way as was done for $T^{\mu\nu}$, we separate the factors
$1/\tau$ from $\pi^{\mu\nu}$ by defining $\tilde{\pi}^{\mu\nu}$
as follows:\begin{equation}
\begin{cases}
\pi^{\mu\nu}=\tilde{\pi}^{\mu\nu},\ \quad\,\mu,\nu\neq\eta\,,\\
\pi^{\mu\eta}=\tilde{\pi}^{\mu\eta}/\tau,\ \;\mu\neq\eta\,\,,\\
\pi^{\eta\eta}=\tilde{\pi}^{\eta\eta}/\tau^{2}\quad.\end{cases}\end{equation}
We rewrite (\ref{evolutionShear},\ref{evolutionBulk}) in the form
used for the numerical solution: \begin{align}
\tilde{\gamma}\left(\dd_{\tau}+\tilde{v}^{i}\tilde{\dd}_{i}\right)\tilde{\pi}^{\mu\nu} & =-\frac{\tilde{\pi}^{\mu\nu}-\tilde{\pi}_{\text{NS}}^{\mu\nu}}{\tau_{\pi}}+I_{\pi}^{\mu\nu}\label{evolShearFinal}\\
\tilde{\gamma}\left(\dd_{\tau}+\tilde{v}^{i}\tilde{\dd}_{i}\right)\Pi & =-\frac{\Pi-\Pi_{\text{NS}}}{\tau_{\Pi}}+I_{\Pi}\label{evolBulkFinal}\end{align}
 where $\tilde{\gamma}=u^{0}$ and $\tilde{v}^{i}=\tilde{u}^{i}/u^{0}$
($i=x,y,\eta$) are the components of 3-velocity. The additional source
terms are:\begin{equation}
I_{\pi}^{\mu\nu}\!=\!-\frac{4}{3}\tilde{\pi}^{\mu\nu}\tilde{\dd}_{;\gamma}\tilde{u}^{\gamma}-[\tilde{u}^{\nu}\tilde{\pi}^{\mu\beta}\!+\!\tilde{u}^{\mu}\tilde{\pi}^{\nu\beta}]\tilde{u}^{\lambda}\tilde{\dd}_{;\lambda}\tilde{u}_{\beta}-I_{\pi,G}^{\mu\nu}\label{Ipi}\end{equation}
\vspace{-1cm}
\begin{equation}
I_{\Pi}=-\frac{4}{3}\Pi\tilde{\dd}_{;\gamma}\tilde{u}^{\gamma}\end{equation}
 with $\tilde{\dd}_{;\gamma}\tilde{u}^{\gamma}=\tilde{\dd}_{\gamma}\tilde{u}^{\gamma}+u^{\tau}/\tau$.
The terms $I_{\pi,G}^{\mu\nu}$ denote geometrical source terms (coming
from Christoffel symbols), given as \begin{align*}
 & I_{\pi,G}^{\tau\tau}=2\tilde{u}^{\eta}\tilde{\pi}^{\tau\eta}/\tau & I_{\pi,G}^{\tau x}=\tilde{u}^{\eta}\tilde{\pi}^{\eta x}/\tau\\
 & I_{\pi,G}^{\tau y}=\tilde{u}^{\eta}\tilde{\pi}^{\eta y}/\tau & \qquad I_{\pi,G}^{\tau\eta}=\tilde{u}^{\eta}(\tilde{\pi}^{\tau\tau}+\tilde{\pi}^{\eta\eta})/\tau\\
 & I_{\pi,G}^{\eta x}=\tilde{u}^{\eta}\tilde{\pi}^{\tau x}/\tau & I_{\pi,G}^{\eta y}=\tilde{u}^{\eta}\tilde{\pi}^{\tau y}/\tau\\
 & I_{\pi,G}^{\eta\eta}=2\tilde{u}^{\eta}\tilde{\pi}^{\tau\eta}/\tau & I_{\pi,G}^{xx}=I_{\pi,G}^{xy}=I_{\pi,G}^{yy}=0\end{align*}

To solve the energy-momentum conservation equations (\ref{eqns1})
in viscous case, we use the technique of ideal-viscous splitting \citet{TakamotoInutsuka}.
This allows us to solve the ideal part very accurately using Godunov
method, employing relativistic HLLE approximation for the solution
of Riemann problem. Israel-Stewart equations (\ref{evolutionShear},\ref{evolutionBulk})
are solved in parallel, and then the evolution of energy/momentum
is corrected according to viscous fluxes and source terms in (\ref{eqns1}).

\section{Testing the hydro procedure}

For the purpose of current paper we skip presenting technical details
of the code and the results of basic tests against analytical hydro
solutions, leaving this to a separate publication. However, in what
follows, we compare our hydro simulations to the results of {}``open
TECHQM'' \citet{techqm}, using the same initial conditions for heavy
ion collisions as they propose, namely \begin{align}
 & \epsilon(\tau_{0},r_{x},r_{y})=C\cdot n_{WN}(r_{x},r_{y})=\nonumber \\
 & \: CT_{A}(r_{x}+b/2,r_{y})\left\{ 1-\left[1-T_{A}(r_{x}-b/2,r_{y})\sigma_{NN}/A\right]^{A}\right\} +\nonumber \\
 & \, CT_{A}(r_{x}-b/2,r_{y})\left\{ 1-\left[1-T_{A}(r_{x}+b/2,r_{y})\sigma_{NN}/A\right]^{A}\right\} \,,\end{align}
where the nuclear thickness function $T_{A}(x,y)=\int dr_{z}\rho(r_{x},r_{y},r_{z})$
is normalized such that $\int T_{A}(x,y)dxdy=A$, and $\rho(r_{x},r_{y},r_{z})=c/(exp[(r-R_{A})/\delta]+1)$
is the density distribution for nucleons in the nucleus. For Au-Au
collision the parameters are $A=197$, $R_{A}=6.37$~fm, $\delta=0.54$~fm,
$\sigma_{NN}=40$~mb is the inelastic nucleon-nucleon cross section
and $C$ is chosen so that $\epsilon_{0}(0,0;b=0)=30$~GeV/fm$^{3}$.
The EoS for relativistic massless gas $p=\epsilon/3$ is used. To
extract the temperature in this EoS we assume 2.5 massless quark degrees
of freedom and $g_{q}=2\cdot2\cdot3=12$ degeneracy factor and $g_{g}=16$
for massless gluons. For viscous hydro runs we fix bulk viscosity
to zero, initialize $\pi^{\mu\nu}$ at $\tau_{0}$ with their Navier-Stokes
values, which yields $\pi^{xx}=\pi^{yy}=-\tau^{2}\pi^{\eta\eta}/2=2\eta/(3\tau_{0})$.

\begin{figure}[tb]
\includegraphics[width=0.4\textwidth]{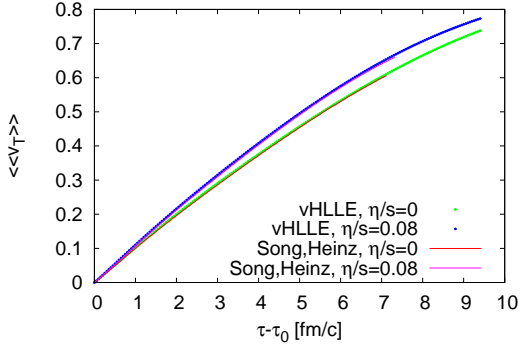} 

\caption{\label{figRadFlowSong}Averaged radial flow as a function of proper
time for our hydro code (vHLLE) compared to \texttt{VISH2+1} by H.~Song}

\end{figure}

\begin{figure}[tb]
\includegraphics[width=0.4\textwidth]{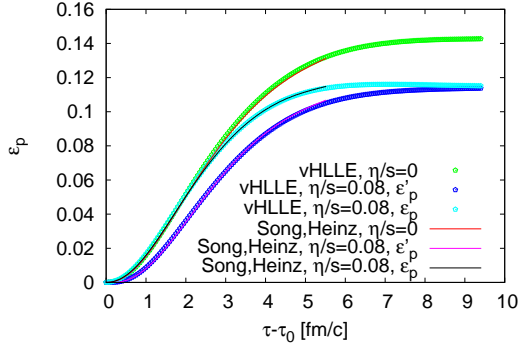} 

\caption{\label{figEpsilonpSong} Flow anisotropies $\epsilon_{p}$ and $\epsilon'_{p}$
(see text for explanation) as a function of proper time for our hydro
code (vHLLE) compared to \texttt{VISH2+1} by H.~Song}

\end{figure}

In Fig. \ref{figRadFlowSong} we show the comparison for average transverse
velocity as a function of evolution time $\tau$ for initial conditions
with impact parameter $b=0$. The average is defined as \[
<<v_{T}>>=\int\frac{v_{T}\cdot\epsilon}{\sqrt{1-v_{T}^{2}}}d^{2}r_{T}\]
with $v_{T}=\sqrt{v_{x}^{2}+v_{y}^{2}}$, and where the integration
is made for a slice of system (cells) with rapidity $y=0$. Shear
viscosity works to equalize the expansion in different directions,
thus decelerating the initially strong longitudinal expansion and
accelerating transverse one, which results in additional acceleration
of transverse radial flow. One can see that the magnitude of the effect
in our results is consistent with those from \texttt{VISH2+1} code.

\begin{figure}[tb]
\includegraphics[width=0.4\textwidth]{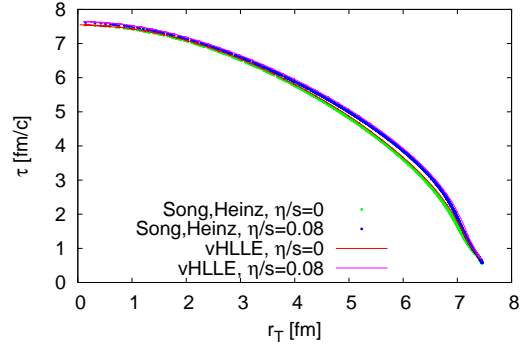} 

\caption{\label{figFreezeoutSong} Iso-thermal surface corresponding to $T_{f}=130$~MeV
obtained with our hydro code (vHLLE) compared to the results from
\texttt{VISH2+1} by H.~Song}

\end{figure}
In the same way shear viscosity suppresses the development of flow
anisotropy in the transverse plane, the latter being generated by
anisotropic pressure gradients in hydrodynamics. To check this effect,
we set the initial conditions for $b=7$~fm and in Fig. \ref{figEpsilonpSong}
we show the corresponding time evolution of flow anisotropy, defined
as \begin{equation}
\epsilon_{p}=\frac{<T_{\text{id}}^{xx}-T_{\text{id}}^{yy}>}{<T_{\text{id}}^{xx}+T_{\text{id}}^{yy}>},\:\epsilon'_{p}=\frac{<T^{xx}-T^{yy}>}{<T^{xx}+T^{yy}>},\end{equation}
 with $<\dots>=\int\dots d^{2}r_{T}$; the quantities $\epsilon_{p}$
and $\epsilon'_{p}$ are calculated using the ideal part of the energy-momentum
tensor and full energy-momentum tensor respectively. The suppression
of $\epsilon_{p}$ comes solely from the rearrangement of collective
flow, while $\epsilon'_{p}$ is suppressed stronger due to contributions
from $\pi^{\mu\nu}$. The results are as well consistent with the
ones from \texttt{VISH2+1} code. 

Finally, in Fig. \ref{figFreezeoutSong}, we show the iso-thermal
surfaces for the case $b=0$ corresponding to temperature $T_{f}=130$~MeV
(or $\epsilon_{f}=0.516$~GeV/fm$^{3}$). Here, slight differences
may come from the details (interpolation scheme) to determine the
position of freeze-out surface, so the discrepancies less than $~\Delta x/2=0.1$~fm
are justified.

\section{Parameters and basic tests }

Our basic framework for the initial conditions is {}``Parton-Based
Gribov-Regge Theory'', described in detail in \citet{hajo}. In that
paper we dedicate a whole chapter to the discussion of parameters,
still being valid now. All {}``basic'' parameters (see table 8.2
in \citet{hajo}) related to define the multiple scattering amplitude
discussed earlier are fixed by comparing to proton-proton data. 

There are very few parameters related to the {}``new features''
discussed earlier. There are first of all the two coefficients $A_{\mathrm{sat}}$
and $B_{\mathrm{sat}}$ in the saturation scale formulas: they are
used to assure binary scaling in p-A and AA scattering at large $p_{t}$
and a correct energy dependence of the total cross section in p-p.
It should be said that the former {}``screening'' procedure introduced
in \citet{kwladsplit}, had at the end more than 20 parameters, without
satisfying results in p-A scattering. The new method (with two parameters)
gives a much more consistent picture for p-p and p-A at LHC energies.
Also Pb-Pb results look promising, but in this paper we concentrate
on p-A. %
\begin{figure}[b]
\begin{centering}
\includegraphics[angle=270,scale=0.24]{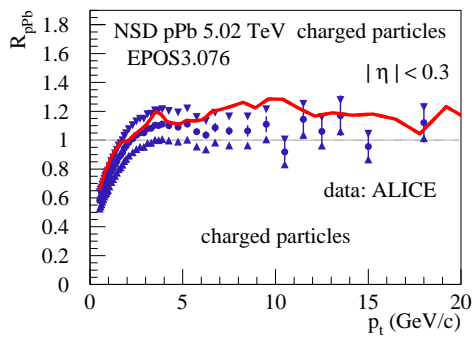}
\par\end{centering}

\caption{(Color online) The nuclear modification factor for NSD p-Pb scattering
at 5.02 TeV. We show our full calculation, EPOS3 with hydro and hadronic
cascade (solid red line), compared to data from ALICE \citet{aliceraa}
(error bars refer to statistical errors, triangles to systematic errors).
\label{z-selb05.ps}}

\end{figure}

The new core-corona procedure depends on the coefficient $f_{\mathrm{Eloss}}$,
which is the most important parameter for the analysis presented in
this paper: increasing this constant will increase the core contribution.
Putting $f_{\mathrm{Eloss}}=0$ would completely suppress the core,
we have a pure string model. Making $f_{\mathrm{Eloss}}$ very big
leads to a purely hydrodynamic expansion. The reality seems to be
between these two extremes: all the results shown in this paper refer
to $f_{\mathrm{Eloss}}=0.137\,\mathrm{fm\, GeV}/c$.

Finally, we use in our approach for the first time viscous hydrodynamics
for the collective expansion. All calculations shown are done for
shear viscosity over entropy density ($\eta/s$) of 0.08. We fix bulk
viscosity to zero, and we do not include the baryon / electric / strange
charge diffusion either. Furthermore, we initialize $\pi^{\mu\nu}$
at $\tau_{0}$ with their Navier-Stokes values, which yields $\pi^{xx}=\pi^{yy}=-\tau^{2}\pi^{\eta\eta}/2=2\eta/(3\tau_{0})$,
and the relaxation time for the shear stress tensor is taken as $\tau_{\pi}=3\eta/(sT)$.

\begin{figure}[tb]
\begin{centering}
\vspace*{-0cm}\includegraphics[angle=270,scale=0.37]{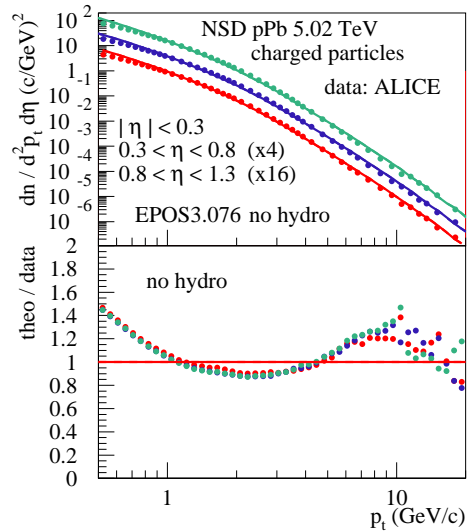}\\

\par\end{centering}

\caption{(Color online) Transverse momentum spectra of charged particles for
NSD p-Pb scattering at 5.02 TeV, for three different pseudorapidity
windows. We show data from ALICE \citet{aliceraa}, compared to our
{}``basic'' calculation without hydro, without cascade, as well
as the ratio theory over data. \label{z-selb08.ps}}

\end{figure}
\begin{figure}[b]
\begin{centering}
\vspace*{-0.4cm}\includegraphics[angle=270,scale=0.37]{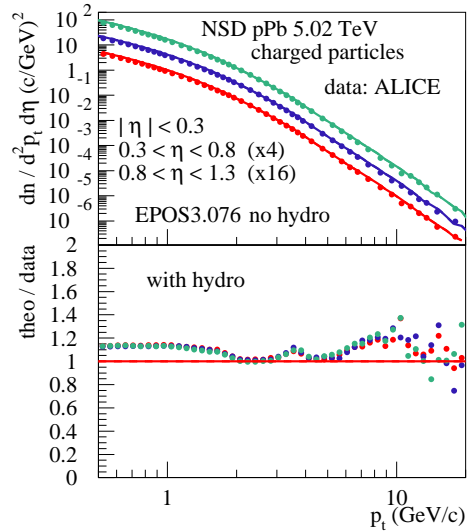}\\

\par\end{centering}

\caption{(Color online) Same as fig.\ref{z-selb08.ps}, but showing the full
calculation, with hydro and hadronic cascade. \label{z-selb09.ps}}

\end{figure}
Before analysing in very detail identified particle spectra, we first
show in figs. \ref{z-selb05.ps}-\ref{z-selb09.ps} some basic tests
of our approach. In fig. \ref{z-selb05.ps}, we compare our calculations
to the experimental nuclear modification factor $R_{pPb}$ (rescaled
p-Pb spectra over p-p). Beyond 2-3 GeV/c, both curves are compatible
with unity. 

In fig. \ref{z-selb08.ps}, we show $p_{t}$ spectra of charged particles
for NSD p-Pb scattering at 5.02 TeV, for three different pseudorapidity
windows. We compare our calculation without hydro to experimental
ALICE data. The simulation results are significantly below the data
in the $p_{t}$ range of 1 to 5 GeV/c. There is nothing one can do
(via parameter change) to significantly improve the agreement -- without
hydro. In fig. \ref{z-selb09.ps}, we compare the same experimental
data to our full calculation, with hydro and hadronic cascade. Taking
into account the hydro evolution, improves the situation considerably.

\section{Identified particle results for p-Pb}

In the following, we will compare experimental data on identified
particle production with our simulation results (referred to as EPOS3),
and in addition to some other models, as there are QGSJETII \citet{qgsjet},
AMPT \citet{ampt}, and EPOS$\,$LHC \citet{eposlhc}. The QGSJETII
model is also based on Gribov-Regge multiple scattering, but there
is no fluid component. The main ingredients of the AMPT model are
a partonic cascade and then a hadronic cascade, providing in this
way some {}``collectivity''. EPOS$\,$LHC is a tune (using LHC data)
of EPOS1.99. As all EPOS1 models, it contains flow, put in by hand,
parametrizing the collective flow at freeze-out. Finally, the approach
discussed in this paper (EPOS3) contains a full viscous hydrodynamical
simulation. So it is interesting to compare these four models, since
they differ considerably concerning the implementation of flow, from
full hydrodynamical flow in EPOS3 to no flow in QGSJETII.

\begin{figure}[tb]
\begin{minipage}[t][1\totalheight]{1\columnwidth}%
\begin{minipage}[c][1\totalheight]{1\columnwidth}%
\vspace*{-0.2cm}

\hspace*{-0.3cm}\includegraphics[angle=270,scale=0.2]{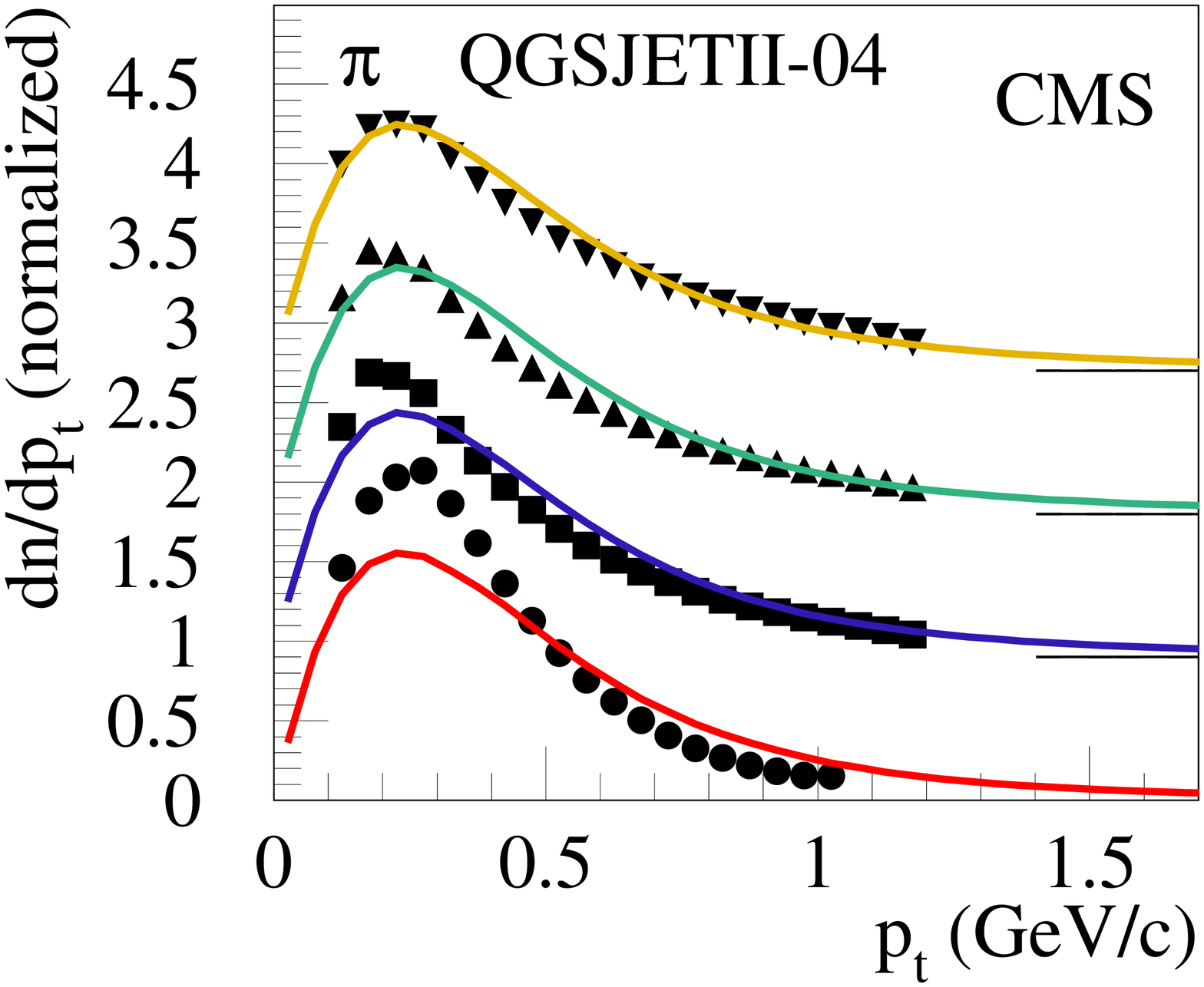}\hspace*{-0.5cm}\includegraphics[angle=270,scale=0.2]{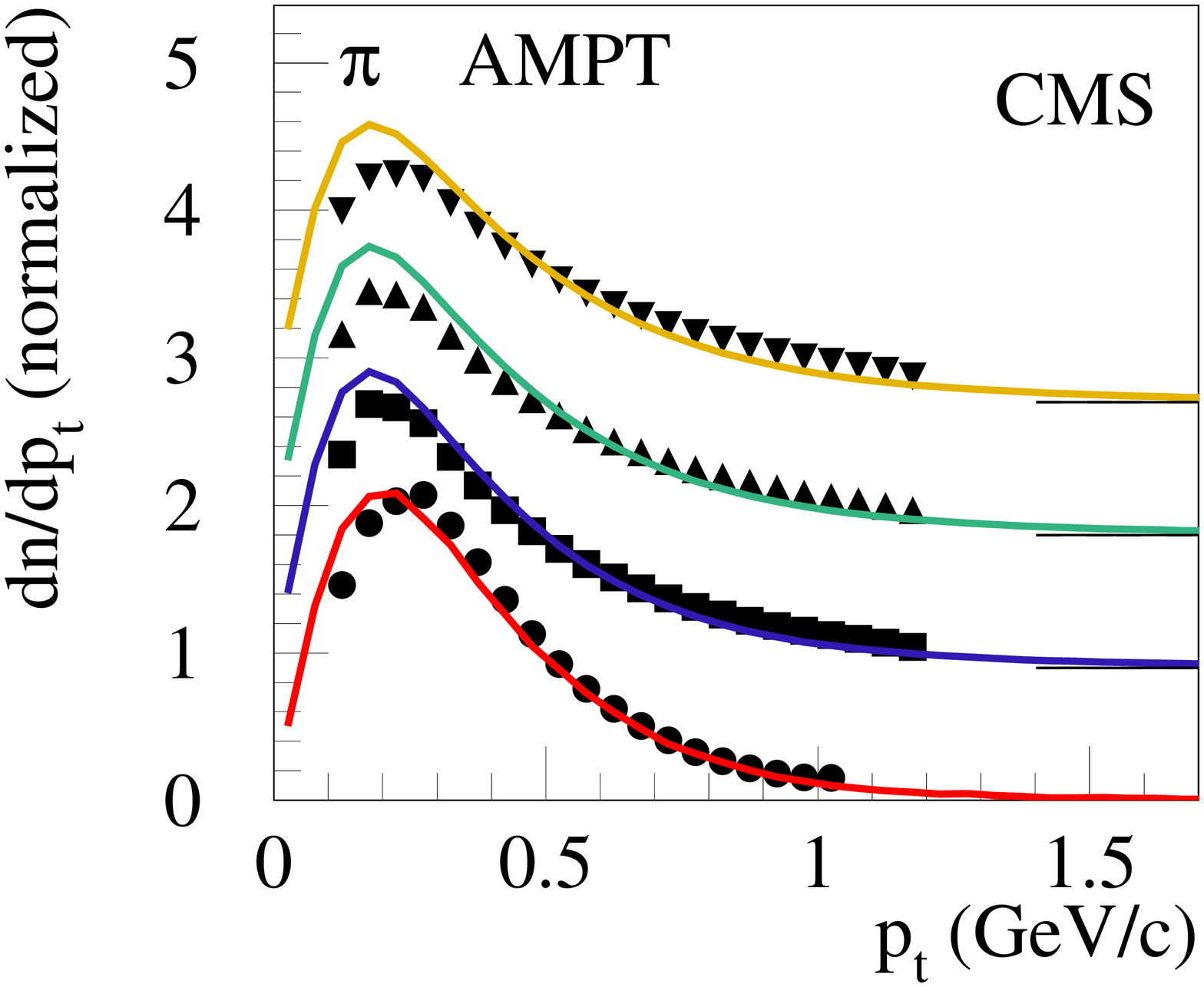}\vspace*{-0.2cm}

\hspace*{-0.3cm}\includegraphics[angle=270,scale=0.2]{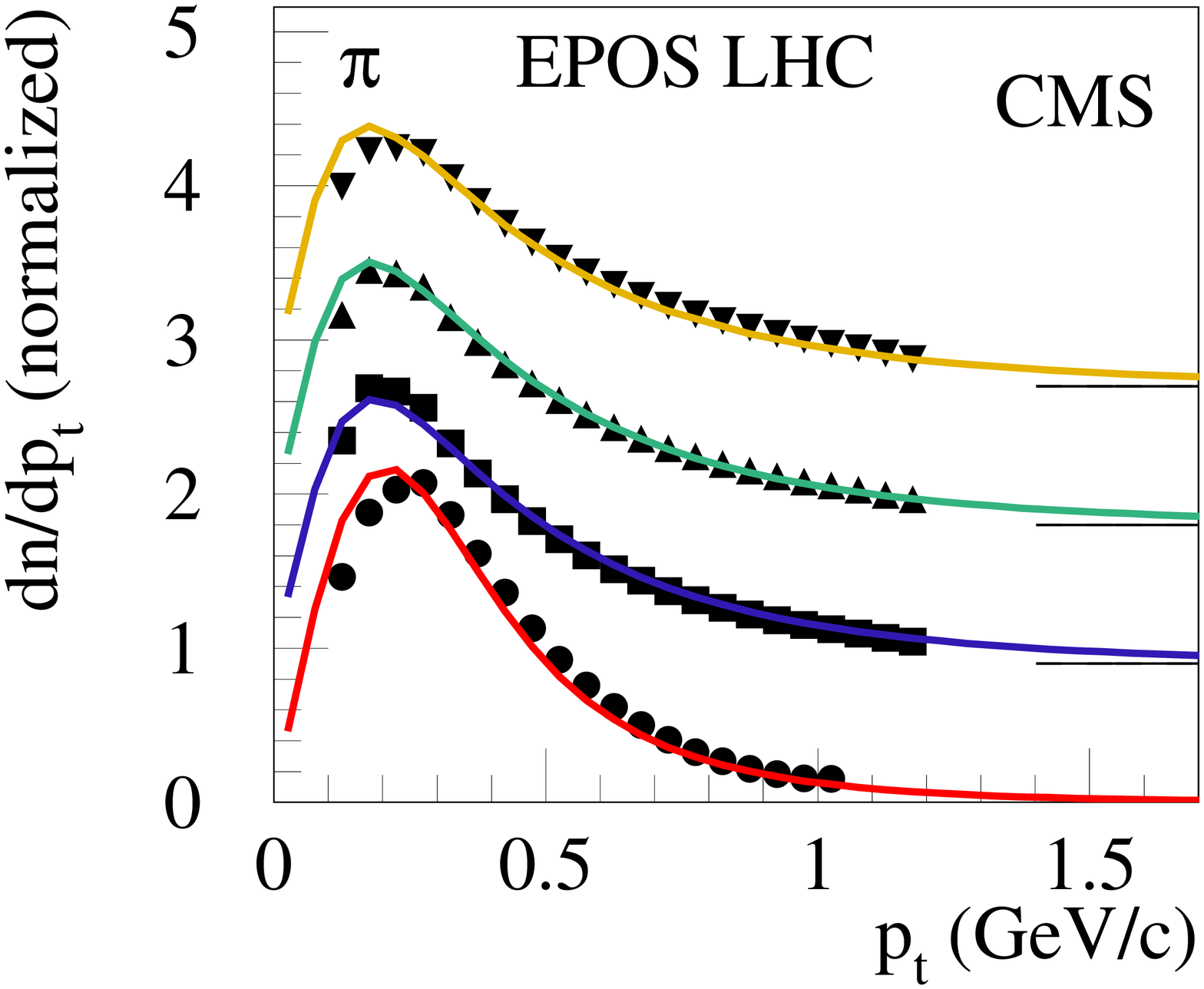}\hspace*{-0.5cm}\includegraphics[angle=270,scale=0.2]{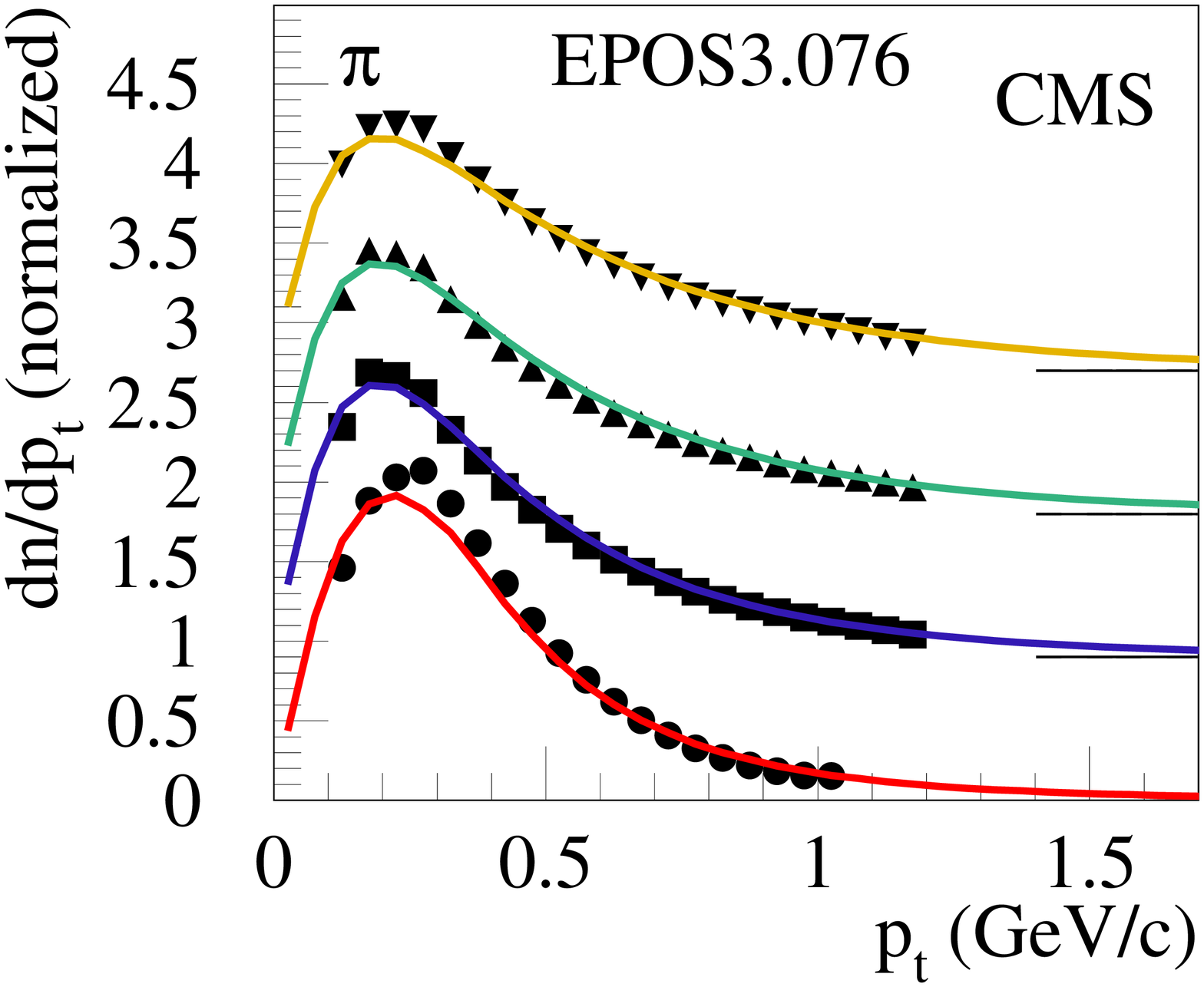}%
\end{minipage}%
\end{minipage}

\noindent \caption{(Color online) Transverse momentum spectra of pions in p-Pb scattering
at 5.02 TeV, for four different multiplicity classes with mean values
(from bottom to top) of 8, 84, 160, and 235 charged tracks. We show
data from CMS \citet{cms} (symbols) and simulations from QGSJETII,
AMPT, EPOS$\,$LHC, and EPOS3, as indicated in the figures. \label{fig:cms1}}

\end{figure}
The CMS collaboration published a detailed study \citet{cms} of the
multiplicity dependence of (normalized) transverse momentum spectra
in p-Pb scattering at 5.02 TeV. The multiplicity (referred to as $N_{\mathrm{track}}$)
counts the number of charged particles in the range $|\eta|<2.4$.
Many multiplicity classes have been considered, but in our analysis
we consider only four, in order not to overload the figures. The mean
values of the four multiplicity classes are 8, 84, 160, and 235. 

In fig. \ref{fig:cms1}, we compare experimental data \citet{cms}
for pions (black symbols) with the simulations from QGSJETII (upper
left figure), AMPT (upper right), EPOS$\,$LHC (lower left), and EPOS3
(lower right). The different curves in each figure refer to different
centralities, with mean values (from bottom to top) of 8, 84, 160,
and 235 charged tracks. They are shifted relative to each other by
a constant amount. Concerning the models, QGSJETII is the easiest
to discuss, since here there are no flow features at all, and the
curves for the different multiplicities are identical. The data, however,
show a slight centrality dependence: the spectra get somewhat harder
with increasing multiplicity. The other models, AMPT, EPOS$\,$LHC,
and EPOS3 are close to the data. %
\begin{figure}[b]
\begin{minipage}[t][1\totalheight]{1\columnwidth}%
\begin{minipage}[c][1\totalheight]{1\columnwidth}%
\vspace*{-0.2cm}

\hspace*{-0.3cm}\includegraphics[angle=270,scale=0.2]{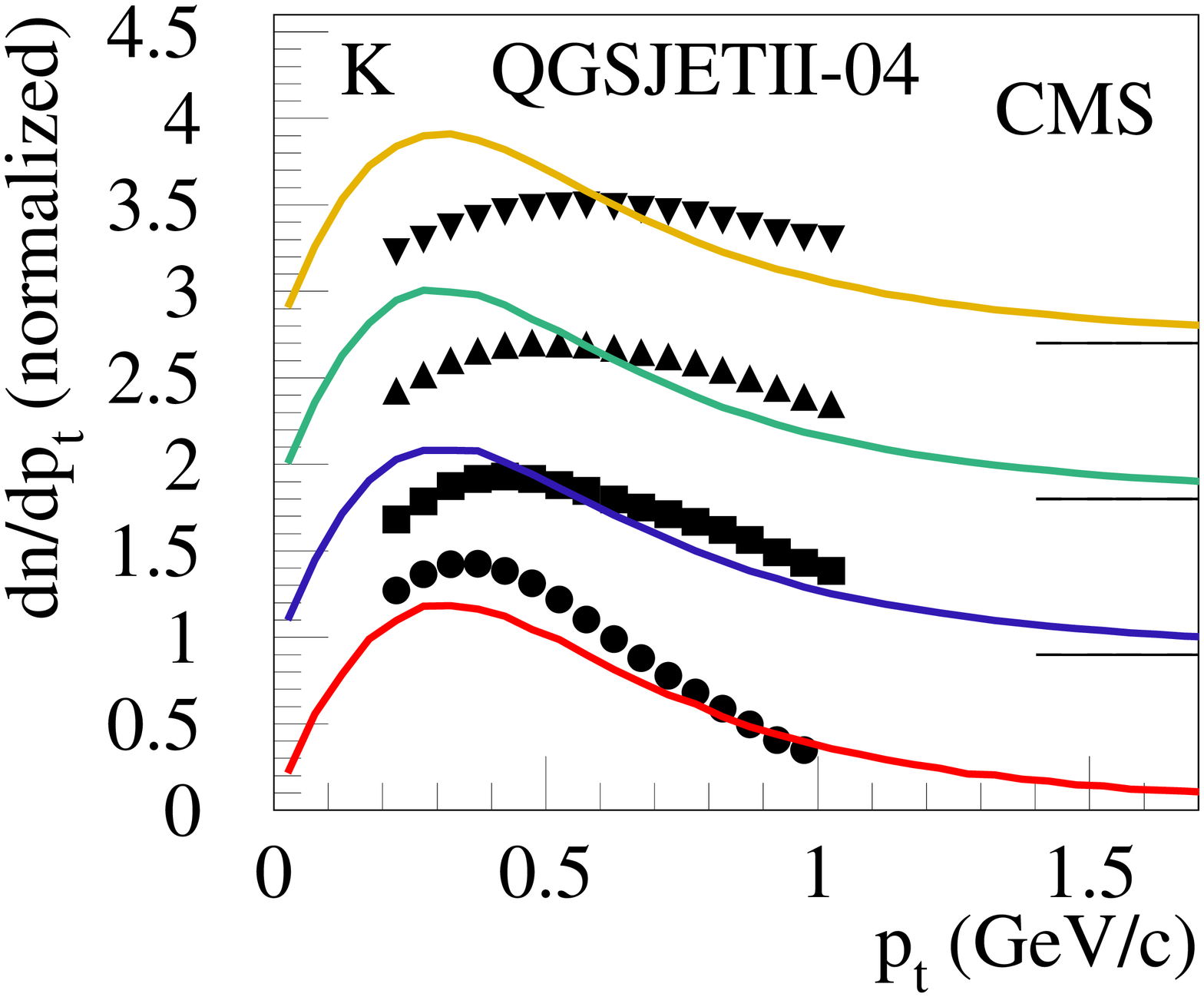}\hspace*{-0.5cm}\includegraphics[angle=270,scale=0.2]{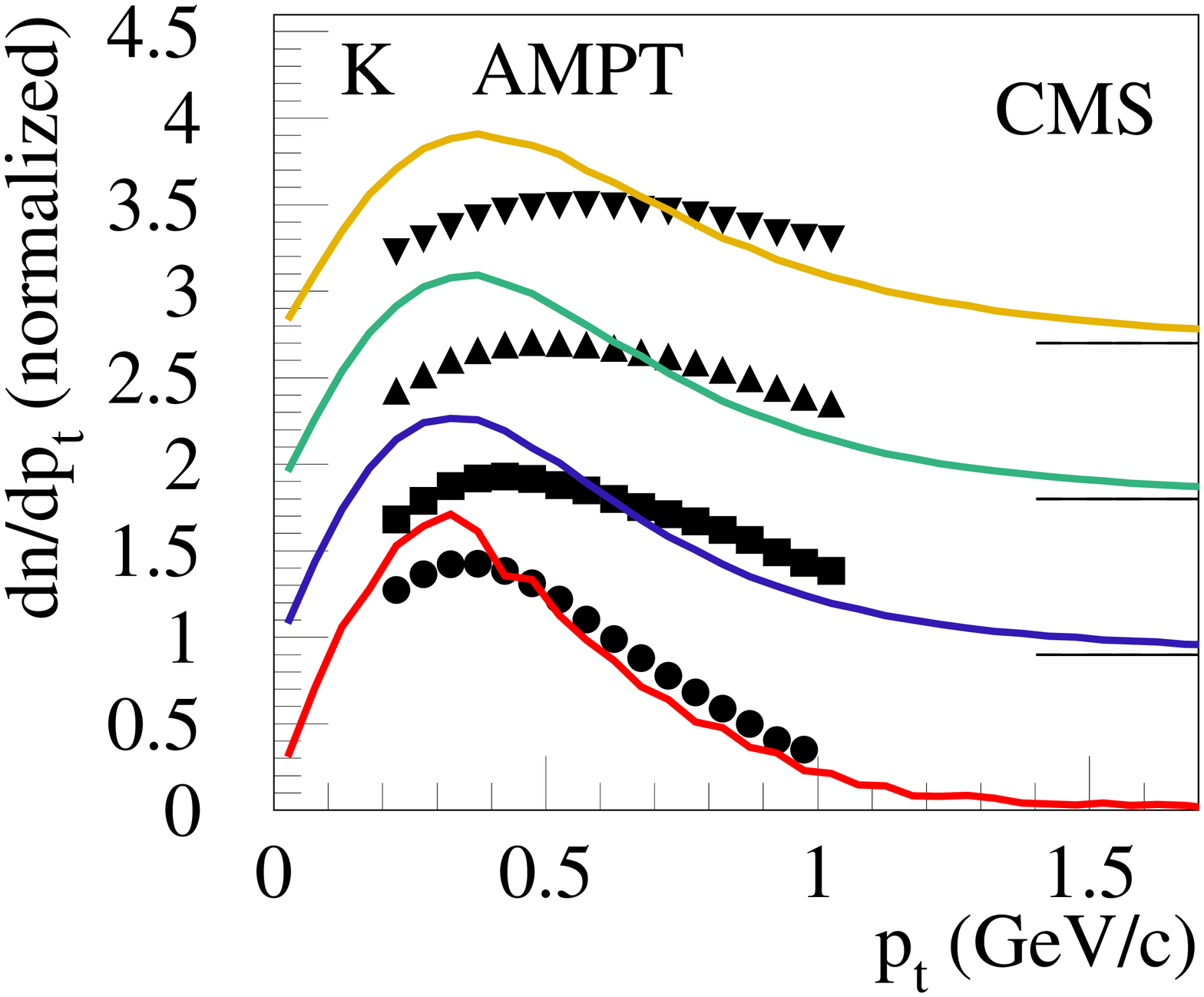}\vspace*{-0.2cm}

\hspace*{-0.3cm}\includegraphics[angle=270,scale=0.2]{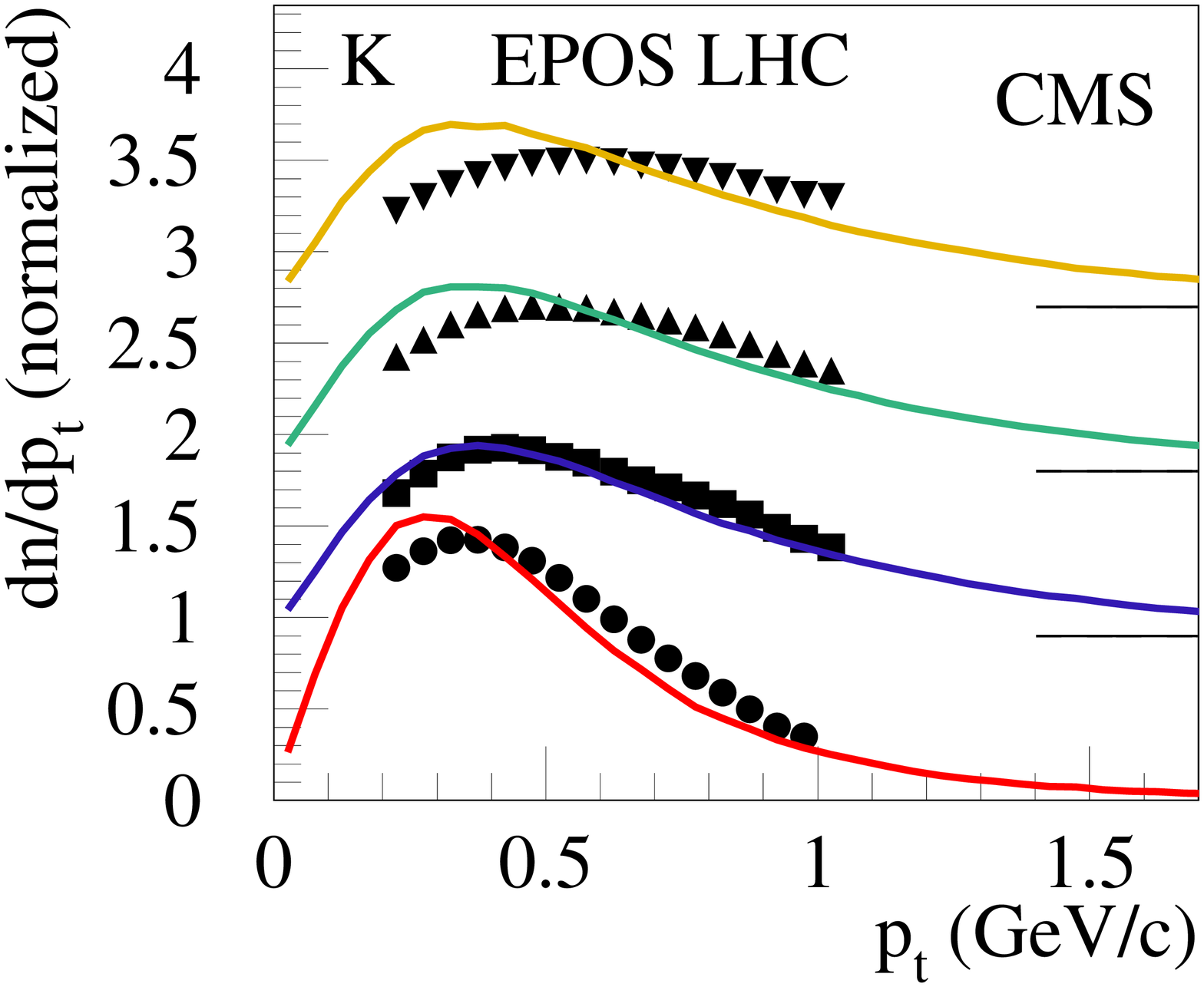}\hspace*{-0.5cm}\includegraphics[angle=270,scale=0.2]{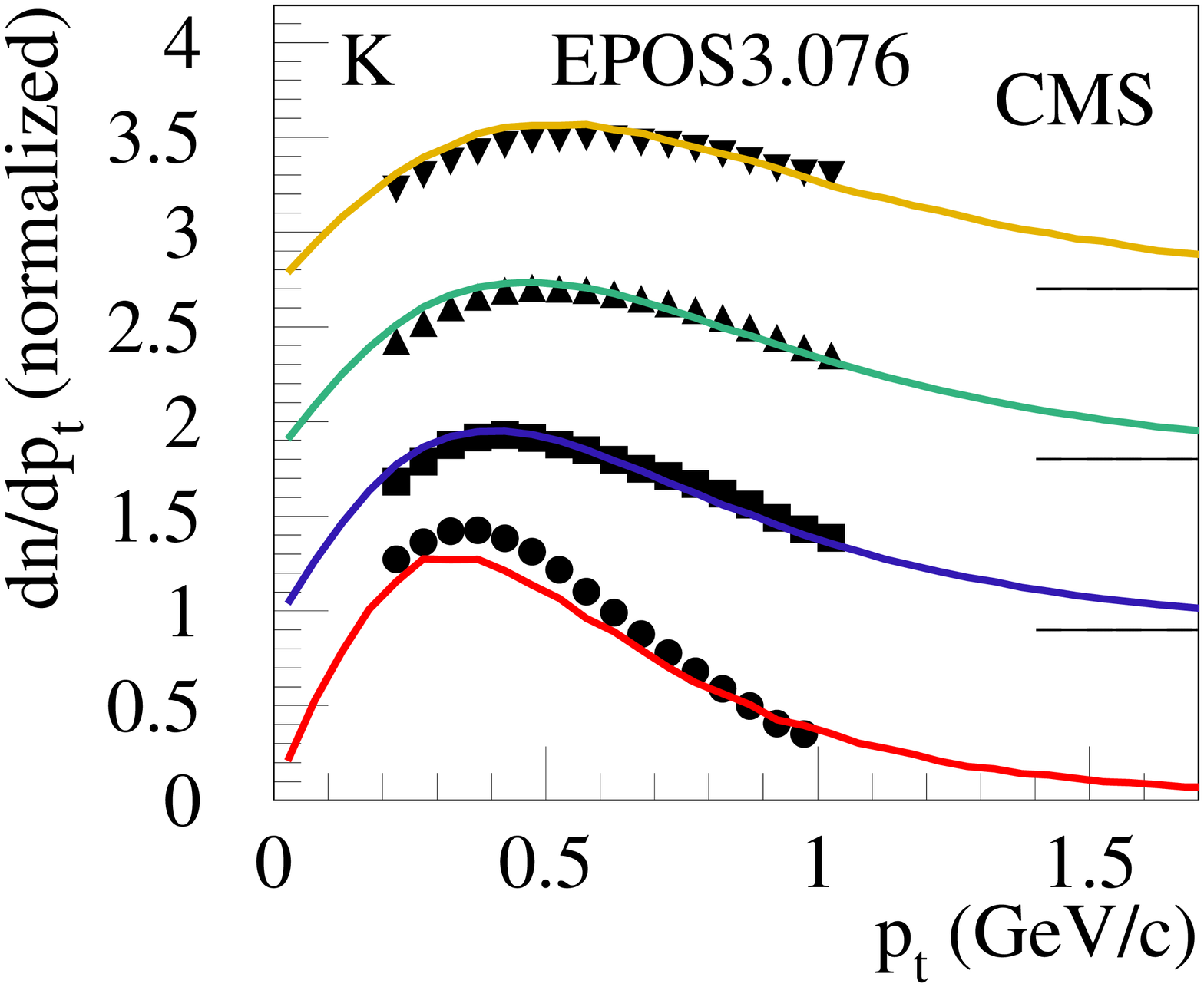}%
\end{minipage}%
\end{minipage}

\noindent \caption{(Color online) Same as fig. \ref{fig:cms1}, but for kaons. \label{fig:cms2}}

\end{figure}
\begin{figure}[tb]
\begin{minipage}[t][1\totalheight]{1\columnwidth}%
\begin{minipage}[c][1\totalheight]{1\columnwidth}%
\vspace*{-0.2cm}

\hspace*{-0.3cm}\includegraphics[angle=270,scale=0.2]{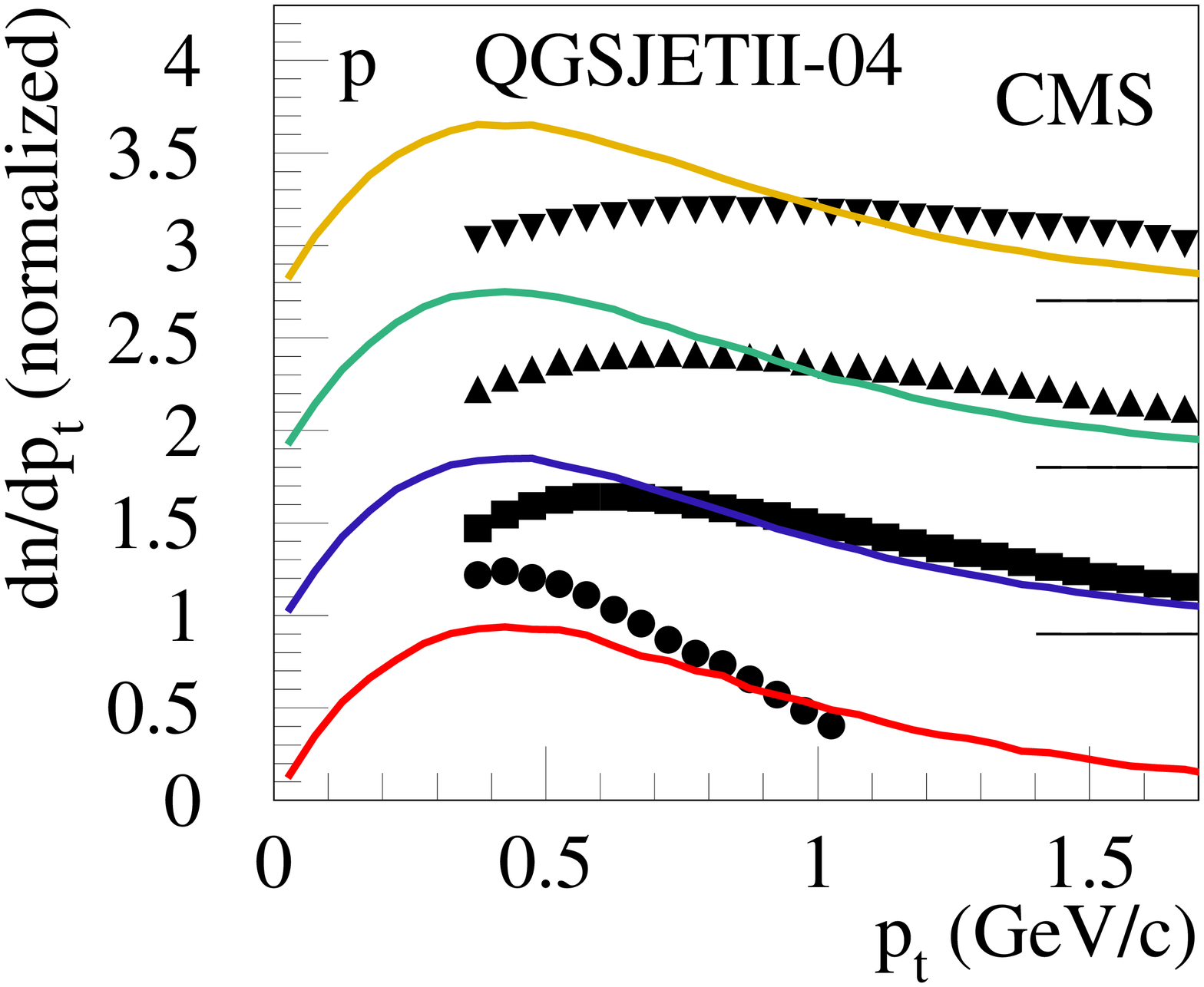}\hspace*{-0.5cm}\includegraphics[angle=270,scale=0.2]{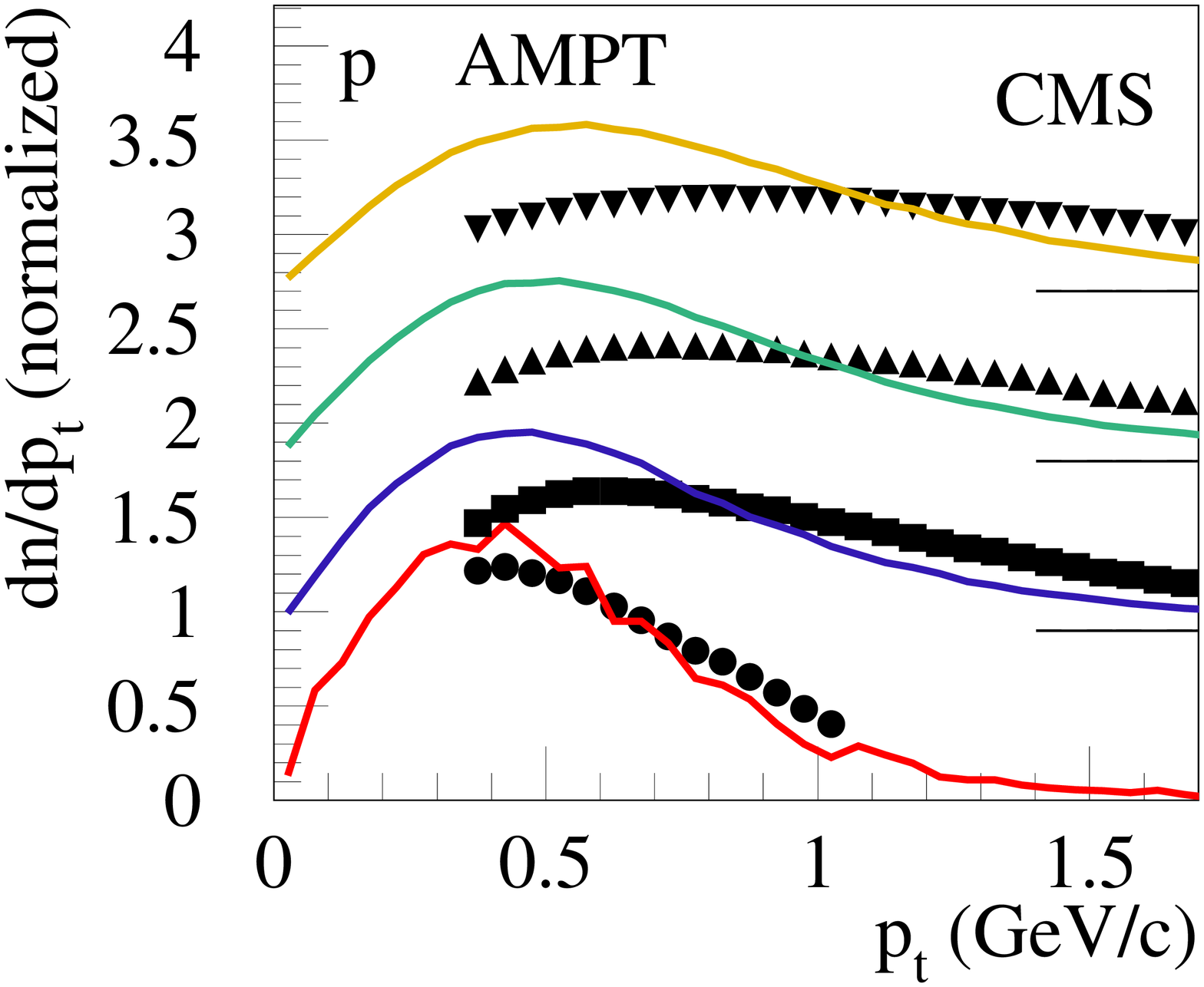}\vspace*{-0.2cm}

\hspace*{-0.3cm}\includegraphics[angle=270,scale=0.2]{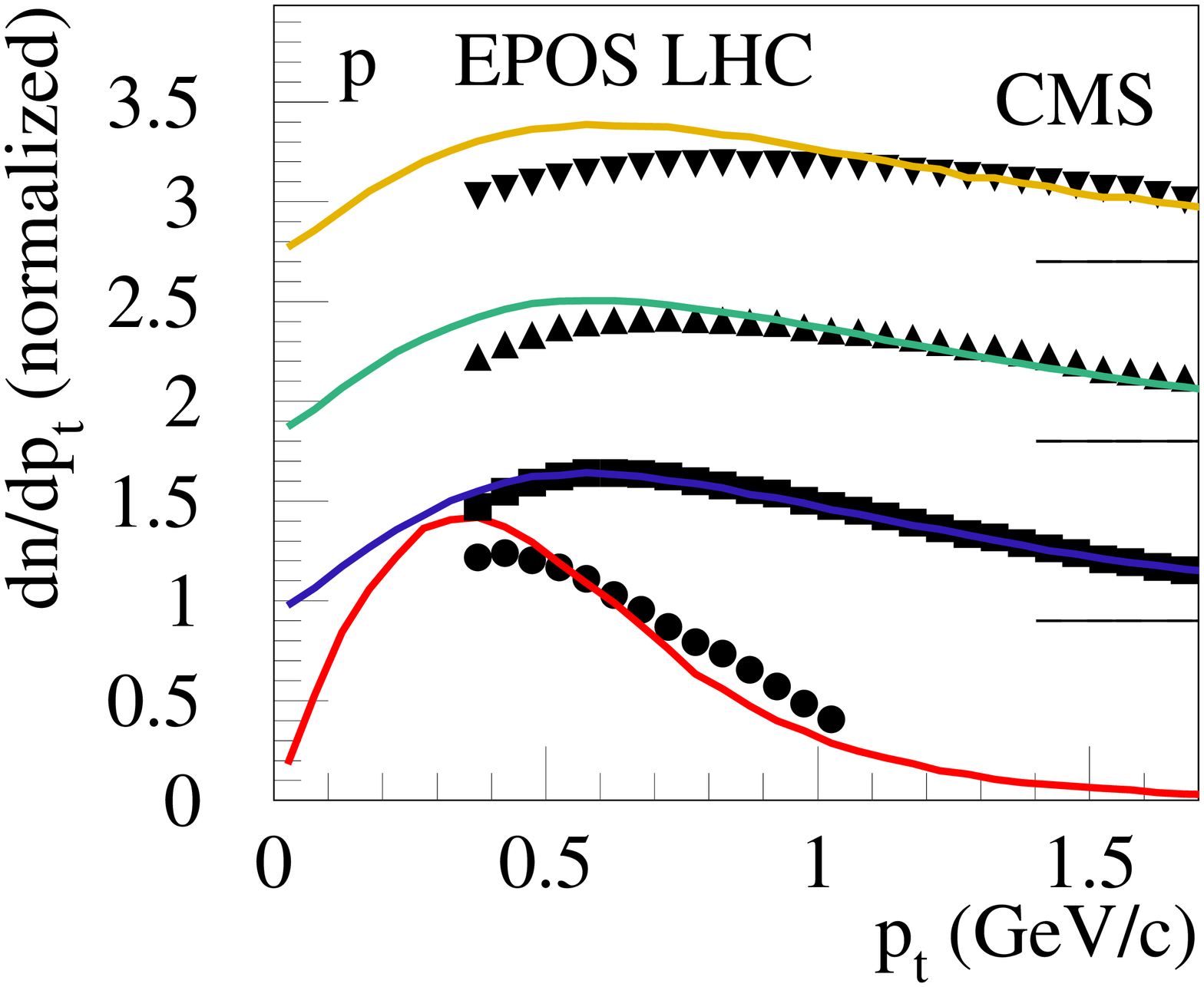}\hspace*{-0.5cm}\includegraphics[angle=270,scale=0.2]{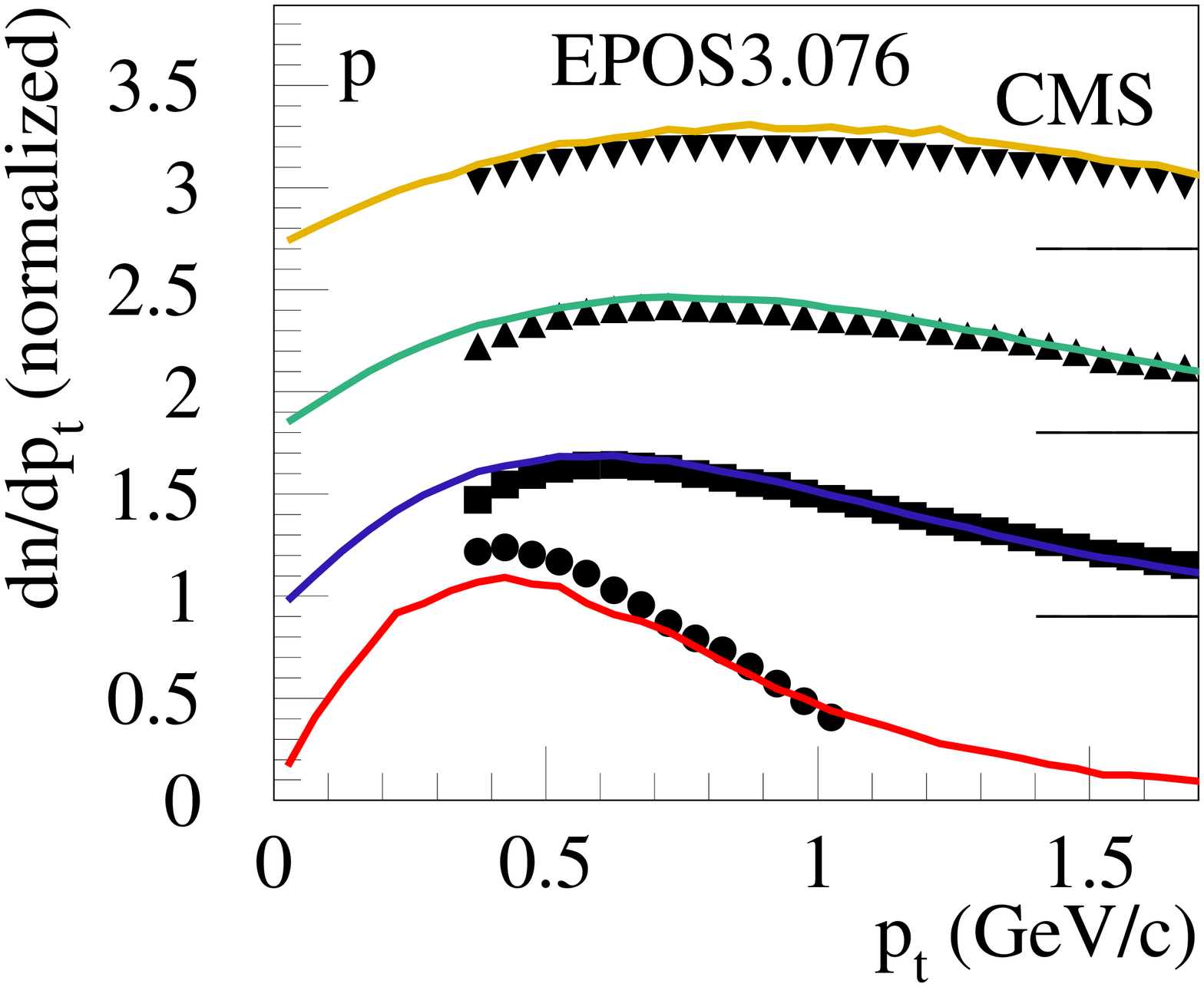}%
\end{minipage}%
\end{minipage}

\noindent \caption{(Color online) Same as fig. \ref{fig:cms1}, but for protons.\label{fig:cms3}}

\end{figure}
\begin{figure}[b]
\begin{centering}
\hspace*{-0.4cm}\includegraphics[angle=270,scale=0.2]{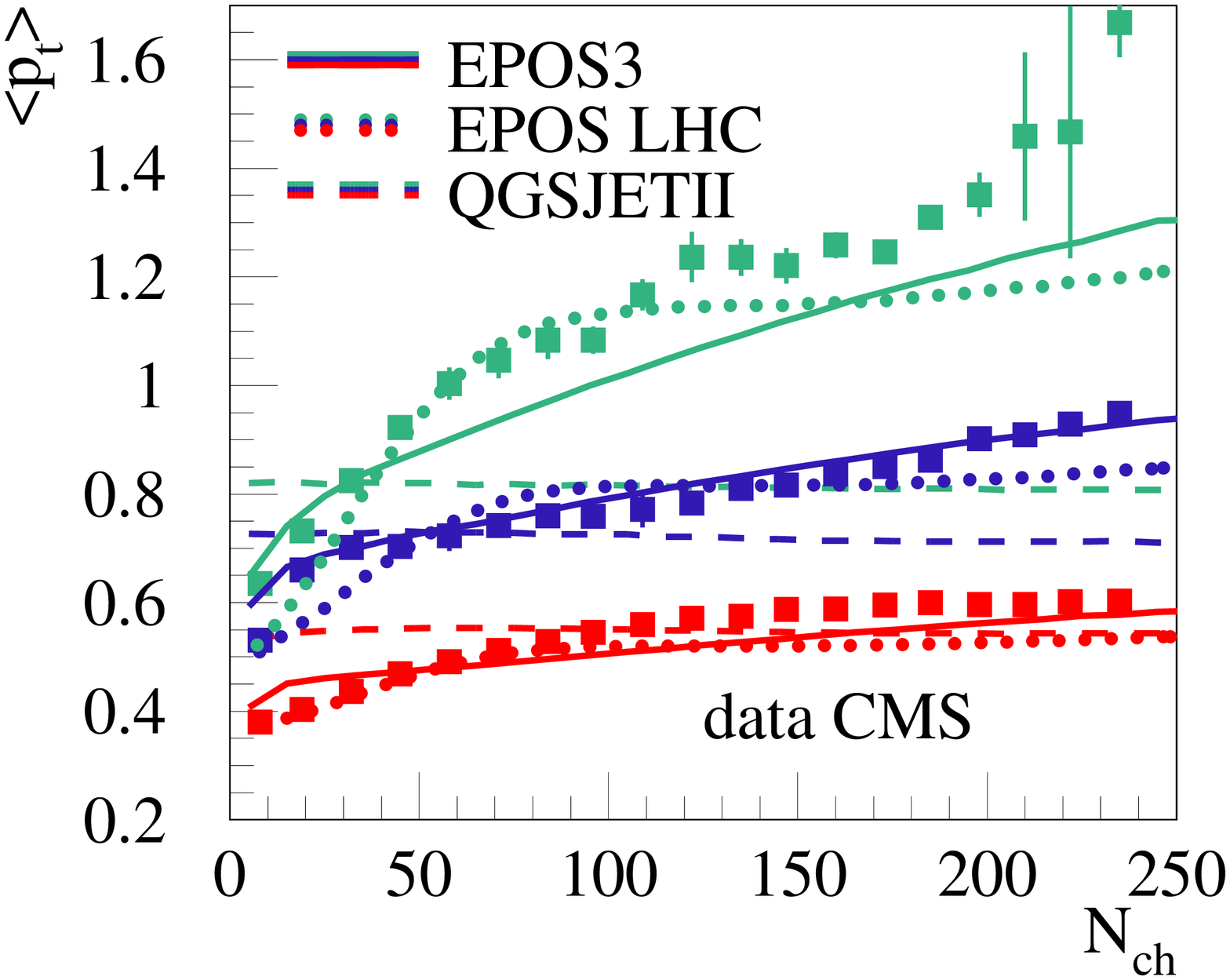}\hspace*{-0.7cm}\includegraphics[angle=270,scale=0.2]{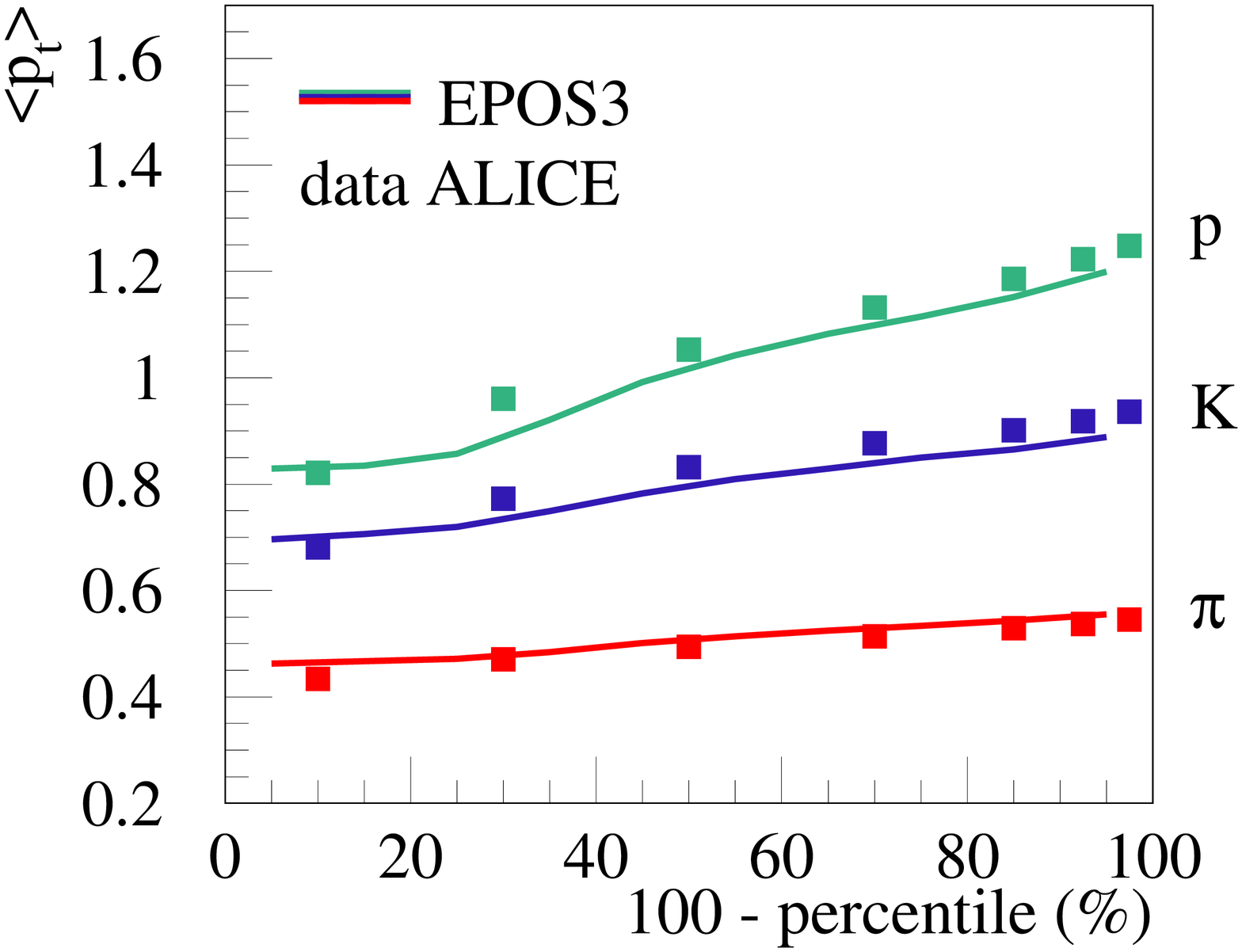}\\

\par\end{centering}

\caption{(Color online) Multiplicity dependence of the average transverse momentum
of protons (green), kaons (blue), and pions (red) in p-Pb scattering
at 5.02 TeV. Left: data from CMS \citet{cms} (symbols) and simulations
from QGSJETII (dashed lines), EPOS$\,$LHC (dotted lines), and EPOS3
(solid lines). Right: data from ALICE and EPOS3 results (percentiles
are defined via the VZERO-A multiplicity). \label{fig:selb22}}

\end{figure}
\begin{figure}[tb]
\begin{centering}
\vspace*{-0.5cm}\includegraphics[angle=270,scale=0.3]{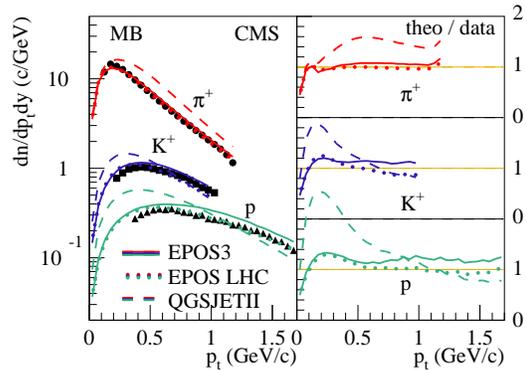}\\

\par\end{centering}

\caption{(Color online) Transverse momentum spectra $dn/dp_{t}dy$ of protons
(green), kaons (blue), and pions (red) in p-Pb scattering at 5.02
TeV. We show data from CMS \citet{cms} (symbols) and simulations
from QGSJETII (dashed lines), EPOS$\,$LHC (dotted lines), and EPOS3
(solid lines). The left panel shows the spectra, the right one the
corresponding ratio theory over experiment. \label{fig:selb21}}

\end{figure}

In fig. \ref{fig:cms2}, we compare experimental data \citet{cms}
for kaons (black symbols) with the simulations. In the data, the shapes
of the $p_{t}$ spectra change considerably with multiplicity: they
get much harder with increasing multiplicity. In QGSJETII, there is
again no change and in AMPT too little change with multiplicity. EPOS$\,$LHC
goes into the right direction, whereas EPOS3 gives a satisfactory
description of the data. In fig. \ref{fig:cms3}, we compare experimental
data \citet{cms} for protons (black symbols) with the simulations.
Again, as for kaons, the experimental shapes of the $p_{t}$ spectra
change considerably, getting much harder with increasing multiplicity.
In QGSJETII, having no flow, the curves for the different multiplicities
are identical. The AMPT model shows some (but too little) change with
multiplicity. EPOS$\,$LHC goes into the right direction, whereas
EPOS3 gives a reasonable description of the data. \textbf{It seems
that hydrodynamical flow helps considerably to reproduce these data}.

Based on these multiplicity dependent $p_{t}$ spectra, one obtains
the multiplicity dependence of the average transverse momentum $\left\langle \right.p_{t}\left.\right\rangle $,
as shown in fig. \ref{fig:selb22}(left), where we plot the multiplicity
dependence of the average transverse momentum of protons (green),
kaons (blue), and pions (red) in p-Pb scattering at 5.02 TeV. We show
data from CMS \citet{cms} (symbols) and simulations from QGSJETII
(dashed lines), EPOS$\,$LHC (dotted lines), and EPOS3 (solid lines).
Whereas QGSJETII shows no multiplicity dependence, EPOS$\,$LHC and
EPOS3 increase with multiplicity, and this increase is more pronounced
for heavier particles (due to the radial flow). However, EPOS$\,$LHC
reaches a kind of plateau at high multiplicity, whereas data (and
EPOS3) increase continuously. This is (in EPOS3) a core-corona effect:
the core (=flow) fraction increases with multiplicity. In fig. \ref{fig:selb22}(right),
we compare $\left\langle \right.p_{t}\left.\right\rangle $ results
from EPOS3 with data from ALICE \citet{alice}, where we use {}``percentiles''
defined via the VZERO-A multiplicity, as in the experiment. Again
we see (in the data and the simulations) the same trend of increasing
$\left\langle \right.p_{t}\left.\right\rangle $, more pronounced
for heavier particles (a more detailed analysis of ALICE data will
be given below).

All the discussions above are based on normalized (to unity) $p_{t}$
spectra. In order to verify the absolute normalizations of the various
particle yields, we plot in fig.~\ref{fig:selb21} $dn/dp_{t}dy$
(particles per event) for protons (green), kaons (blue), and pions
(red) as a function of $p_{t}$, in p-Pb scattering at 5.02 TeV. We
show data from CMS \citet{cms} (symbols) and simulations from QGSJETII
(dashed lines), EPOS$\,$LHC (dotted lines), and EPOS3 (solid lines).
The left panel shows the spectra, the right one the corresponding
ratio theory over experiment. EPOS$\,$LHC and EPOS3 are compatible
with the data, whereas QGSJETII seriously overpredicts baryon production
at low transverse momentum.

\begin{figure}[b]
\begin{minipage}[c][1\totalheight]{1\columnwidth}%
\begin{flushleft}
(a)\vspace{-1.5cm}

\par\end{flushleft}

\hspace*{-0.1cm}\includegraphics[angle=270,scale=0.3]{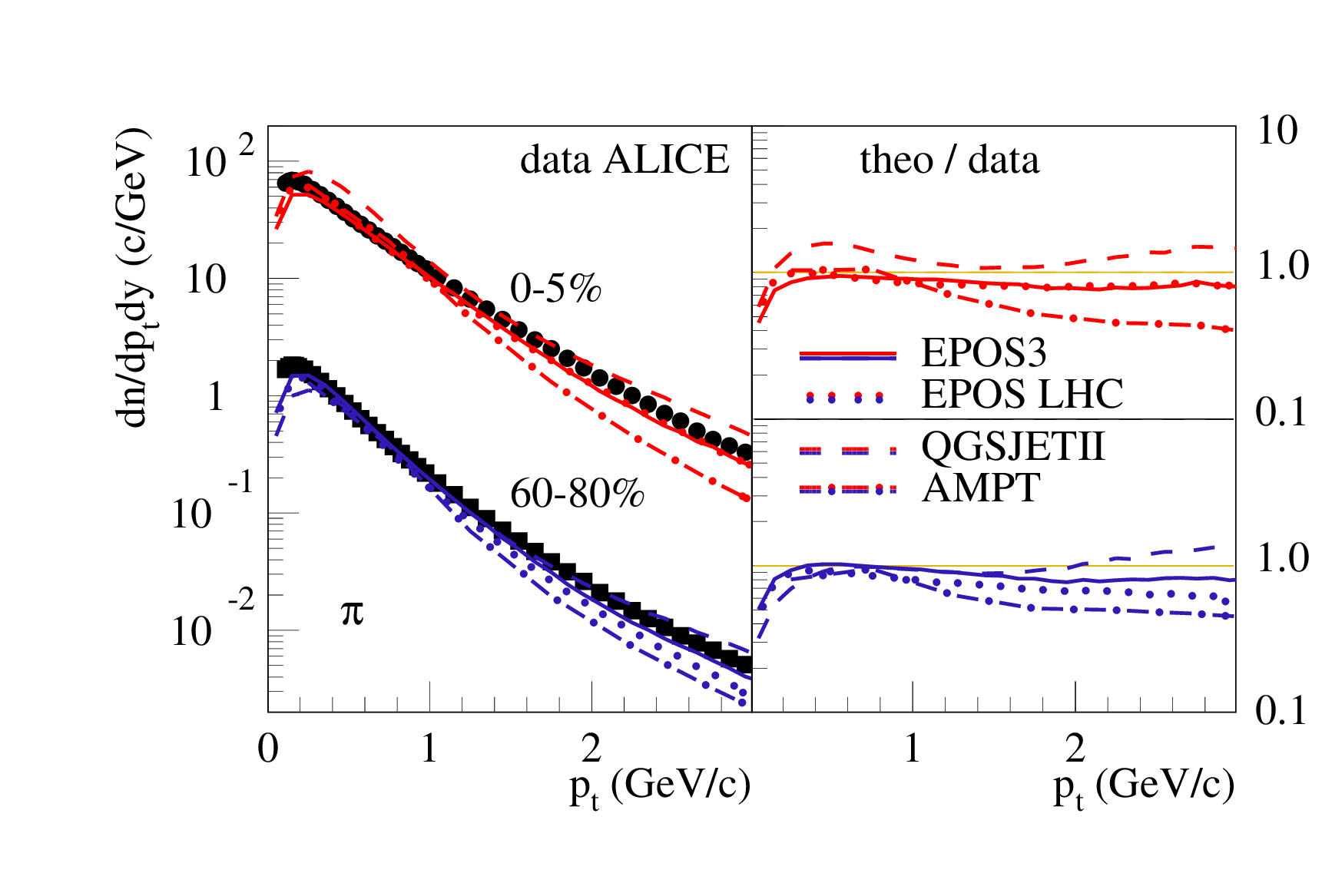}\vspace{-1cm}

\begin{flushleft}
(b)\vspace{-1.5cm}

\par\end{flushleft}

\hspace*{-0.1cm}\includegraphics[angle=270,scale=0.3]{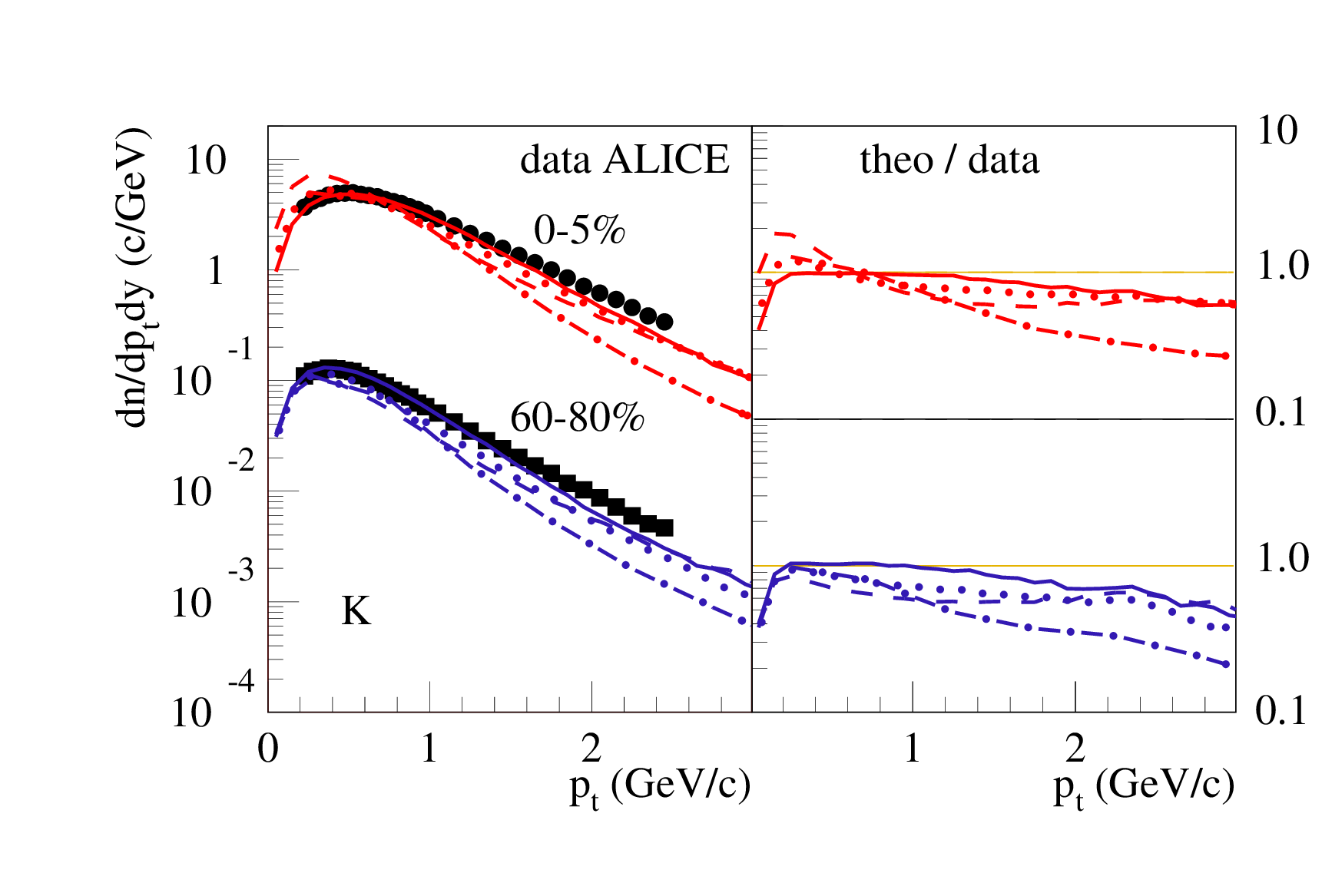}%
\end{minipage}

\noindent \caption{(Color online) (a) Transverse momentum spectra of charged pions in
p-Pb scattering at 5.02 TeV, for two different multiplicity classes:
0-5\% highest multiplicity (red, upper plots) and low multiplicity
events, 60-80\% (blue, lower plots). We show data from ALICE \citet{alice}
(symbols) and simulations from QGSJETII (dashed lines), AMPT (dashed-dotted),
EPOS$\,$LHC (dotted), and EPOS3 (solid). (b) Same as fig. (a), but
for charged kaons. \label{fig:selid01}}

\end{figure}
\begin{figure}[b]
\begin{minipage}[c][1\totalheight]{1\columnwidth}%
\begin{flushleft}
(a)\vspace{-1.5cm}

\par\end{flushleft}

\hspace*{-0.1cm}\includegraphics[angle=270,scale=0.3]{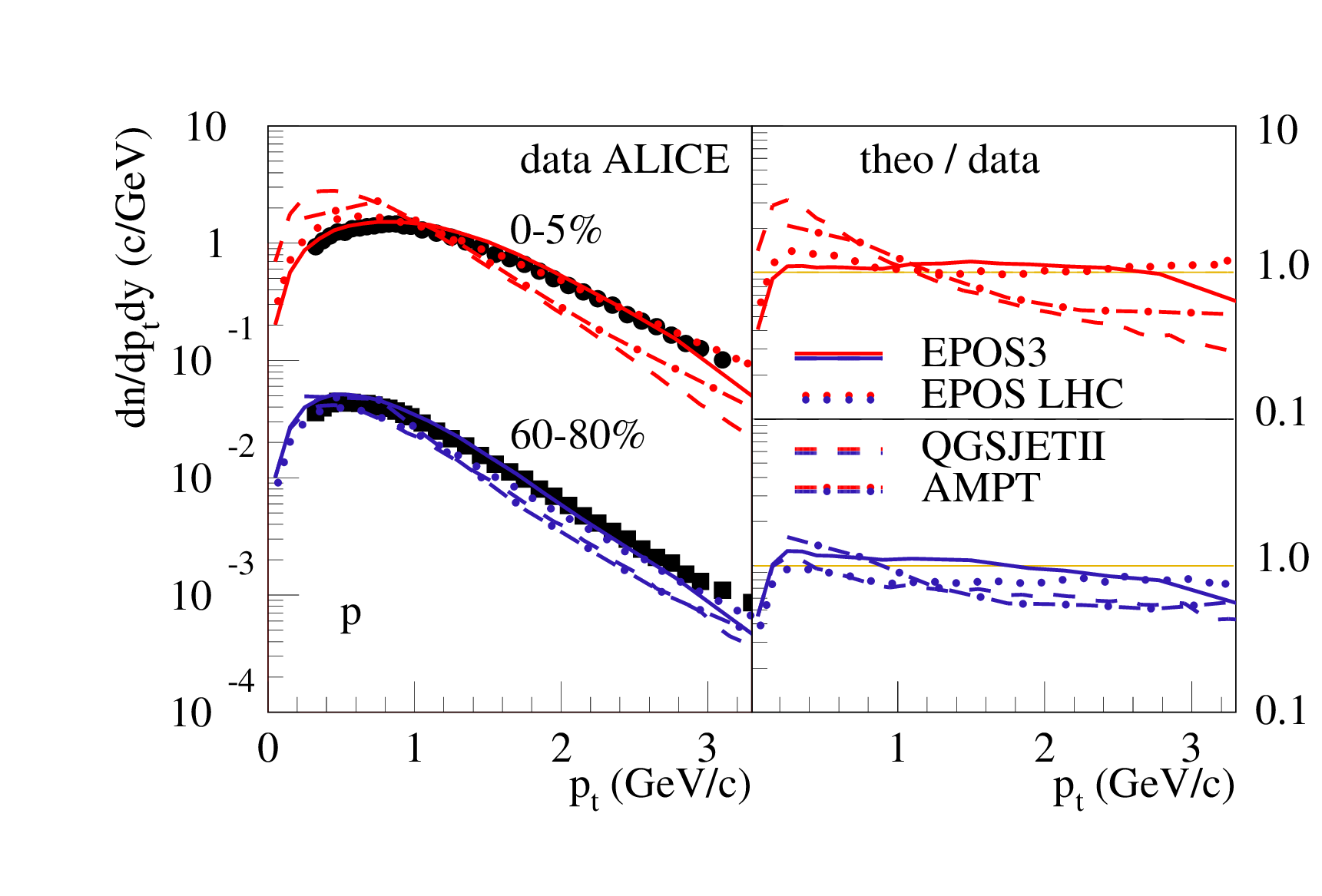}\vspace{-1cm}

\begin{flushleft}
(b)\vspace{-1.5cm}

\par\end{flushleft}

\hspace*{-0.1cm}\includegraphics[angle=270,scale=0.3]{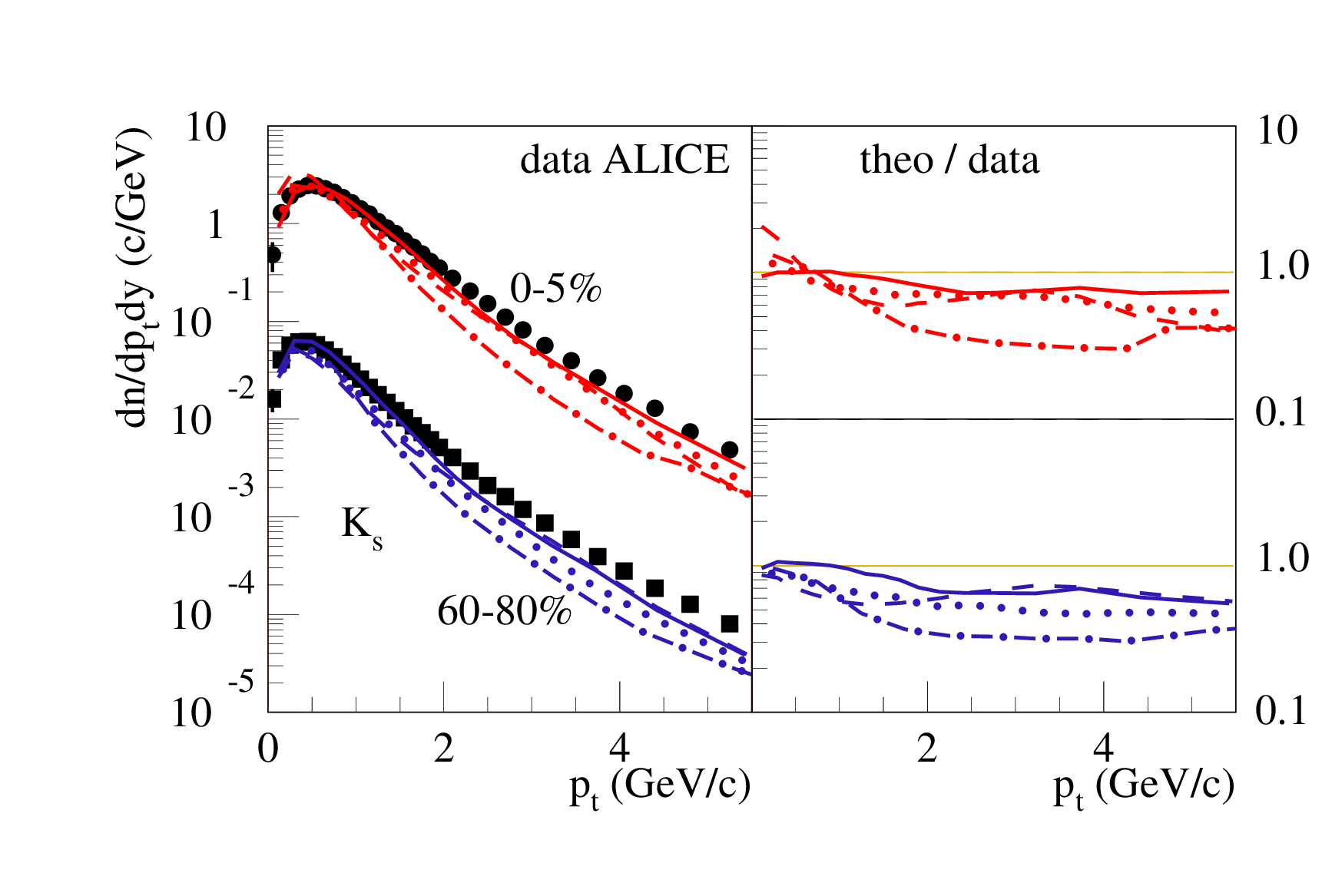}\vspace{-1cm}

\begin{flushleft}
(c)\vspace{-1.5cm}

\par\end{flushleft}

\hspace*{-0.1cm}\includegraphics[angle=270,scale=0.3]{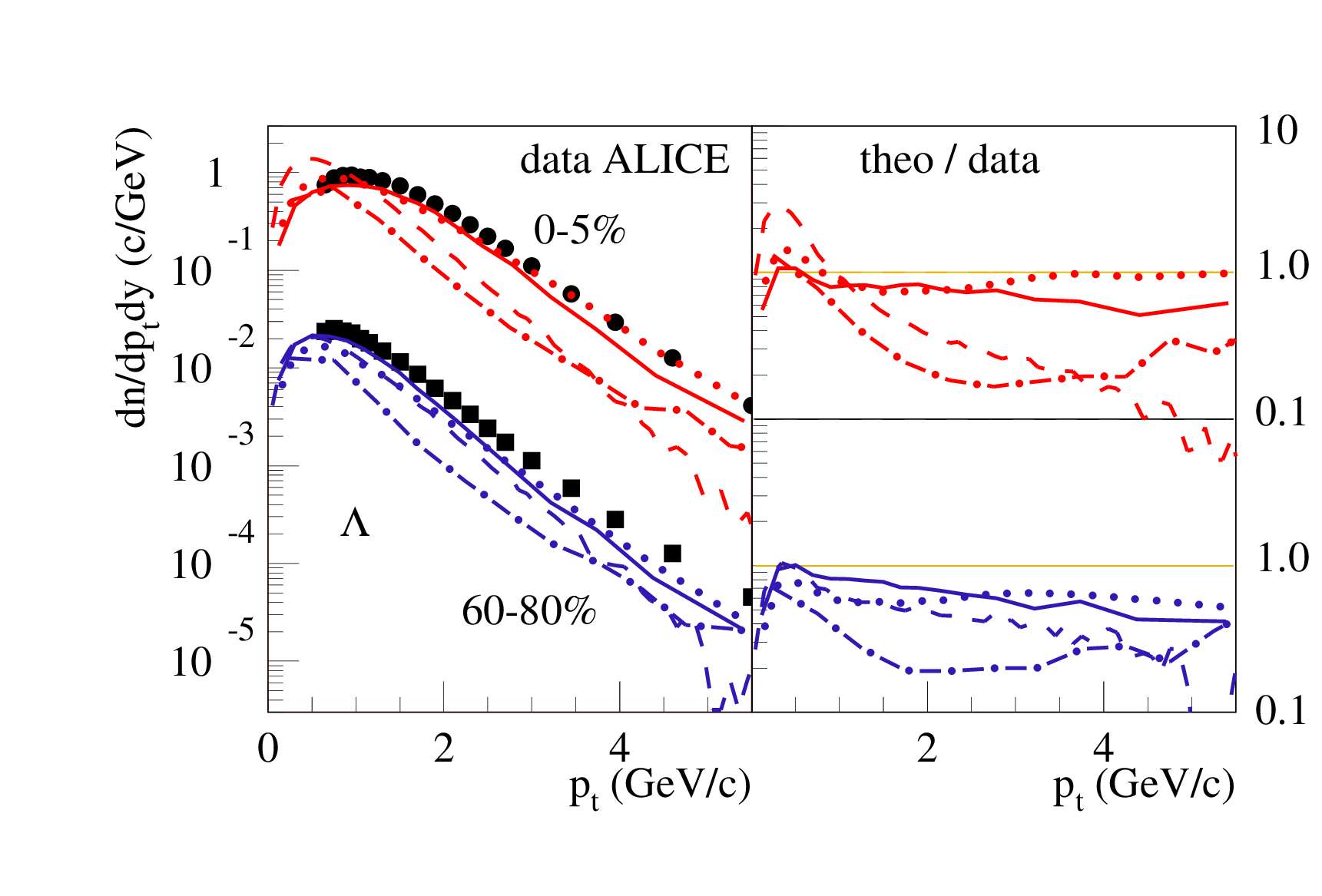}%
\end{minipage}

\noindent \caption{(Color online) Same as fig. \ref{fig:selid01}, but for protons (a),
neutral kaons ($K_{s})$ (b), and lambdas (c). \label{fig:selid03}}

\end{figure}

Also ALICE has measured identified particle spectra for different
multiplicities in p-Pb scattering at 5.02~TeV. The multiplicity counts
the number of charged particles in the range $2.8<\eta_{\mathrm{lab}}<5.1$.
Many classes have been considered, in this work, we consider the high
multiplicity (0-5\%) and the low multiplicity (60-80\%) classes. 

In fig. \ref{fig:selid01}, we show transverse momentum spectra of
charged pions and kaons in p-Pb scattering at 5.02 TeV, for the 0-5\%
and 60-80\% highest multiplicity event classes, referred to as ''high
multiplicity'' and {}``low multiplicity'' events. We show data
from ALICE \citet{alice} (symbols) and simulations from QGSJETII
(dashed lines), AMPT (dashed-dotted), EPOS$\,$LHC (dotted), and EPOS3
(solid). In fig. \ref{fig:selid03}, we show the corresponding results
for protons, neutral kaons, and lambdas. We show always the spectra
for data and theory (left) as well as the ratios theory over data
(right). For the latter ones, we use logarithmic scales (with a range
0.1 to 10), since in some cases the models are off (compared to data)
by more than a factor of 10! 

\begin{figure}[tb]
\begin{minipage}[t][1\totalheight]{1\columnwidth}%
\begin{minipage}[c][1\totalheight]{1\columnwidth}%
\vspace*{-0.2cm}

\hspace*{-0.3cm}\includegraphics[angle=270,scale=0.2]{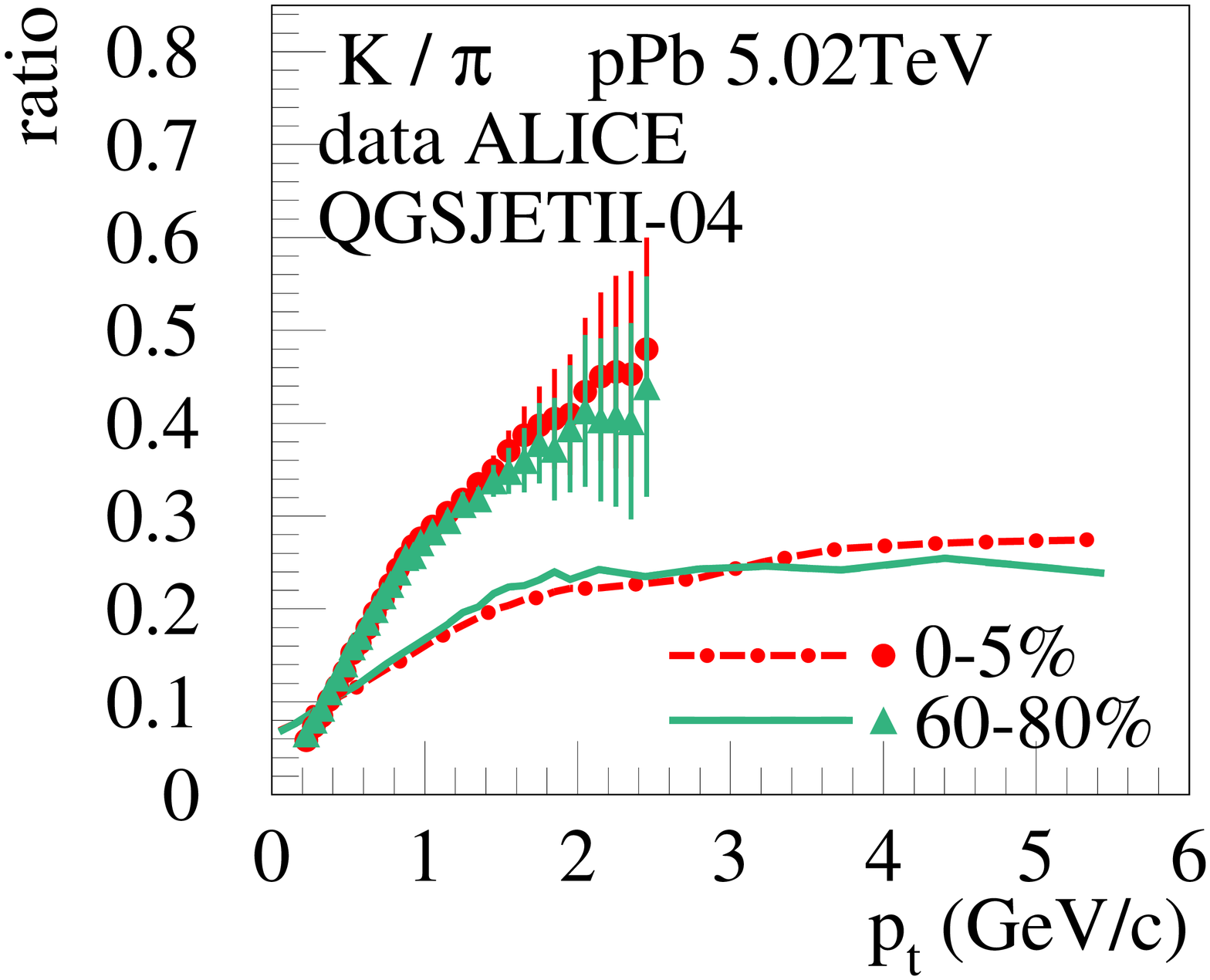}\hspace*{-0.5cm}\includegraphics[angle=270,scale=0.2]{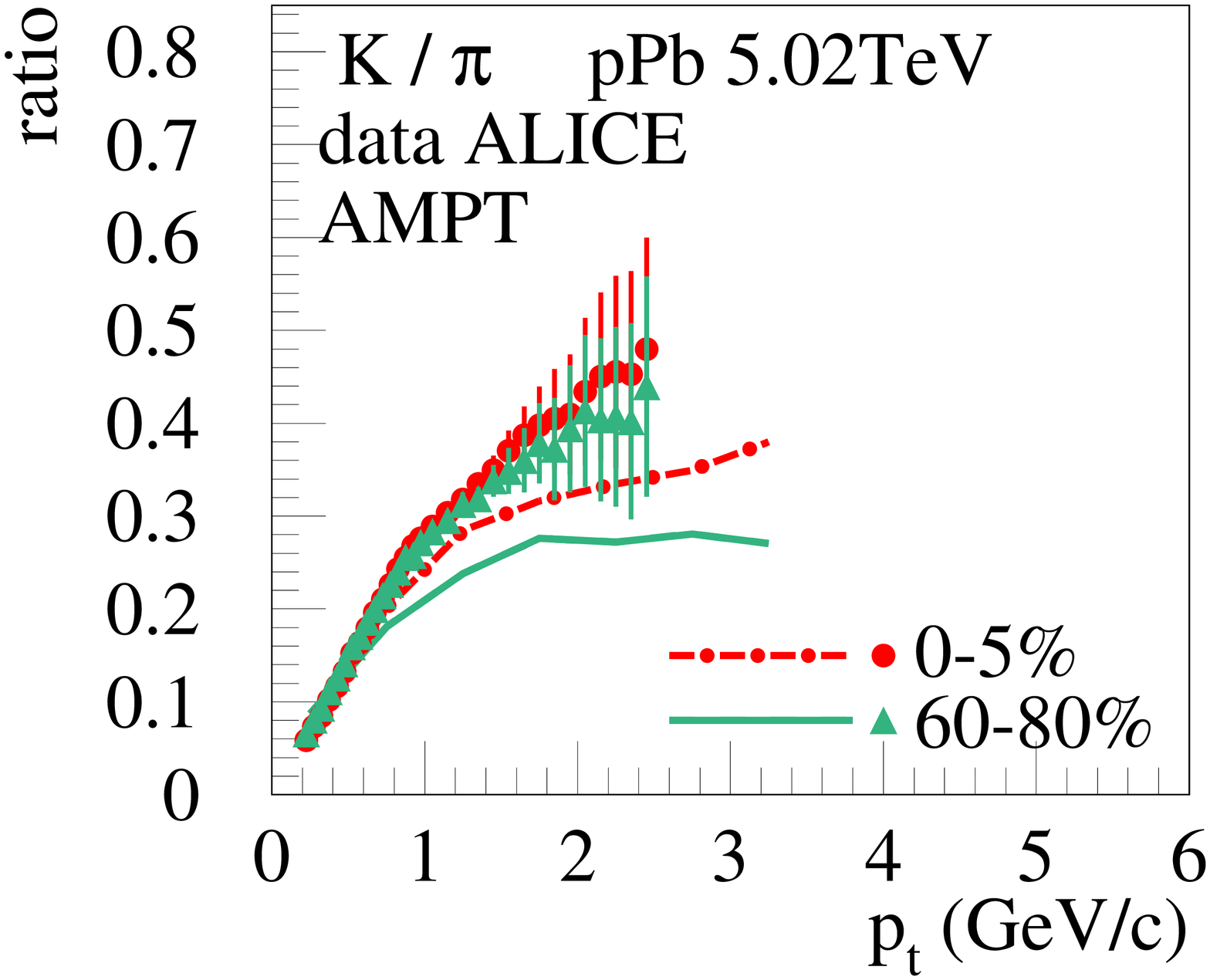}\vspace*{-0.2cm}

\hspace*{-0.3cm}\includegraphics[angle=270,scale=0.2]{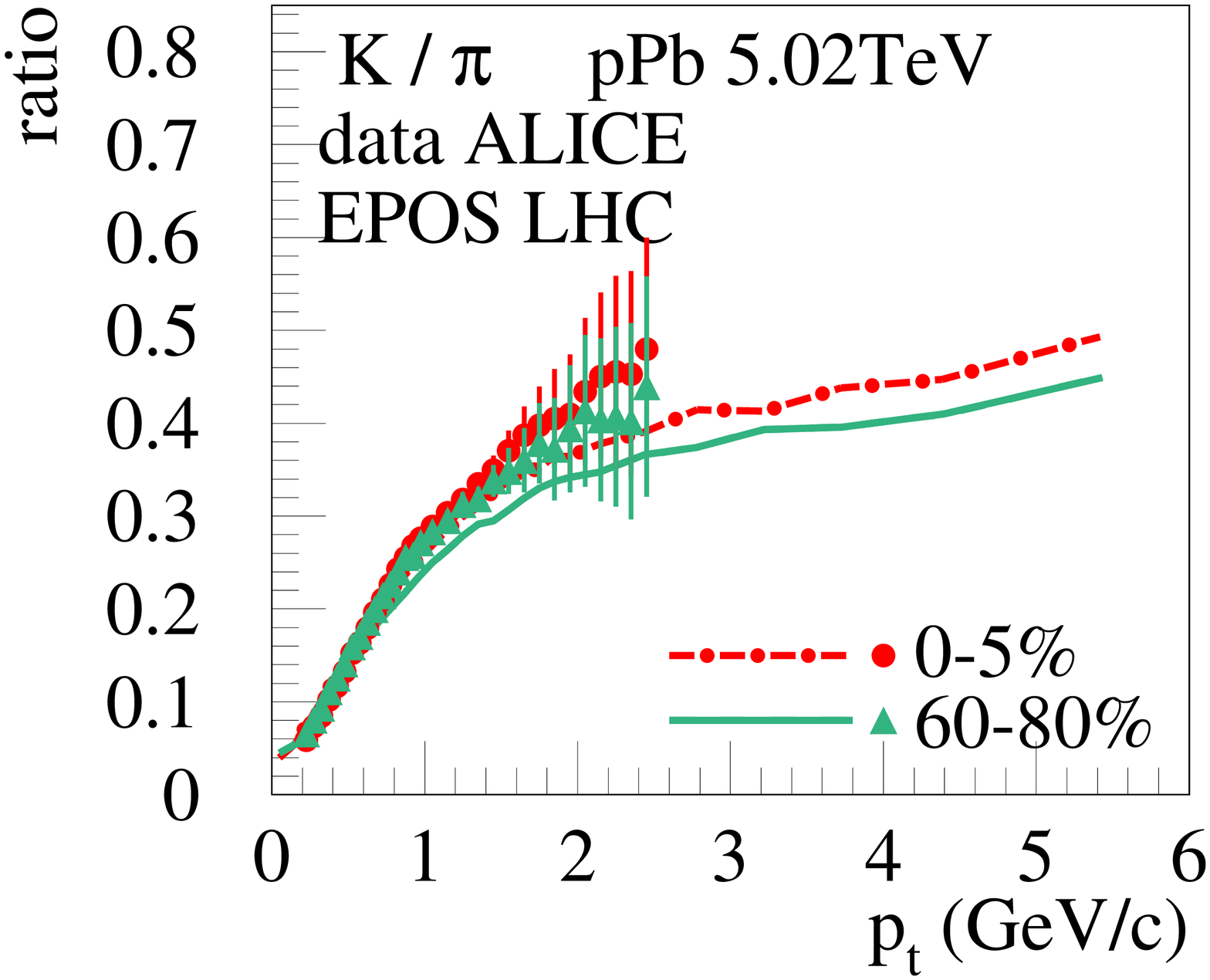}\hspace*{-0.5cm}\includegraphics[angle=270,scale=0.2]{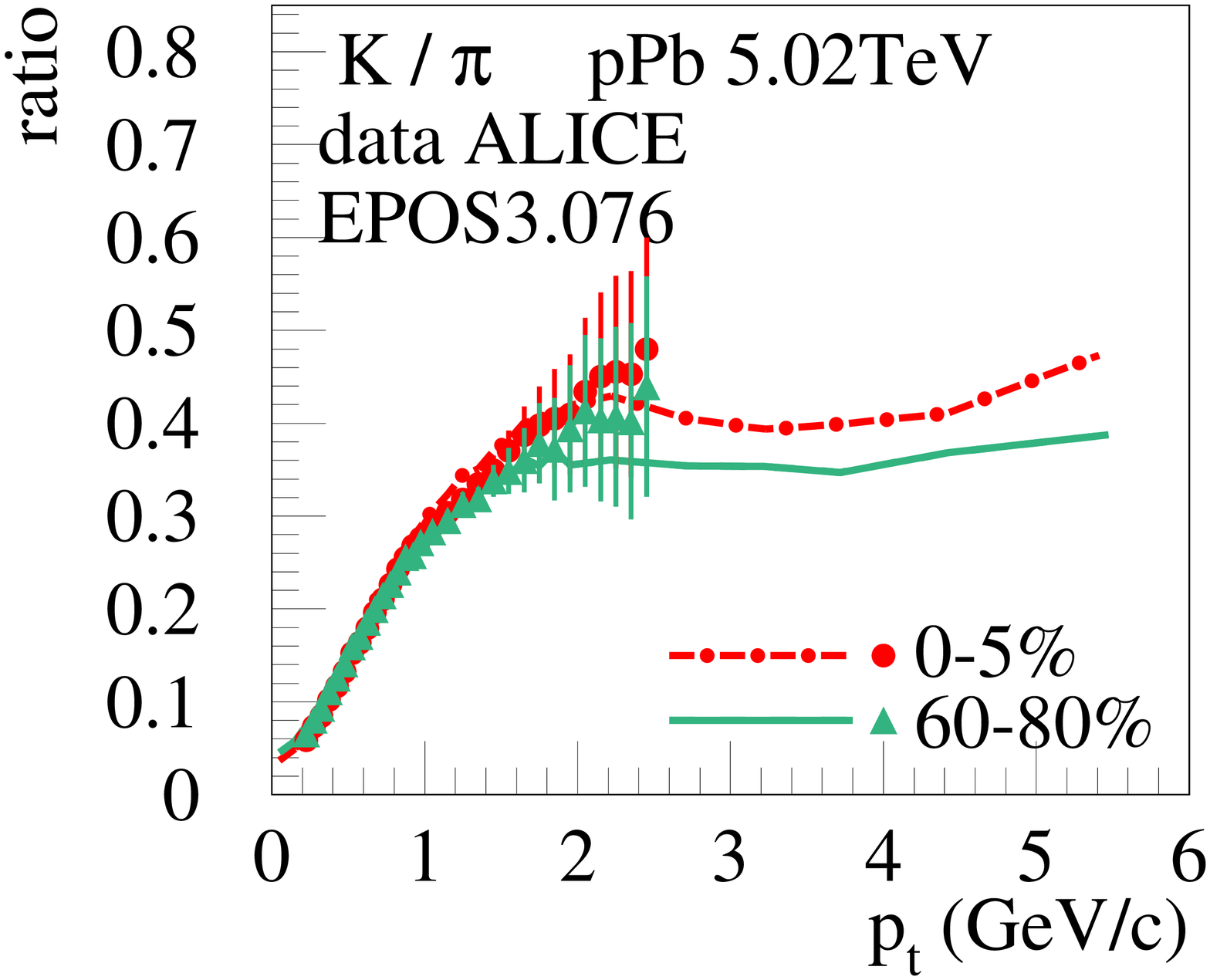}%
\end{minipage}%
\end{minipage}

\noindent \caption{(Color online) Kaon over pion ratio as a function of transverse momentum
in p-Pb scattering at 5.02 TeV, for two different multiplicity classes:
0-5\% highest multiplicity (red dashed-dotted lines, circles) and
low multiplicity events, 60-80\% (green solid lines, triangles). We
show data from ALICE \citet{alice} (symbols) and simulations from
QGSJETII, AMPT, EPOS$\,$LHC, EPOS3 (lines). \label{fig:selid07}}

\end{figure}
\begin{figure}[b]
\begin{minipage}[t][1\totalheight]{1\columnwidth}%
\begin{minipage}[c][1\totalheight]{1\columnwidth}%
\vspace*{-0.2cm}

\hspace*{-0.3cm}\includegraphics[angle=270,scale=0.2]{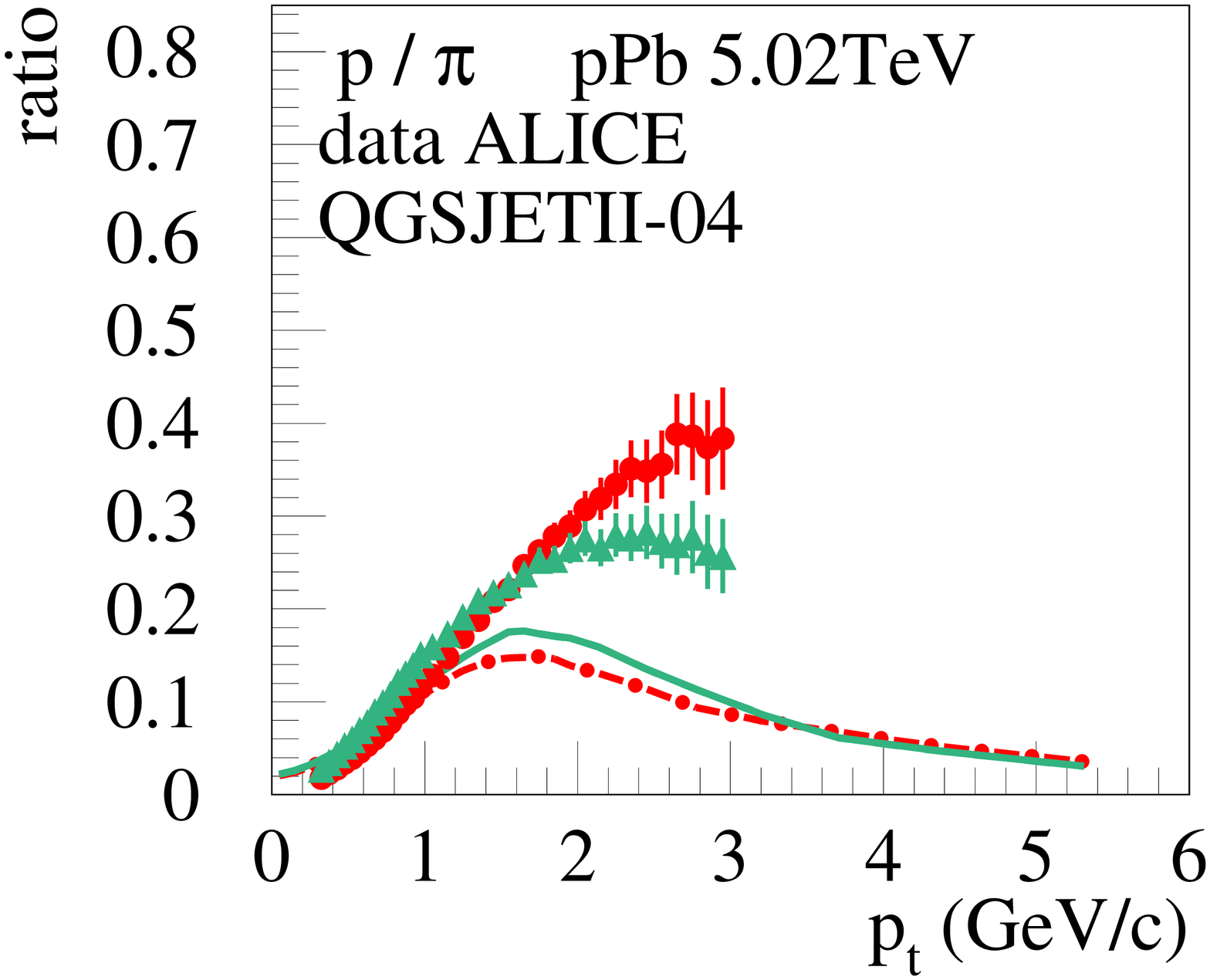}\hspace*{-0.5cm}\includegraphics[angle=270,scale=0.2]{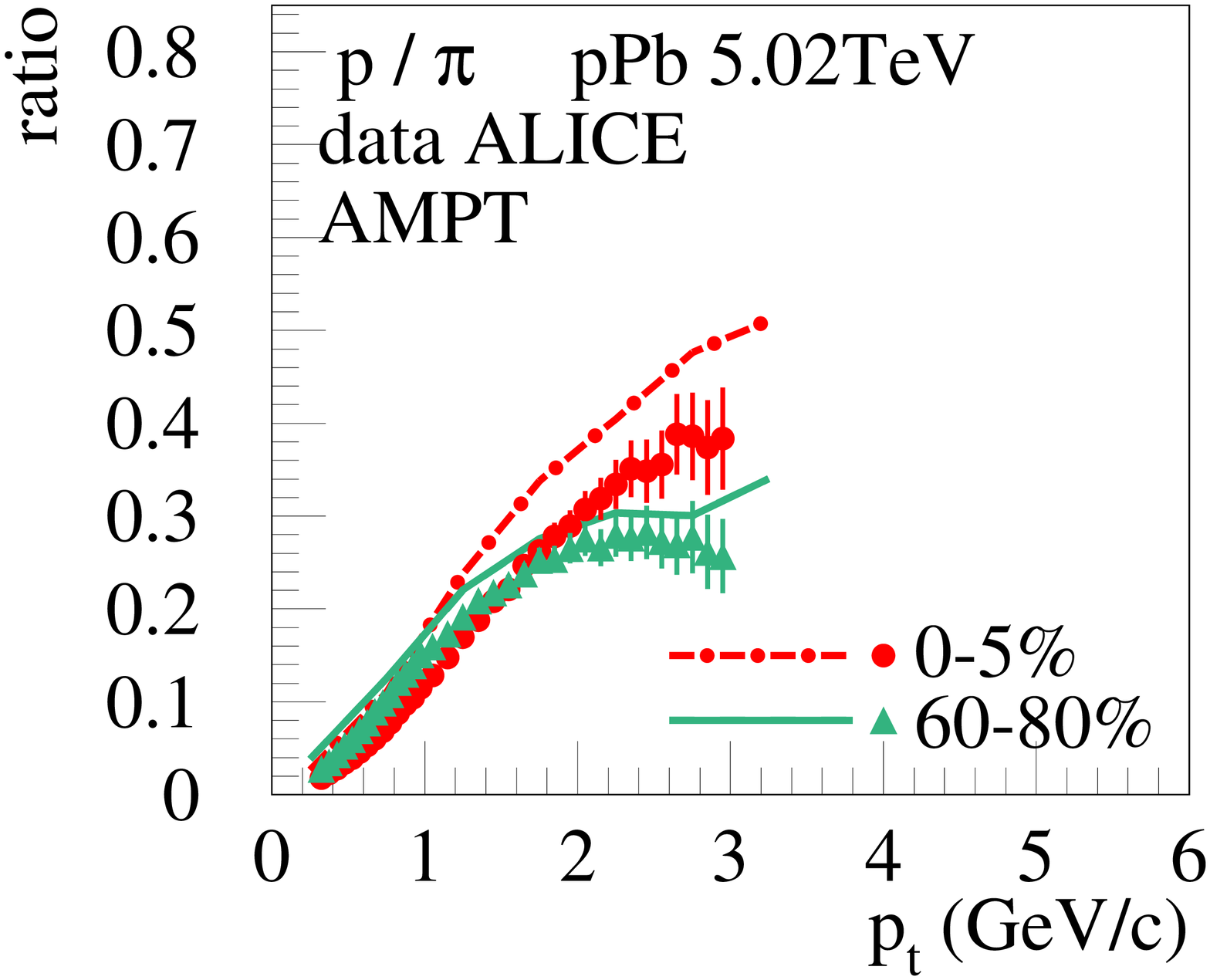}\vspace*{-0.2cm}

\hspace*{-0.3cm}\includegraphics[angle=270,scale=0.2]{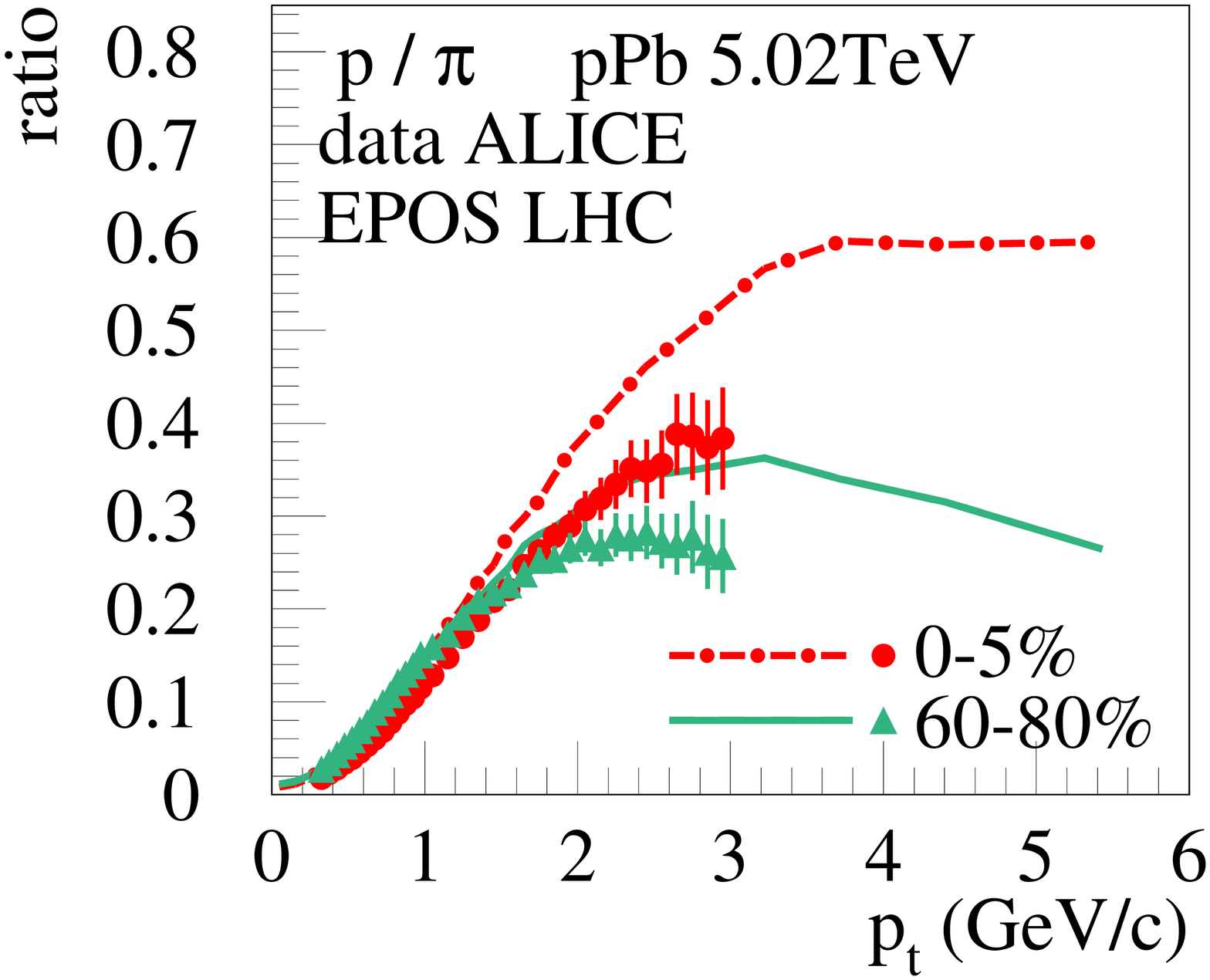}\hspace*{-0.5cm}\includegraphics[angle=270,scale=0.2]{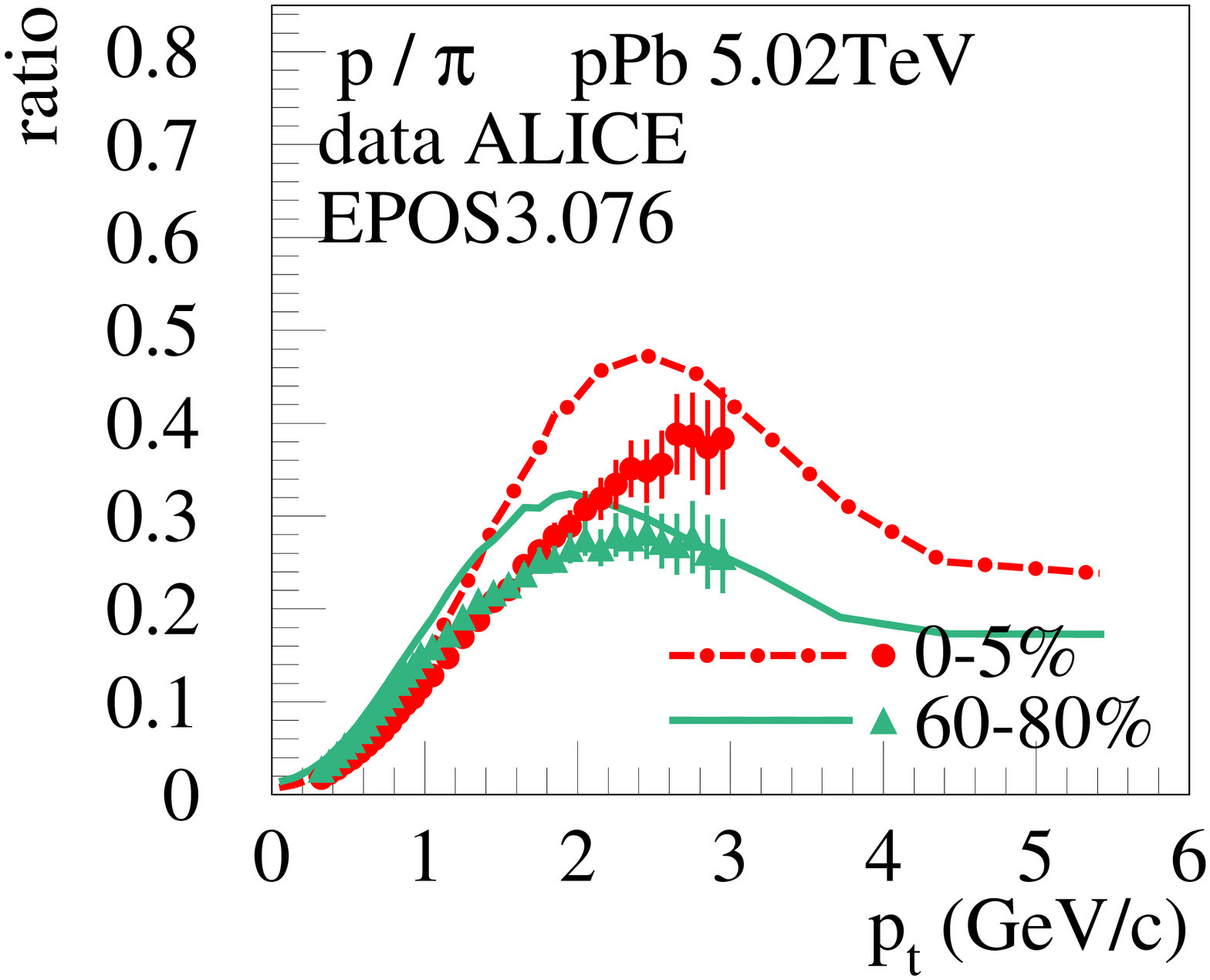}%
\end{minipage}%
\end{minipage}

\noindent \caption{(Color online) Same as fig. \ref{fig:selid07}, but proton over pion
ratio. \label{fig:selid12}}

\end{figure}
\begin{figure}[t]
\begin{minipage}[t][1\totalheight]{1\columnwidth}%
\begin{minipage}[c][1\totalheight]{1\columnwidth}%
\vspace*{-0.2cm}

\hspace*{-0.3cm}\includegraphics[angle=270,scale=0.2]{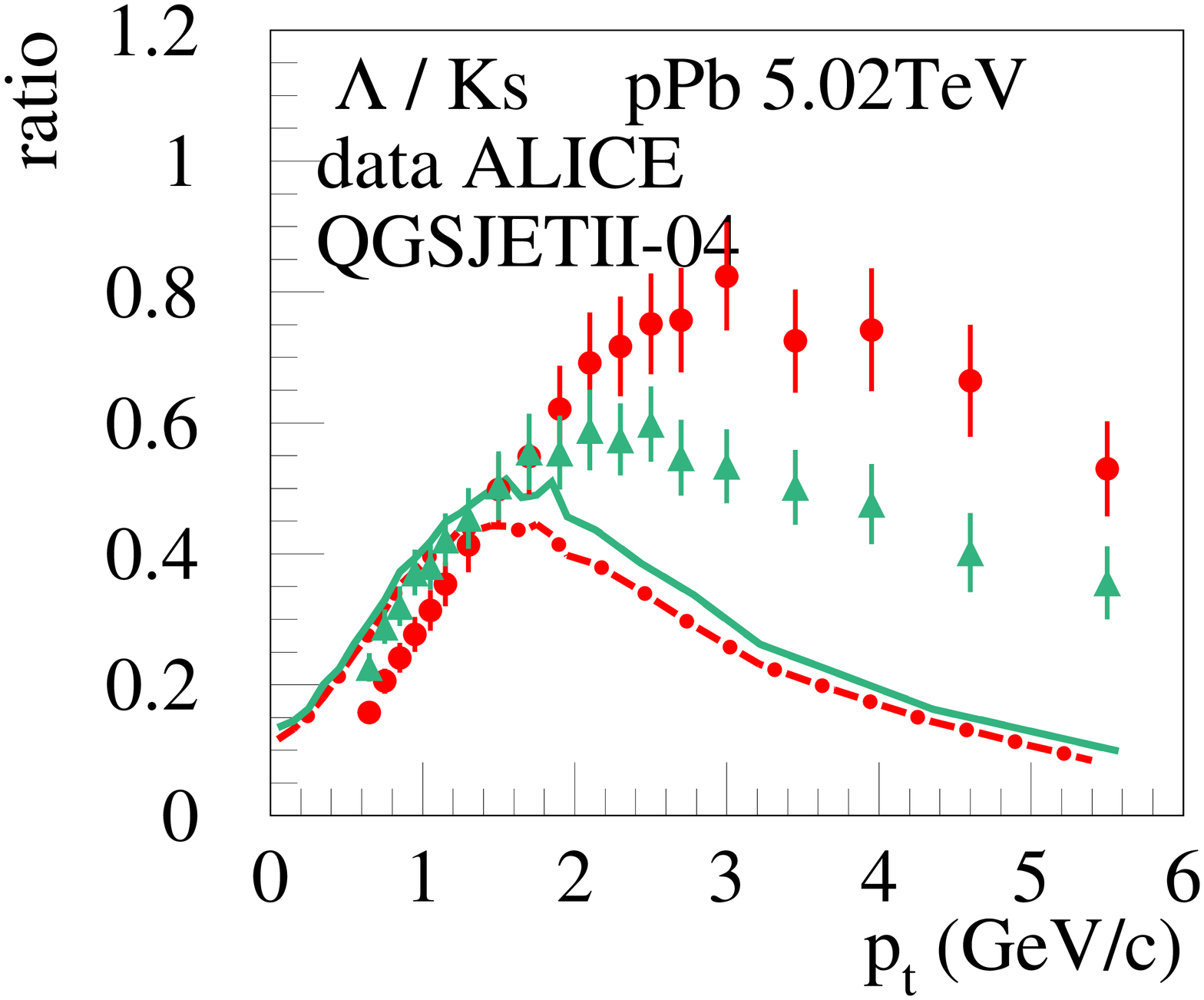}\hspace*{-0.5cm}\includegraphics[angle=270,scale=0.2]{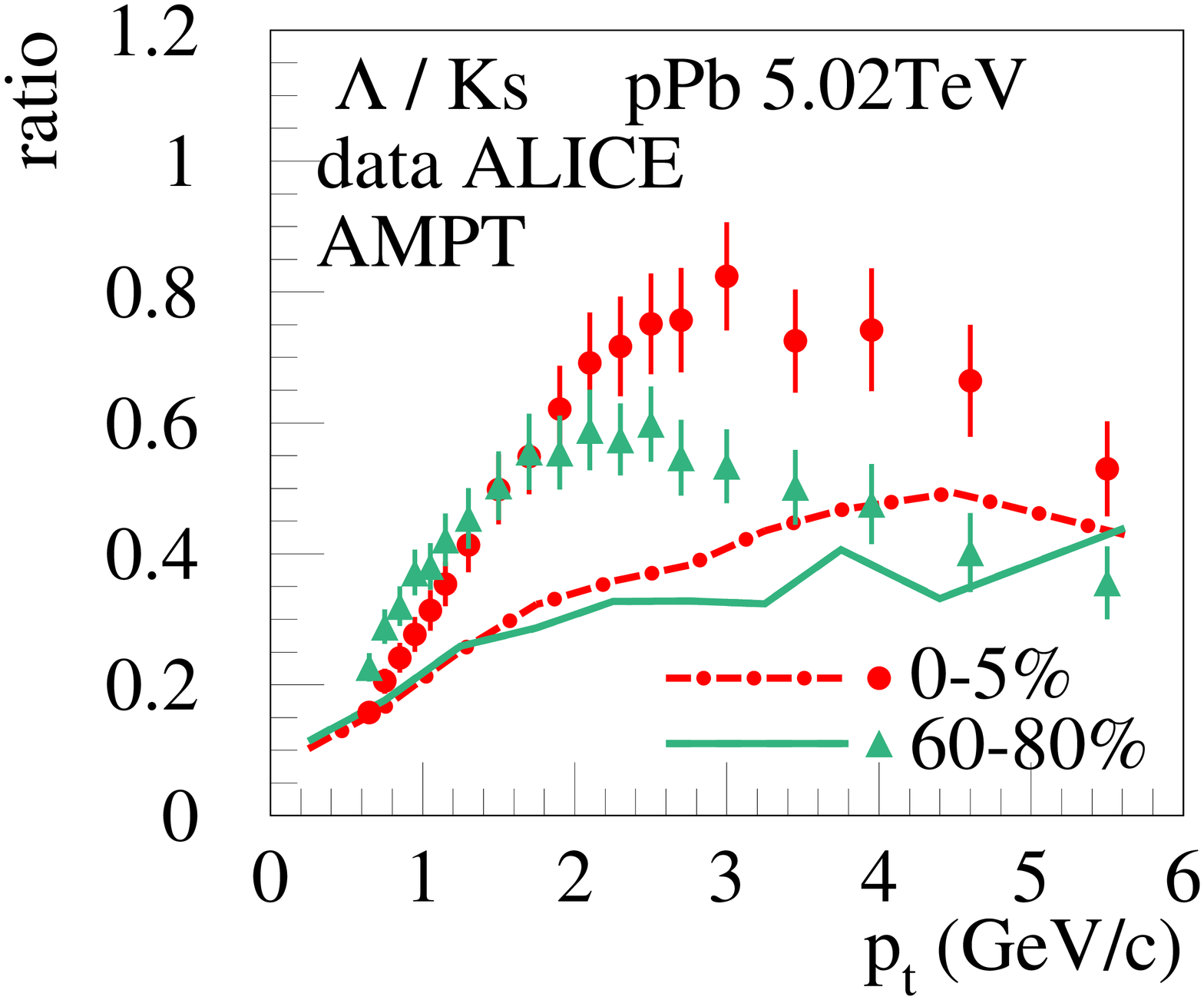}\vspace*{-0.2cm}

\hspace*{-0.3cm}\includegraphics[angle=270,scale=0.2]{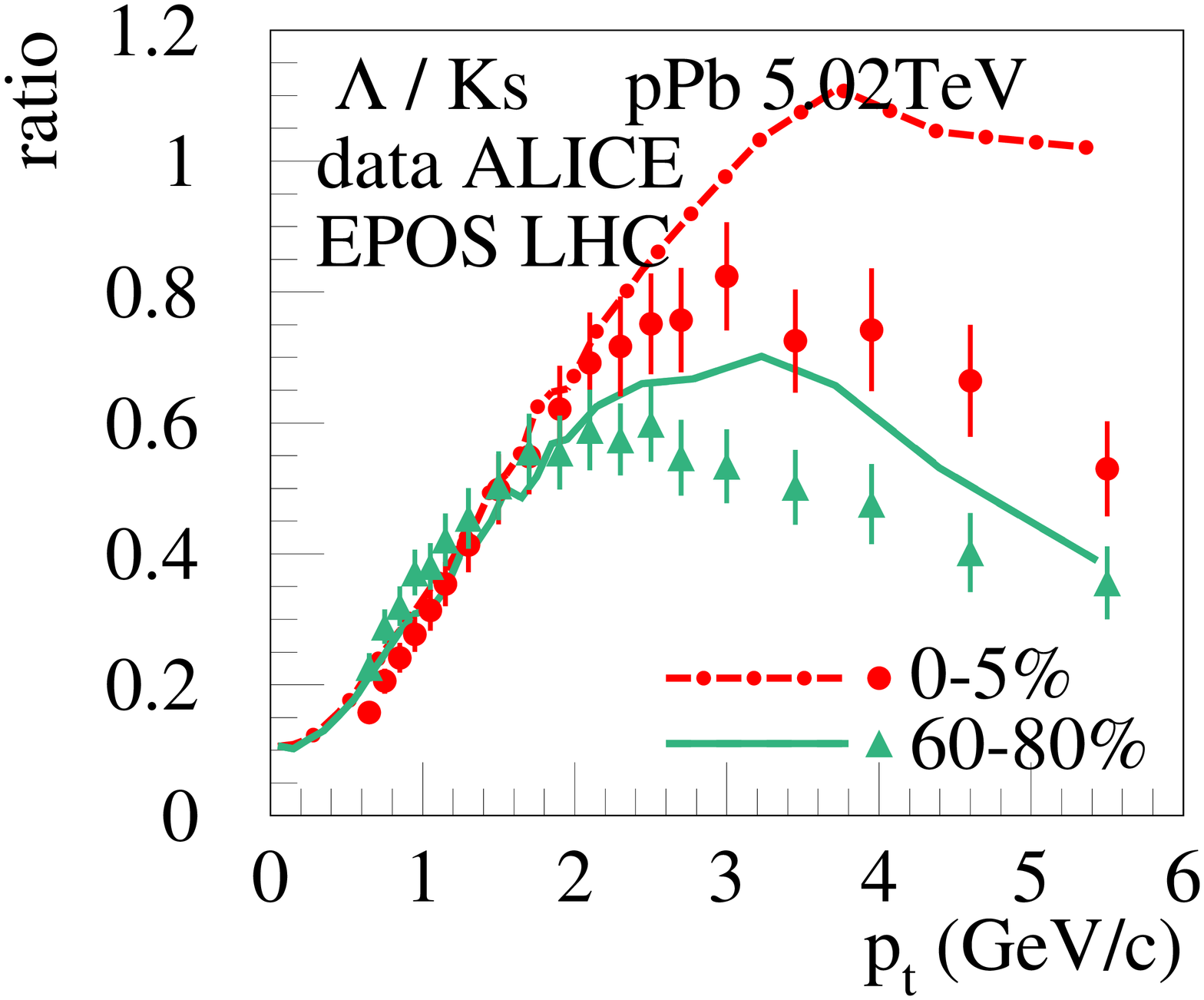}\hspace*{-0.5cm}\includegraphics[angle=270,scale=0.2]{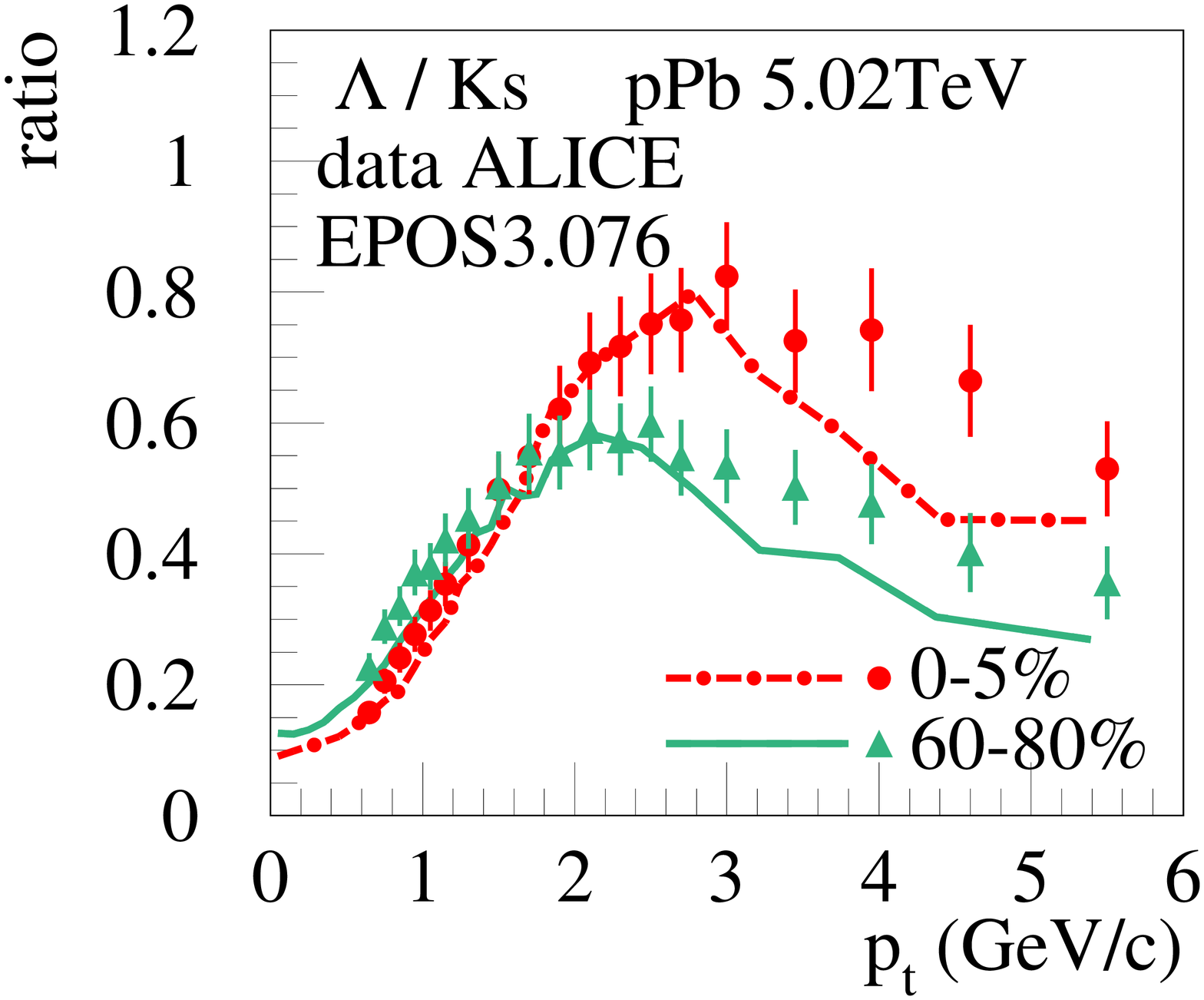}%
\end{minipage}%
\end{minipage}

\noindent \caption{(Color online) Same as fig. \ref{fig:selid07}, but $\Lambda$ over
$K_{s}$  ratio. \label{fig:selid17}}

\end{figure}

In \citet{alice}, the authors discuss in detail the multiplicity
dependence, in particular by investigating particle ratios for {}``high
multiplicity'' and {}``low multiplicity'' events. But what is clear
from the figures \ref{fig:selid01} and \ref{fig:selid03}: the models
without flow or with little flow (QGSJETII, AMPT) considerably underpredict
the intermediate $p_{t}$ range, in particular for the baryons (protons
and lambdas), for both {}``high multiplicity'' and {}``low multiplicity''
events. It is again the ''flow effect'' in EPOS$\,$LHC and EPOS3
which helps, pushing heavier particles to higher $p_{t}$ values (in
the range 2-4 GeV/c). It should be noted that even for the {}``low
multiplicity'' events, flow is needed. As we have shown in fig. \ref{z-selb-10.ps}
for EPOS3, even for {}``peripheral events'' (60-80\%) the core (=flow)
already contributes. 

Flow seems to be always present, for all multiplicities (or centralities),
just the relative importance of the core part (and therefore the flow)
increases (moderately) with multiplicity. This leads immediately to
the question of collective flow in proton-proton scattering -- which
we are going to address later.

Nevertheless, it is useful to study the multiplicity dependence, best
done by looking at ratios. In fig. \ref{fig:selid07}, we show the
pion over kaon ($K/\pi$) ratio as a function of transverse momentum
 in p-Pb scattering at 5.02 TeV, for high multiplicity (red dashed-dotted
lines, circles) and low multiplicity events (green solid lines, triangles),
comparing data from ALICE \citet{alice} (symbols) and simulations
from QGSJETII, AMPT, EPOS$\,$LHC, and EPOS3 (lines). In all models,
as in the data, there is little multiplicity dependence. However,
the QGSJETII model is considerably below the data, for both high and
low multiplicity events. AMPT is slightly below, whereas EPOS$\,$LHC
and EPOS3 do a reasonable job. Concerning the proton over pion ($p/\pi$)
ratio, fig. \ref{fig:selid12}, again QGSJETII is way below the data,
for both high and low multiplicity events, whereas the three other
models show the trend correctly, but being slightly above the data.
Most interesting are the lambdas over kaon ($\Lambda/K_{s}$) ratios,
as shown in fig. \ref{fig:selid17}, because here a wider transverse
momentum range is considered, showing a clear peak structure with
a maximum around 2-3 GeV/c and a slightly more pronounced peak for
the higher multiplicities. QGSJETII and AMPT cannot (even qualitatively)
reproduce this structure. EPOS$\,$LHC shows the right trend, but
the peak is much too high for the high multiplicities. EPOS3 is close
to the data. 

To summarize these ratio plots (keeping in mind that the QGSJETII
model has no flow, AMPT {}``some'' flow, EPOS$\,$LHC a parametrized
flow, and EPOS3 hydrodynamic flow): Flow seems to help considerably.
However, from the $\Lambda/K_{s}$ ratios, we conclude that EPOS$\,$LHC
uses a too strong radial flow for high multiplicity events. The hydrodynamic
flow employed in EPOS3 seems to get the experimental features reasonably
well. Crucial is the core-corona procedure discussed earlier: there
is more core (compared to corona) in more central collisions, but
the centrality (or multiplicity) dependence is not so strong, and
there is already an important core (=flow) contribution in peripheral
events.

\section{Proton-proton scattering at 7 TeV}

From our above studies of p-Pb scattering at 5.02 TeV, we conclude
that hydrodynamical flow seems to play an important role, similar
as in heavy ion collisions, contrary to all expectations. Even more
surprisingly, these hydrodynamical features already appear in peripheral
(or low multiplicity) p-Pb events, being close to proton-proton scatterings.
So after being obliged to give up the common prejudice that proton-nucleus
scattering is a simple {}``base line'' compared to the hydrodynamically
evolving heavy ion collisions, do we have to do so for proton-proton
scattering as well?

To answer this question, we will investigate identified particle production
in EPOS3, compared to experimental data and many other models, see
table \ref{tab:List-of-models}.%
\begin{table}[b]

\begin{centering}
\begin{tabular}{|c|c|c|c|}
\hline 
Model & Theoretical & Flow & Ref.\tabularnewline
 & concept &  & \tabularnewline
\hline
\hline 
EPOS3.076 & GR & hydro & this paper\tabularnewline
\hline 
EPOS$\,$LHC & GR & parametrized & \citet{eposlhc}\tabularnewline
\hline 
QGSJETII-04 & GR & no & \citet{qgsjet}\tabularnewline
\hline 
SIBYLL2.1 & GR & no & \citet{mc-sybill}\tabularnewline
\hline 
PHOJET1.12a & GR & no & \citet{mc-phojet}\tabularnewline
\hline 
AMPT & PHC & no (in pp) & \citet{ampt}\tabularnewline
\hline 
PYTHIA6.4.27 & Fact & no & \citet{mc-pythia6}\tabularnewline
\hline 
PYTHIA8.170 & Fact & no & \citet{mc-pythia8}\tabularnewline
\hline 
HERWIG++2.6.1a & Fact & no & \citet{mc-herwig}\tabularnewline
\hline
SHERPA1.4.1 & Fact & no & \citet{mc-sherpa}\tabularnewline
\hline
\end{tabular}
\par\end{centering}

\caption{List of models used to analyse identified particle production in proton-proton
scattering at 7 TeV. {}``GR'' stands for Gribov-Regge approach,
{}``PHC'' for partonic and hadronic cascade, {}``Fact'' for factorization
approach.\label{tab:List-of-models}}

\end{table}
\begin{table}[tb]
\begin{centering}
\includegraphics[angle=270,scale=0.18]{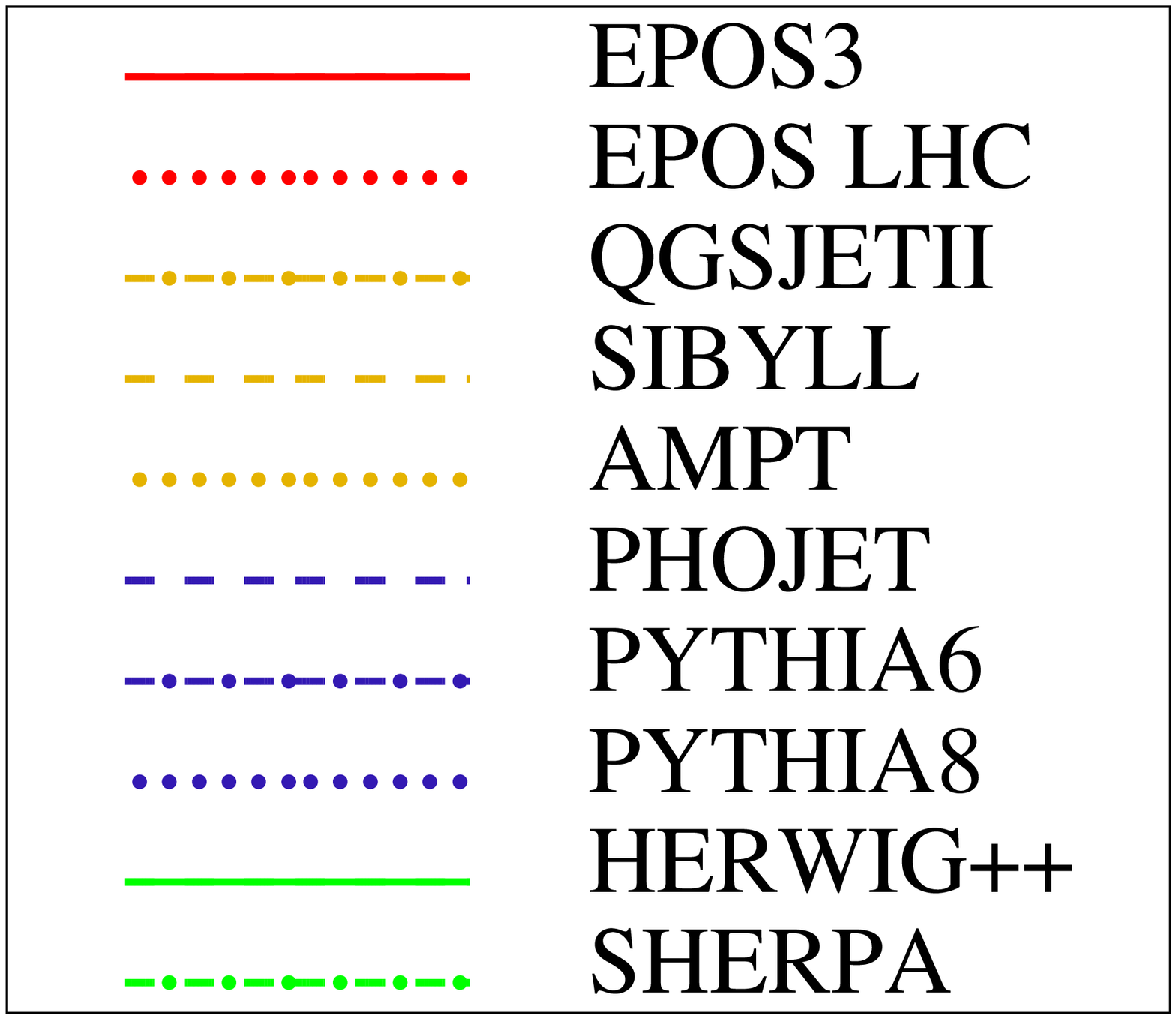}
\par\end{centering}

\caption{(Color online) Line codes for the different models. \label{tab:selpi22}}

\end{table}
The QGSJETII \citet{qgsjet}, SIBYLL \citet{mc-sybill}, and PHOJET
model \citet{mc-phojet} are also based on Gribov-Regge multiple scattering,
but there is no fluid component. The main ingredients of the AMPT
model \citet{ampt} are a partonic cascade and a subsequent hadronic
cascade, providing in this way some {}``collectivity'' in nuclear
collisions, but not in proton-proton as studied here. EPOS$\,$LHC\citet{eposlhc}
is a tune of EPOS1.99, containing flow put in by hand, parametrizing
the collective flow at freeze-out. The EPOS3 approach contains a full
viscous hydrodynamical simulation. In addition, we will also show
results from the so-called {}``general-purpose event generators for
LHC physics'' \citet{mc-11}, as there are PYTHIA6 \citet{mc-pythia6},
PYTHIA8 \citet{mc-pythia8}, HERWIG++ \citet{mc-herwig}, and SHERPA
\citet{mc-sherpa}. All these models are based on the factorization
formula for inclusive cross sections, with a more or less sophisticated
treatment of multiple scattering, whereas Gribov-Regge theory provides
a multiple scattering scheme from the beginning.

\begin{figure}[b]
\begin{centering}
\includegraphics[angle=270,scale=0.3]{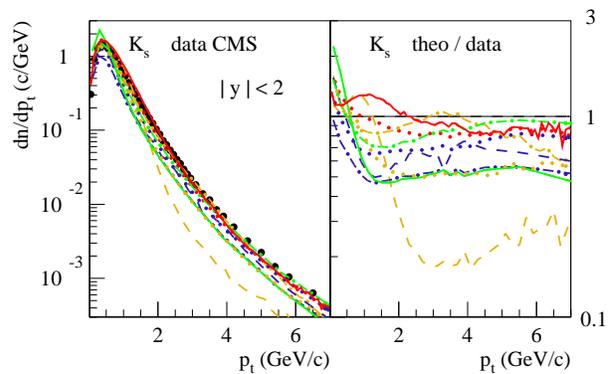}\\

\par\end{centering}

\caption{(Color online) Transverse momentum spectra (left) and ratios {}``theory
over data'' (right) of $K_{s}$ mesons in p-p scattering at 7 TeV.
We show data from CMS \citet{cmspp} (symbols) and simulations from
the different models, using the line codes defined in tab. \ref{tab:selpi22}.
\label{fig:selpi02}}

\end{figure}

We have learned from studying identified particle production in p-Pb
scattering that hydrodynamic flow helps enormously to quantitatively
reproduce experimental data, which show the typical {}``radial flow
effect'' of pushing intermediate $p_{t}$ particles to higher $p_{t}$
values, more and more pronounced with increasing particle mass. Huge
effects are seen for example for lambdas. We will do the corresponding
studies for proton-proton scattering at 7 TeV. In tab. \ref{tab:selpi22},
we show the line codes for the different models used in the following
plots. In fig. \ref{fig:selpi02}, %
\begin{figure}[tb]
\begin{centering}
\begin{minipage}[c][1\totalheight]{1\columnwidth}%
\begin{flushleft}
(a)\vspace{-0.8cm}

\par\end{flushleft}

\hspace*{0.3cm}\includegraphics[angle=270,scale=0.3]{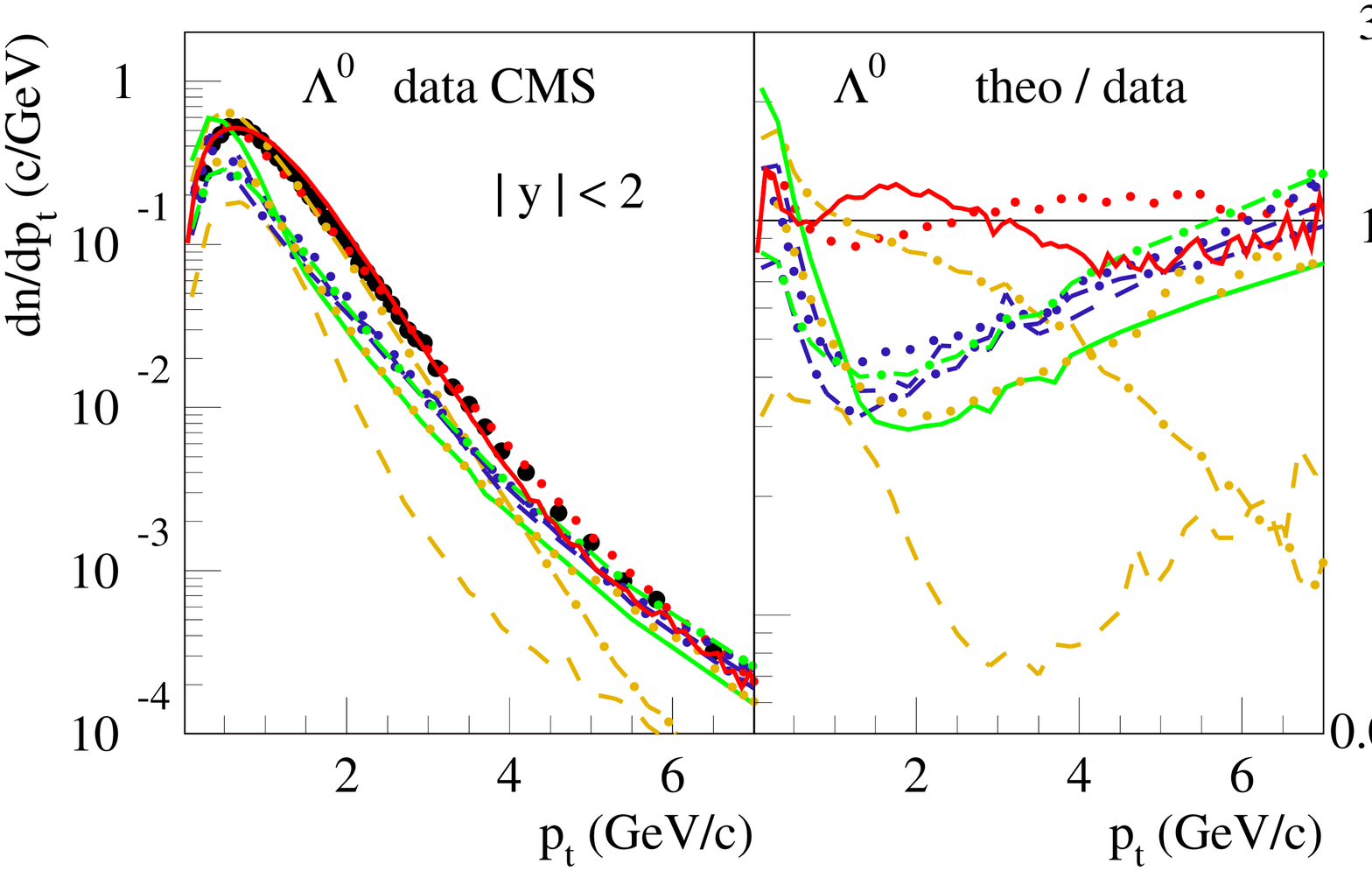}\vspace{-1cm}

\begin{flushleft}
(b)\vspace{-0.4cm}

\par\end{flushleft}

\hspace*{0.3cm}\includegraphics[angle=270,scale=0.3]{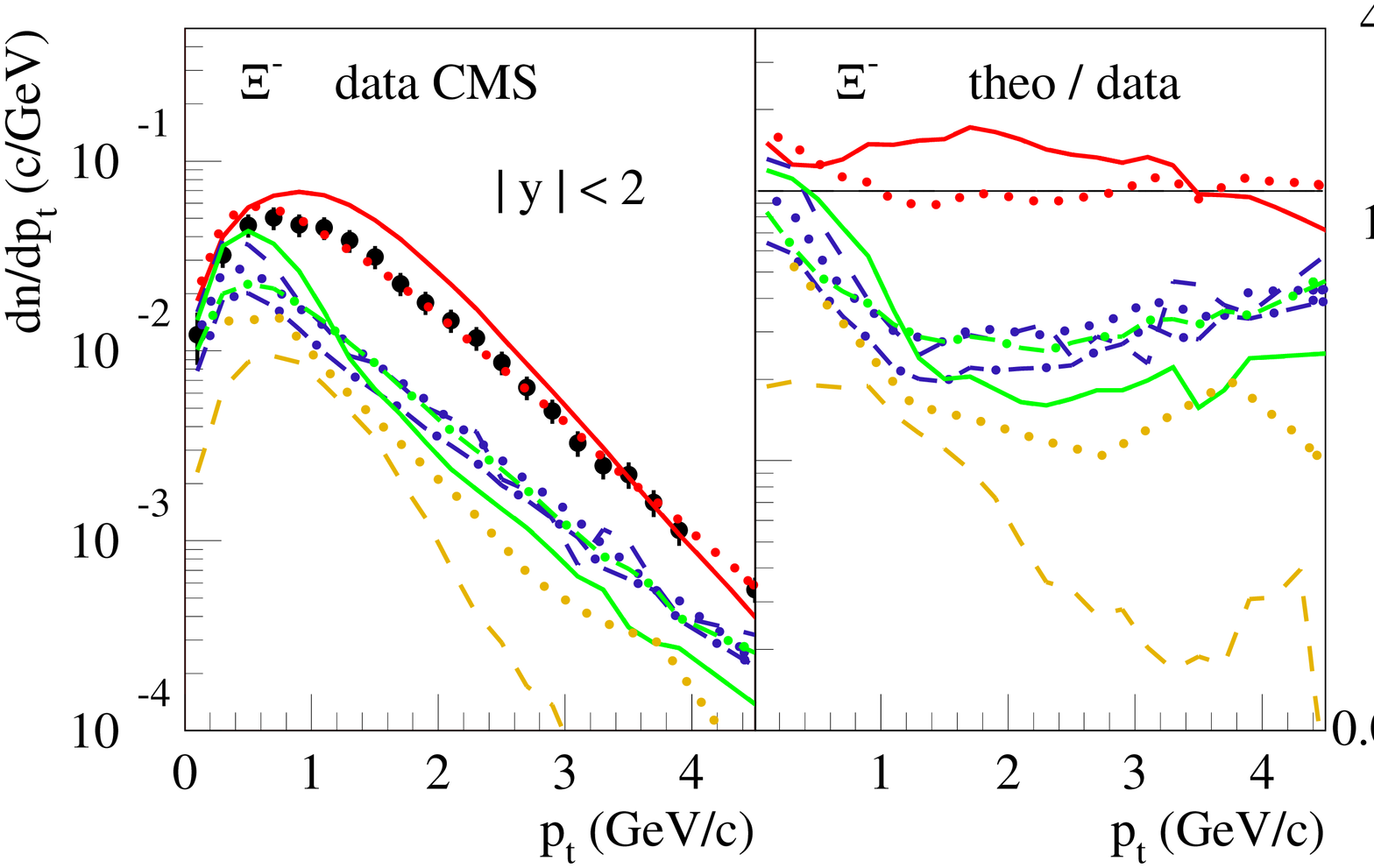}%
\end{minipage}
\par\end{centering}

\caption{(Color online) Same as fig. \ref{fig:selpi02}, but here we plot results
for $\Lambda$ baryons (a) and $\Xi$ baryons (b). CMS data from \citet{cmspp,cmsppmcp}.
\label{fig:selpi03}}

\end{figure}
we show the simulation results for $K_{s}$ production compared to
experimental data from CMS \citet{cmspp}. The best models are within
20\% of the data, others considerable below. In fig. \ref{fig:selpi03},
we show the corresponding results for $\Lambda$ and $\Xi$ baryons,
comparing simulations with data from CMS \citet{cmspp,cmsppmcp}.
Here one can distinguish three groups of models: (1) QGSJETII and
SIBYLL are far off the data, they are simply not constructed to produce
these kind of baryons. (2) The so-called QCD generators like PYTHIA,
HERWIG, SHERPA etc show a profound {}``dip'' in the region between
1 and 5 GeV/c, underpredicting the data by a factor of 4-5 for the
$\Xi$ baryons, and by a factor of around 3 for the $\Lambda$ baryons.
(3) The two EPOS versions are relatively close to the data. We recall
that EPOS$\,$LHC contains collective flow (put in by hand) and EPOS3
hydrodynamic flow. In addition, particles are produced via statistical
hadronization, which gives much higher yields for multi-strange baryons,
compared to string fragmentation, where these particles are hightly
suppressed.

\begin{figure}[tb]
\begin{centering}
\begin{minipage}[c][1\totalheight]{1\columnwidth}%
\begin{flushleft}
(a)\vspace{-0.8cm}

\par\end{flushleft}

\hspace*{0.3cm}\includegraphics[angle=270,scale=0.24]{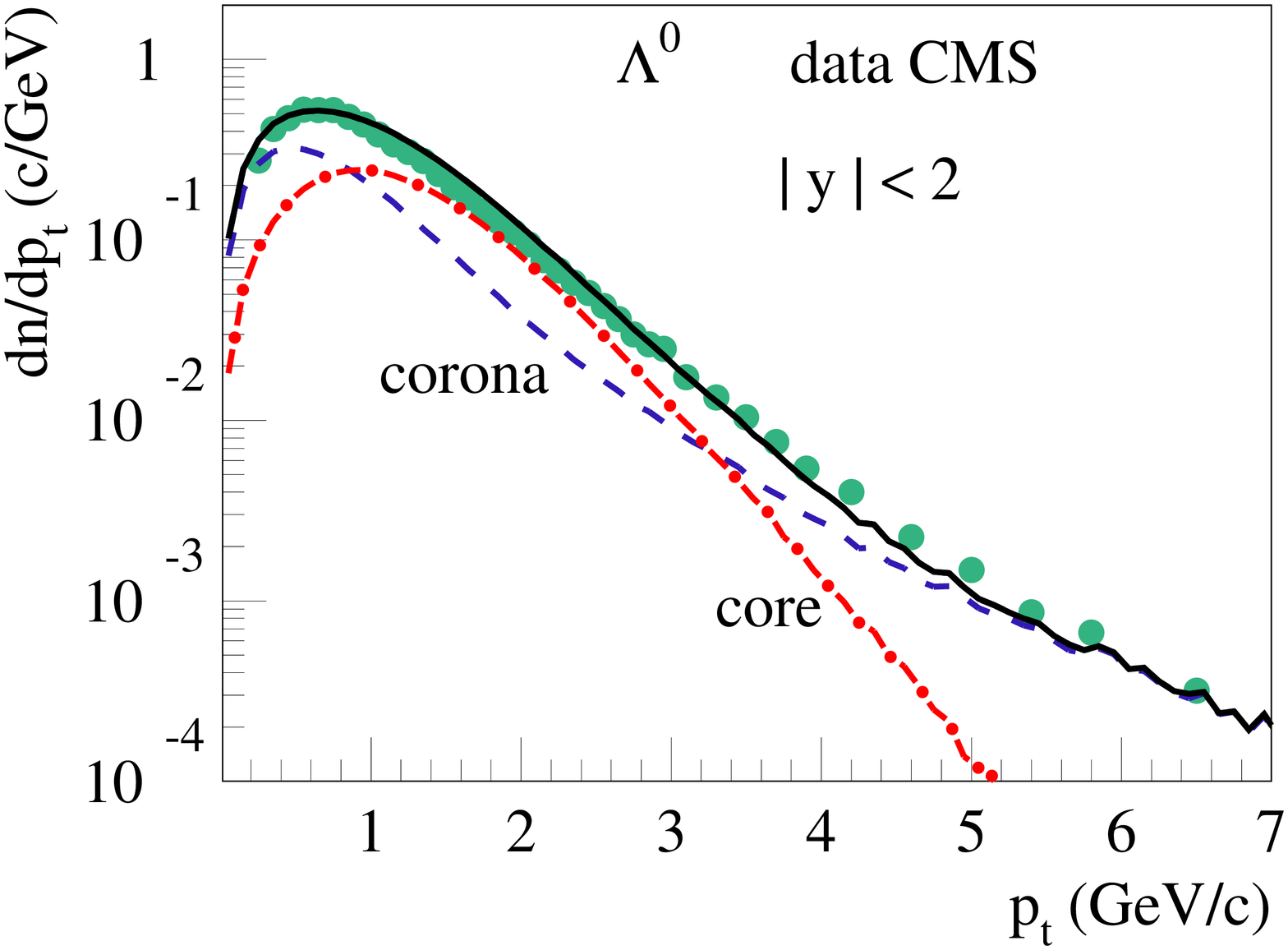}\vspace{-1cm}

\begin{flushleft}
(b)\vspace{-0.4cm}

\par\end{flushleft}

\hspace*{0.3cm}\includegraphics[angle=270,scale=0.24]{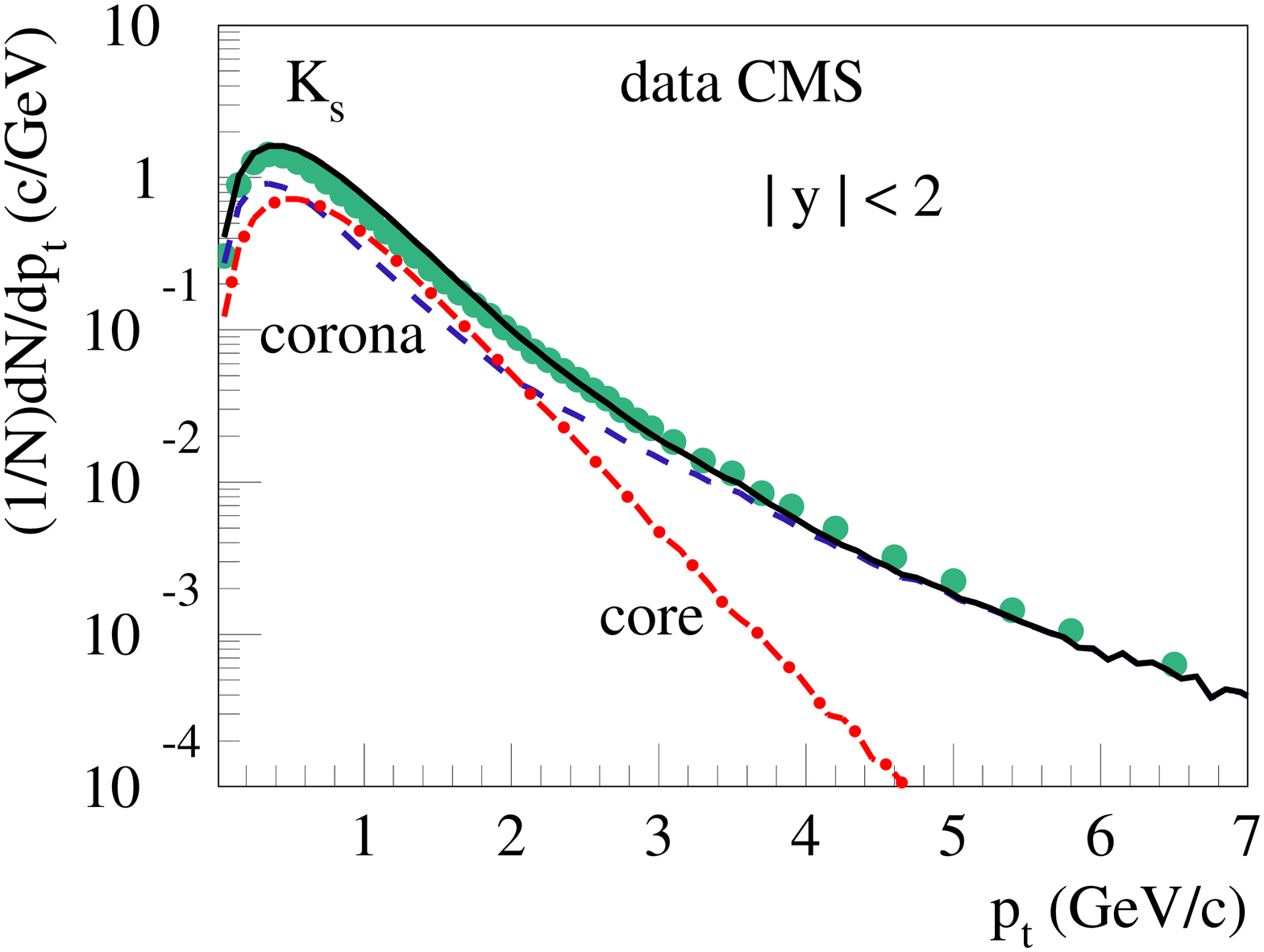}%
\end{minipage}\\

\par\end{centering}

\caption{(Color online) Transverse momentum spectra of $\Lambda$ baryons (a)
and $K_{s}$ mesons (b) in p-p scattering at 7 TeV. We show data from
CMS \citet{cmspp} (symbols) and simulations from EPOS3. The dashed
lines are the corona contributions, the dashed-dotted ones the core
contributions, the full lines are the sums of all contributions. \label{fig:selpi06}}

\end{figure}
\begin{figure}[b]
\begin{centering}
\begin{minipage}[c][1\totalheight]{1\columnwidth}%
\hspace*{0.3cm}\includegraphics[angle=270,scale=0.3]{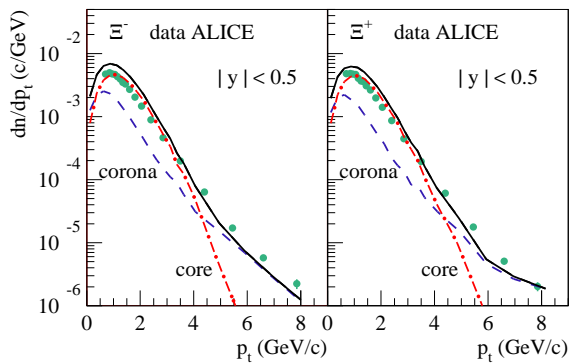}%
\end{minipage}
\par\end{centering}

\caption{(Color online) Same as fig. \ref{fig:selpi06}, but here we plot results
for $\Xi^{-}$ and $\Xi^{+}$ baryons, compared to ALICE data \citet{alicepp}.
\label{fig:selpi05}}

\end{figure}
From the above study we conclude that flow seems to help also in p-p
scattering to explain particle spectra. To understand better the flow
contribution, we plot in fig. \ref{fig:selpi06}(a) again the transverse
momentum spectra of $\Lambda$ baryons, but this time only for EPOS3,
also showing the corona and the core contribution. The core evolved
hydrodynamically, and one can see clearly the intermediate $p_{t}$
enhancement due to flow, as compared to {}``normal'' production
from (kinky) strings in the corona contributions. The same analysis
for kaons in fig. \ref{fig:selpi06}(b) shows that here the flow contribution
is less prominent, due to the typical flow feature that smaller masses
(like kaons compared to lambdas) are less {}``pushed'' to larger
$p_{t}$ values. So we get huge flow effects for heavy particles like
$\Lambda$ and $\Xi$ baryons, as also seen in fig. \ref{fig:selpi05},
where we compare the different contributions to $\Xi$ baryon production
to ALICE data \citet{alicepp}.

\section{Multiplicity dependent particle production in proton-proton scattering
at 7 TeV}

We discussed earlier the multiplicity dependence of particle production
in p-Pb. The $p_{t}$ spectra get systematically harder with multiplicity,
and this effect is more pronounced for heavier particles. This is
precisely what we get in a hydrodynamical scenario, and it even seems
to work on a quantitative level.

Quite similar results (concerning the hardening of $p_{t}$ spectra)
have been obtained by the CMS collaboration for p-p scattering \citet{cmspp2}.
\begin{figure}[b]
\begin{minipage}[t][1\totalheight]{1\columnwidth}%
\begin{minipage}[c][1\totalheight]{1\columnwidth}%
\vspace*{-0.2cm}

\hspace*{-0.3cm}\includegraphics[angle=270,scale=0.2]{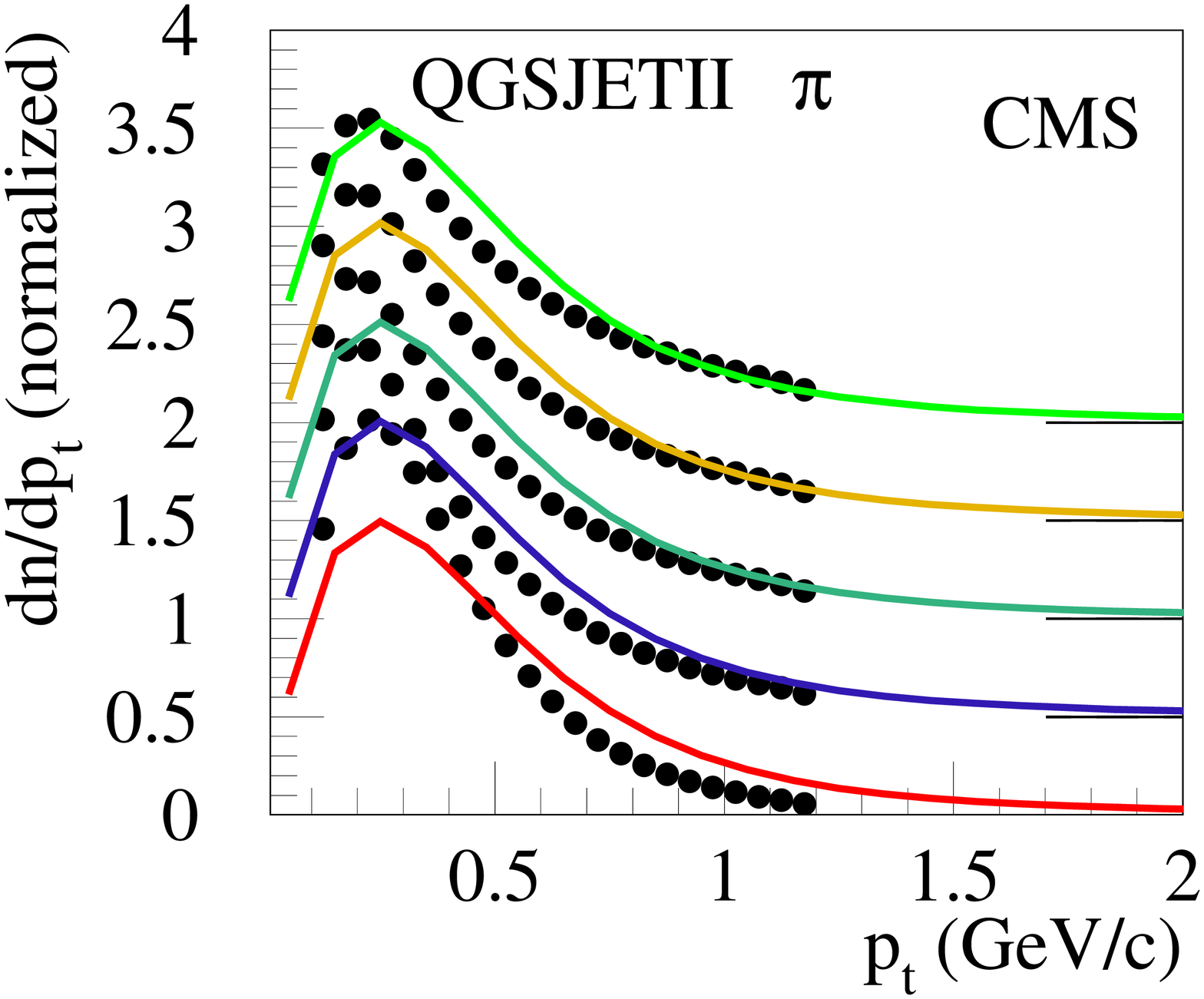}\hspace*{-0.5cm}\includegraphics[angle=270,scale=0.2]{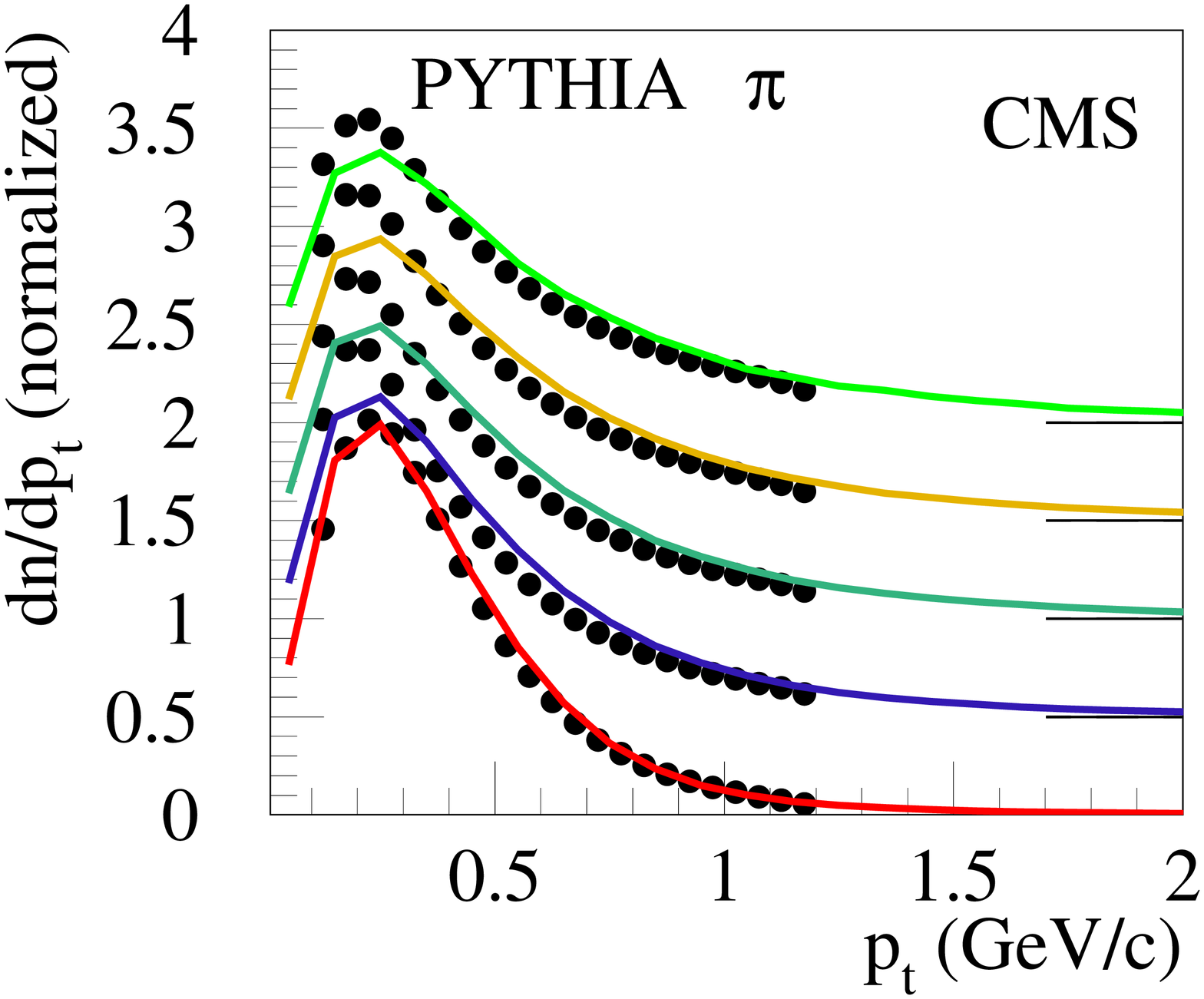}\vspace*{-0.2cm}

\hspace*{-0.3cm}\includegraphics[angle=270,scale=0.2]{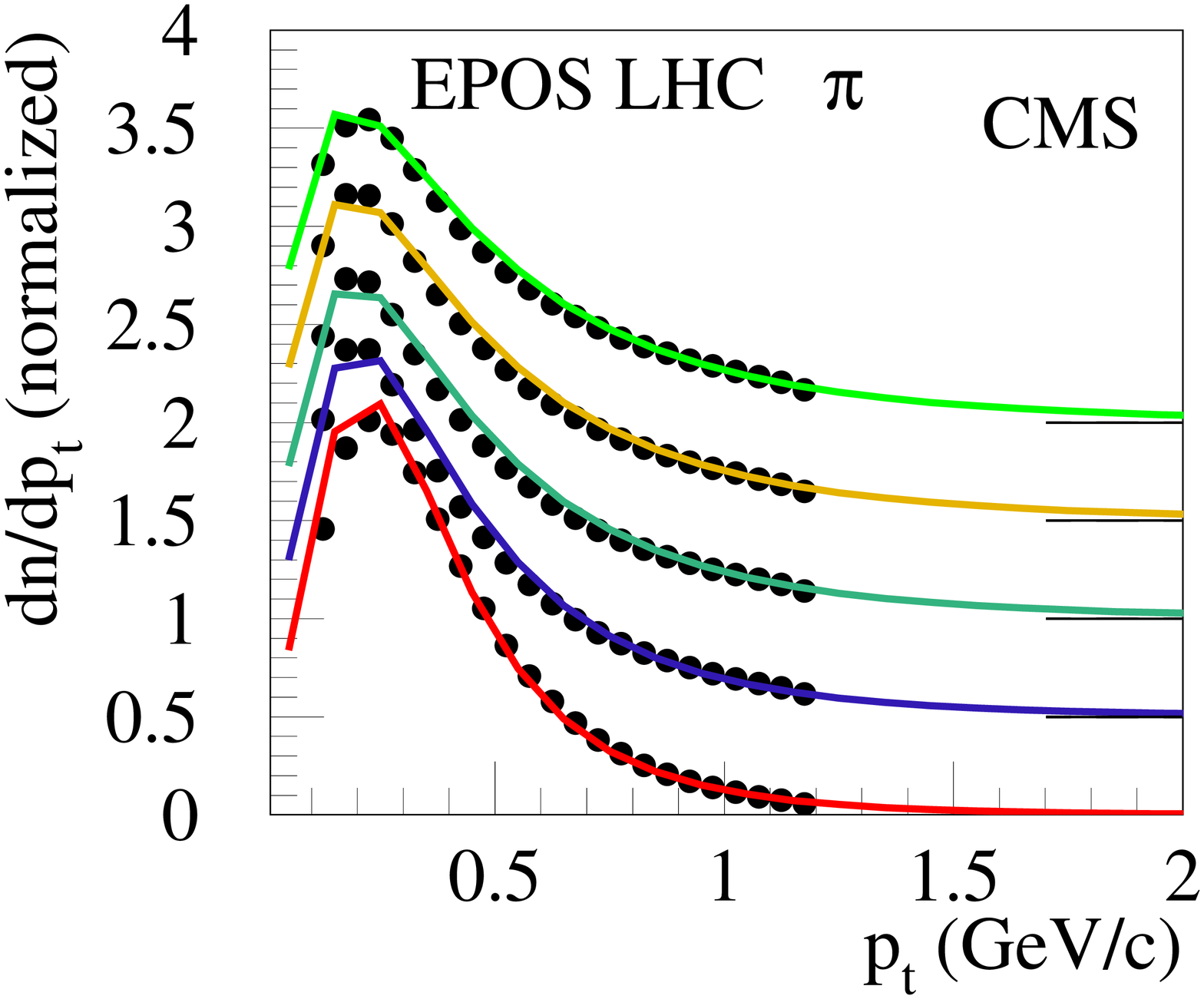}\hspace*{-0.5cm}\includegraphics[angle=270,scale=0.2]{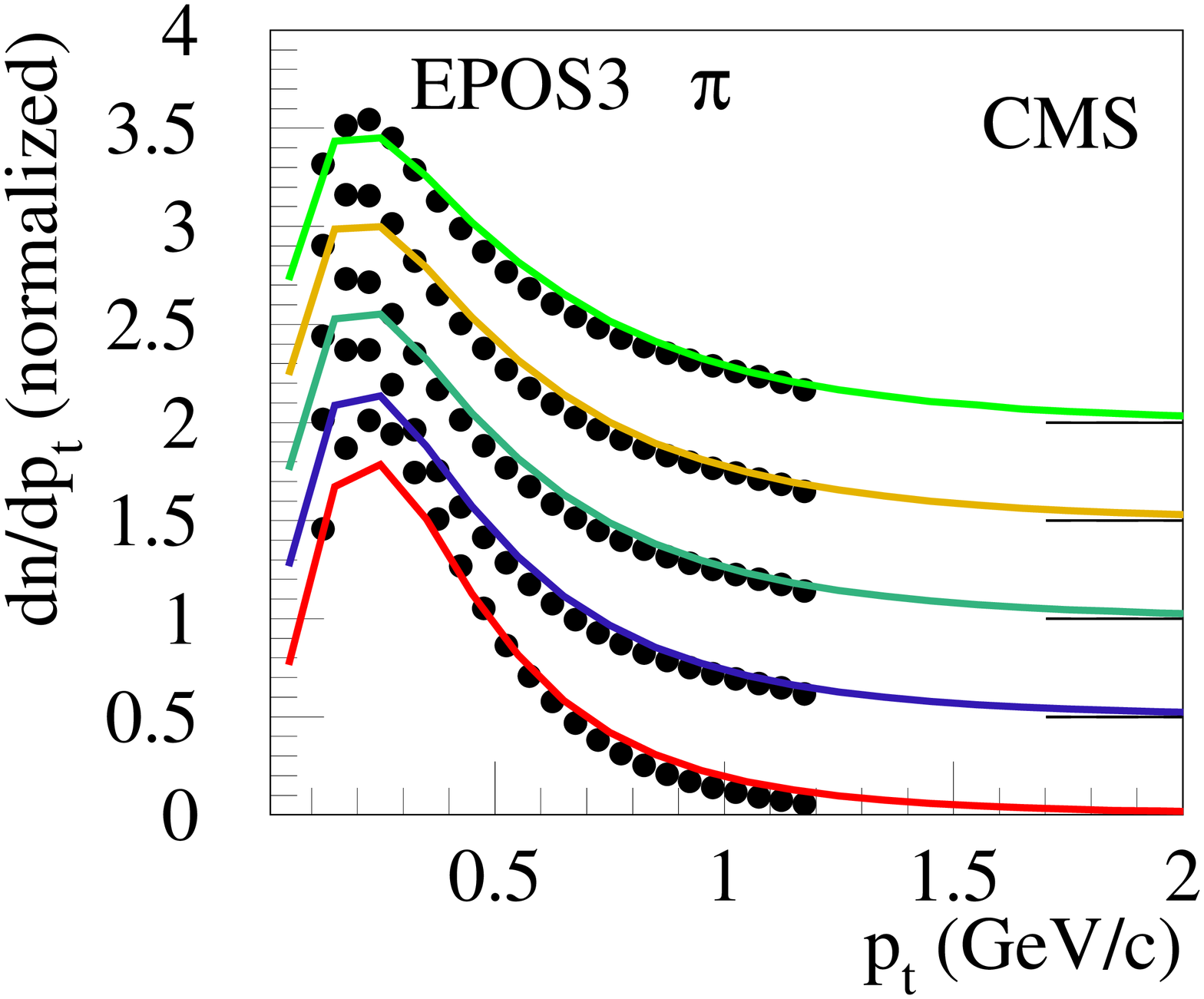}%
\end{minipage}%
\end{minipage}

\noindent \caption{(Color online) Transverse momentum spectra of pions in p-p scattering
at 7 TeV, for five different multiplicity classes with mean values
(from bottom to top) of 7, 40, 75, 98, and 131 charged tracks. We
show data from CMS \citet{cmspp2} (symbols) and simulations from
QGSJETII, PYTHIA6, EPOS$\,$LHC, and EPOS3, as indicated in the figures.
\label{fig:selpi10}}

\end{figure}
\begin{figure}[tb]
\begin{minipage}[t][1\totalheight]{1\columnwidth}%
\begin{minipage}[c][1\totalheight]{1\columnwidth}%
\vspace*{-0.2cm}

\hspace*{-0.3cm}\includegraphics[angle=270,scale=0.2]{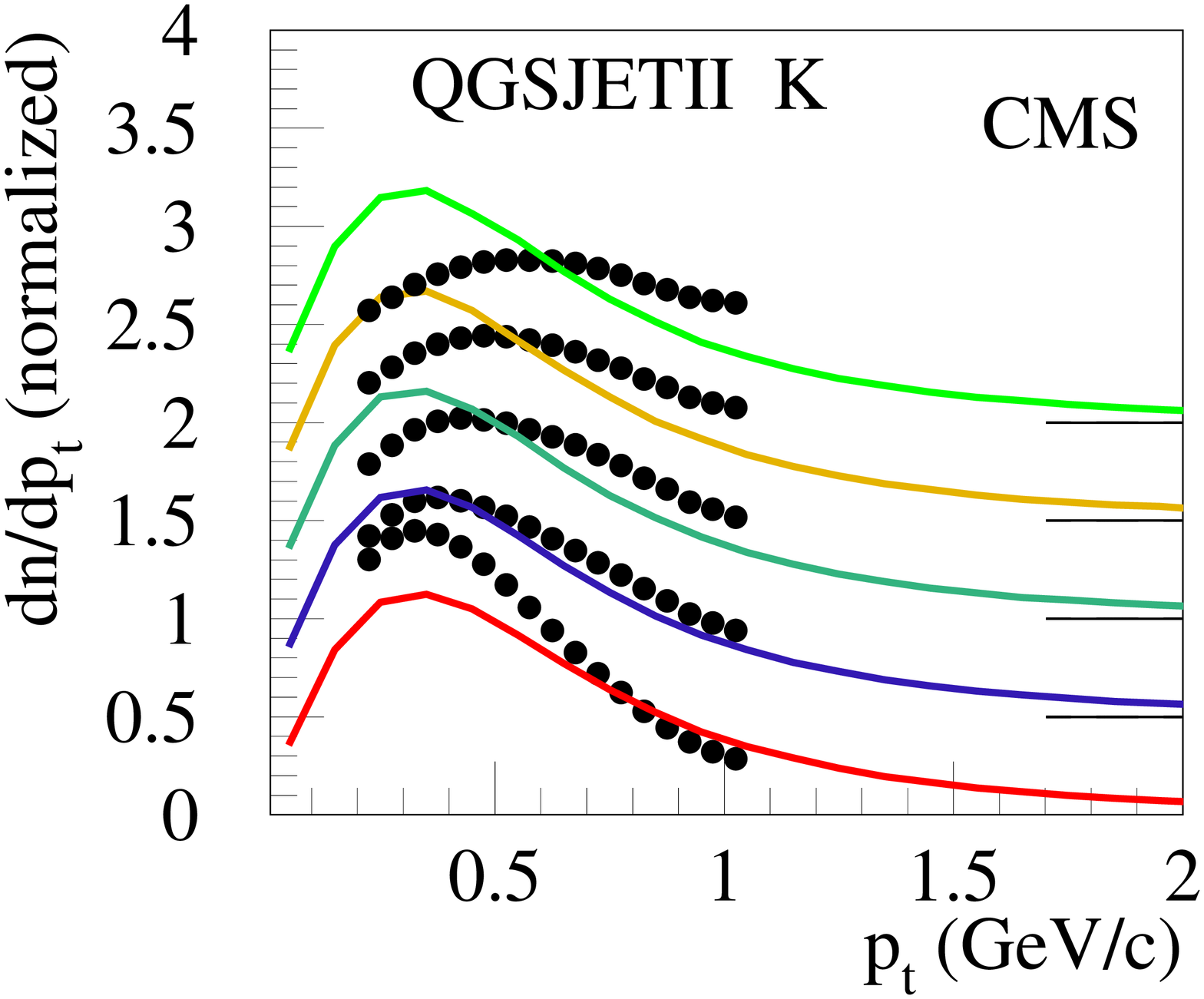}\hspace*{-0.5cm}\includegraphics[angle=270,scale=0.2]{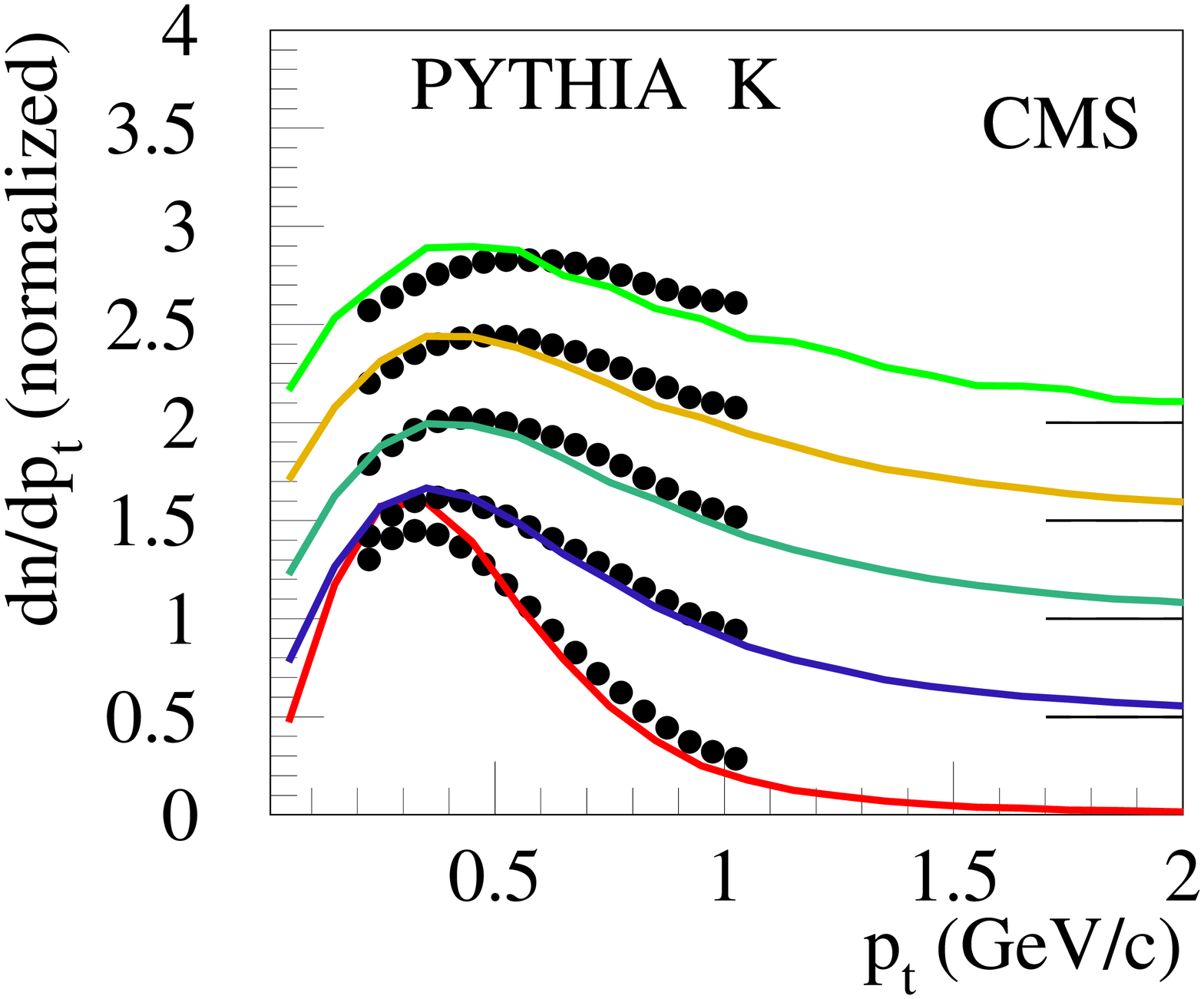}\vspace*{-0.2cm}

\hspace*{-0.3cm}\includegraphics[angle=270,scale=0.2]{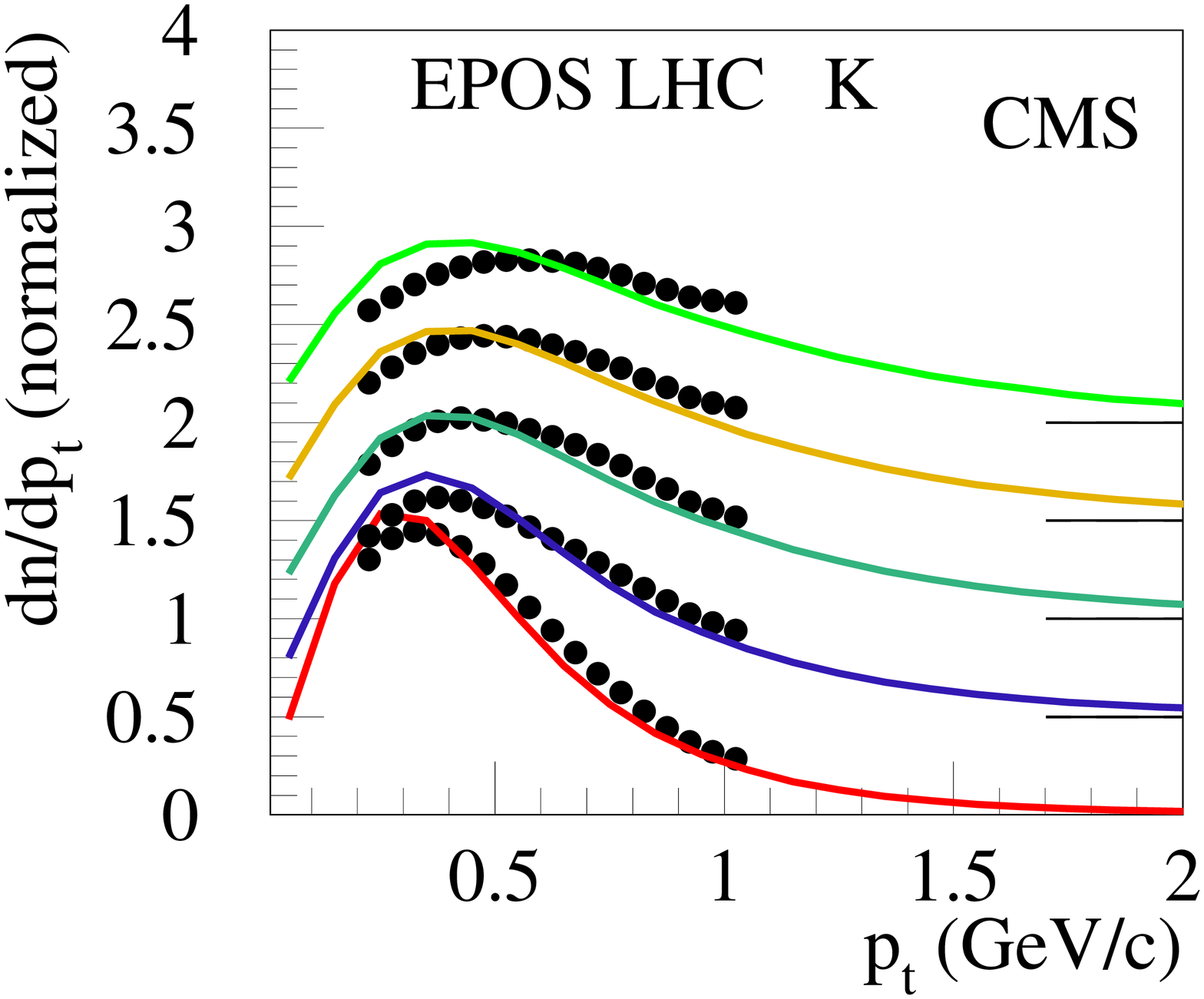}\hspace*{-0.5cm}\includegraphics[angle=270,scale=0.2]{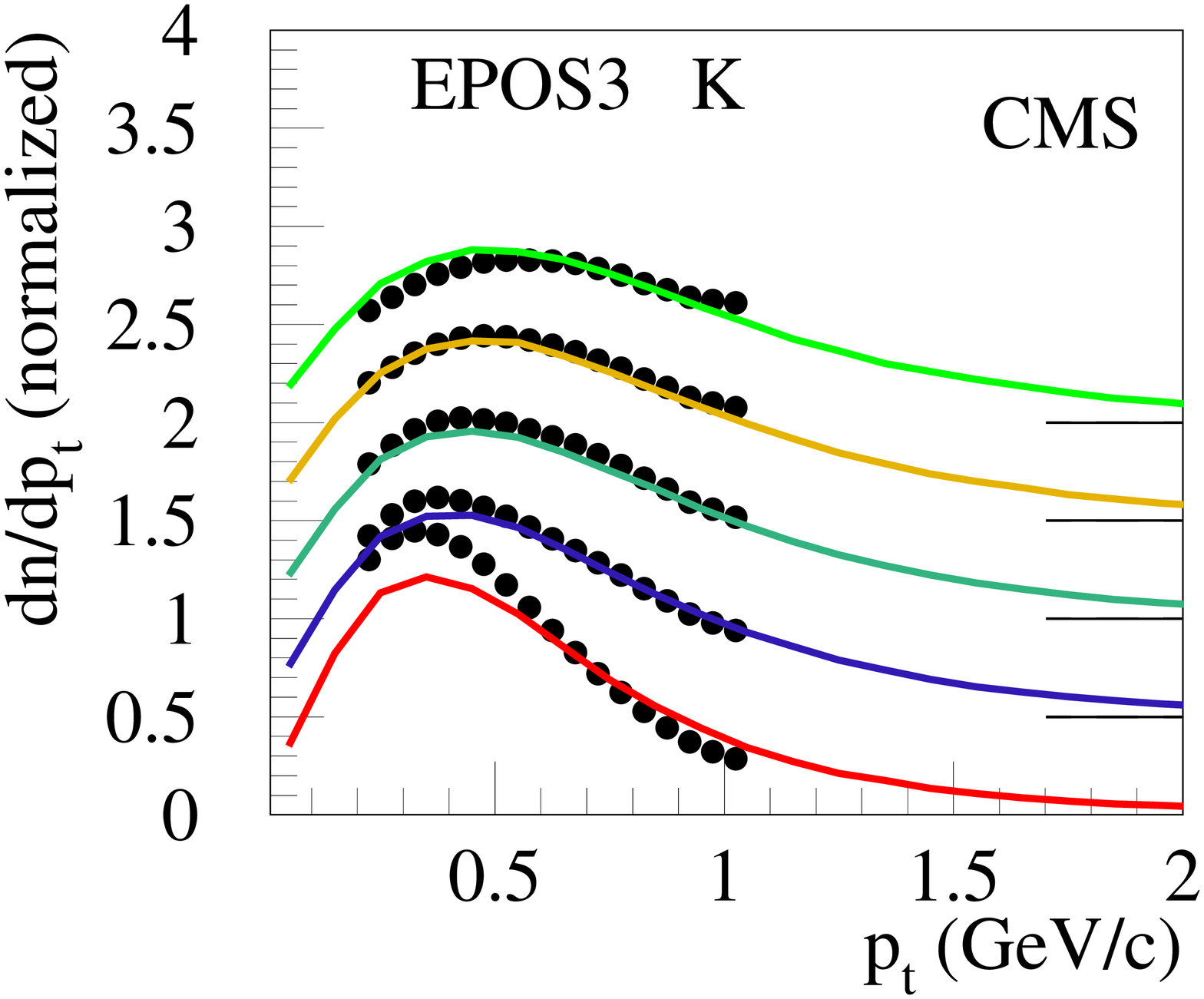}%
\end{minipage}%
\end{minipage}

\noindent \caption{(Color online) Same as fig. \ref{fig:cms1}, but for kaons. \label{fig:selpi14}}

\end{figure}
They performed a detailed study of the multiplicity dependence of
(normalized) transverse momentum spectra. The multiplicity (referred
to as $N_{\mathrm{track}}$ in \citet{cmspp2}) counts the number
of charged particles in the range $|\eta|<2.4$. In our analysis we
consider five multiplicity classes with mean values of 7, 40, 75,
98, and 131 in p-p scattering at 7 TeV.

\begin{figure}[b]
\begin{minipage}[t][1\totalheight]{1\columnwidth}%
\begin{minipage}[c][1\totalheight]{1\columnwidth}%
\vspace*{-0.2cm}

\hspace*{-0.3cm}\includegraphics[angle=270,scale=0.2]{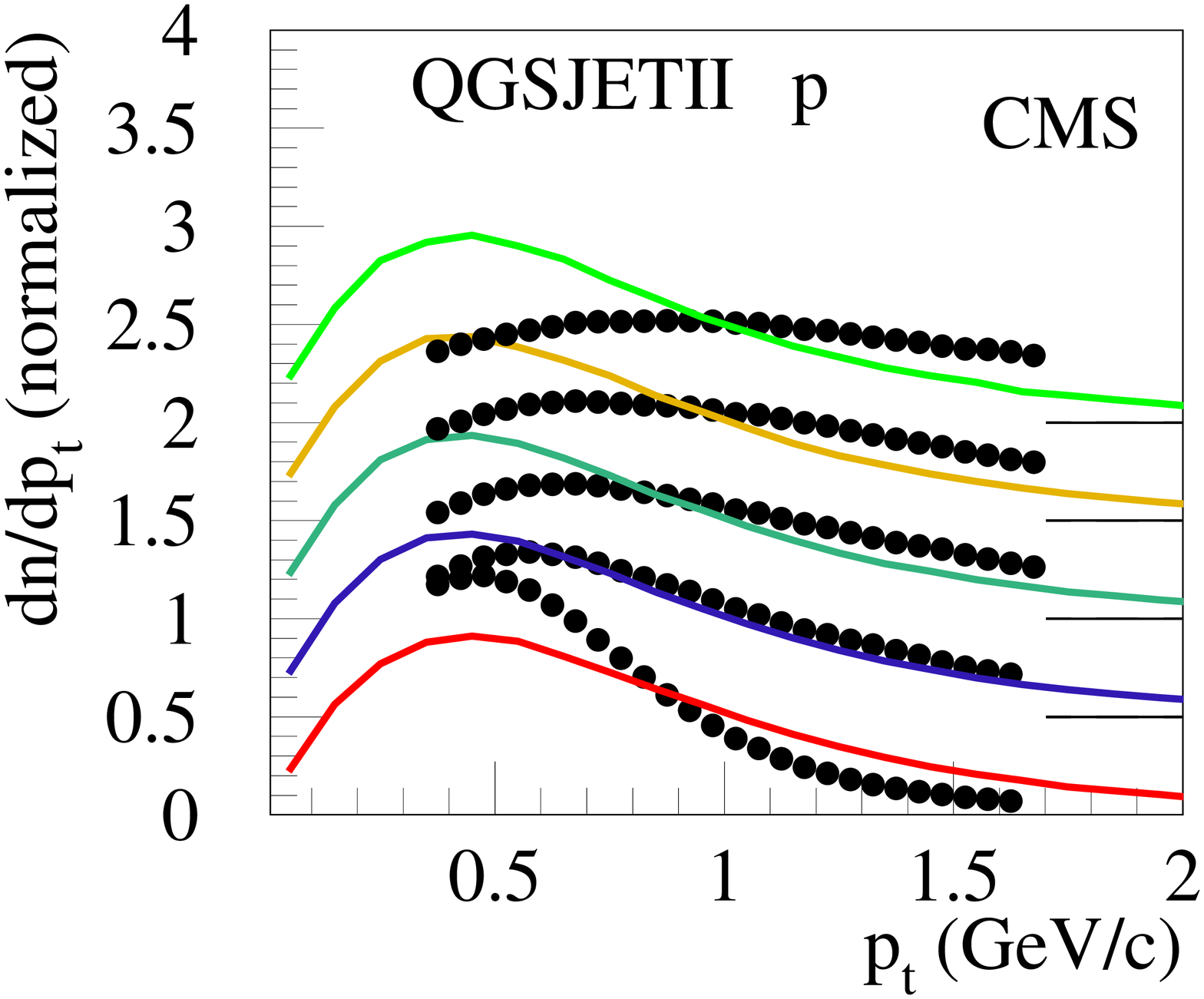}\hspace*{-0.5cm}\includegraphics[angle=270,scale=0.2]{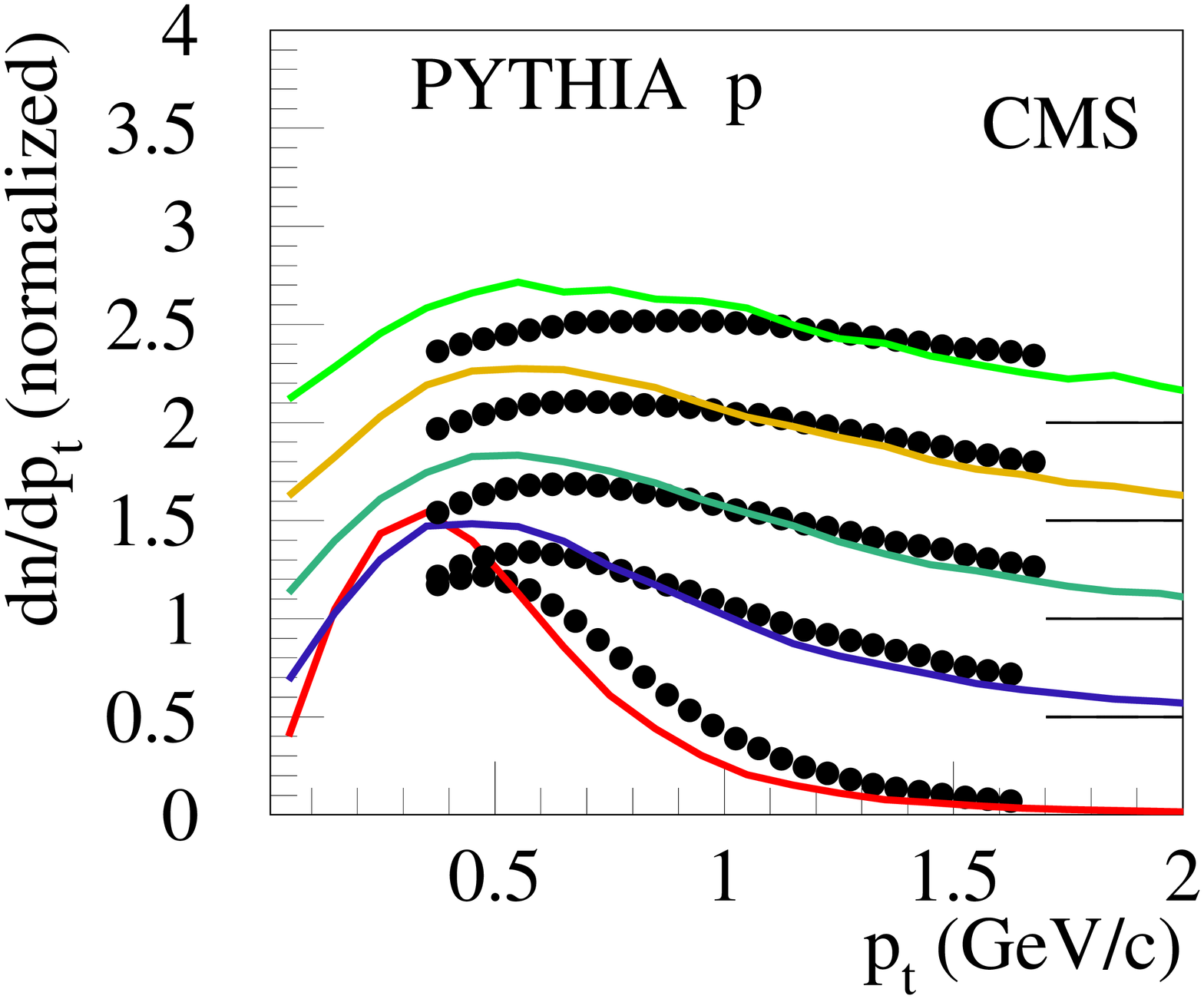}\vspace*{-0.2cm}

\hspace*{-0.3cm}\includegraphics[angle=270,scale=0.2]{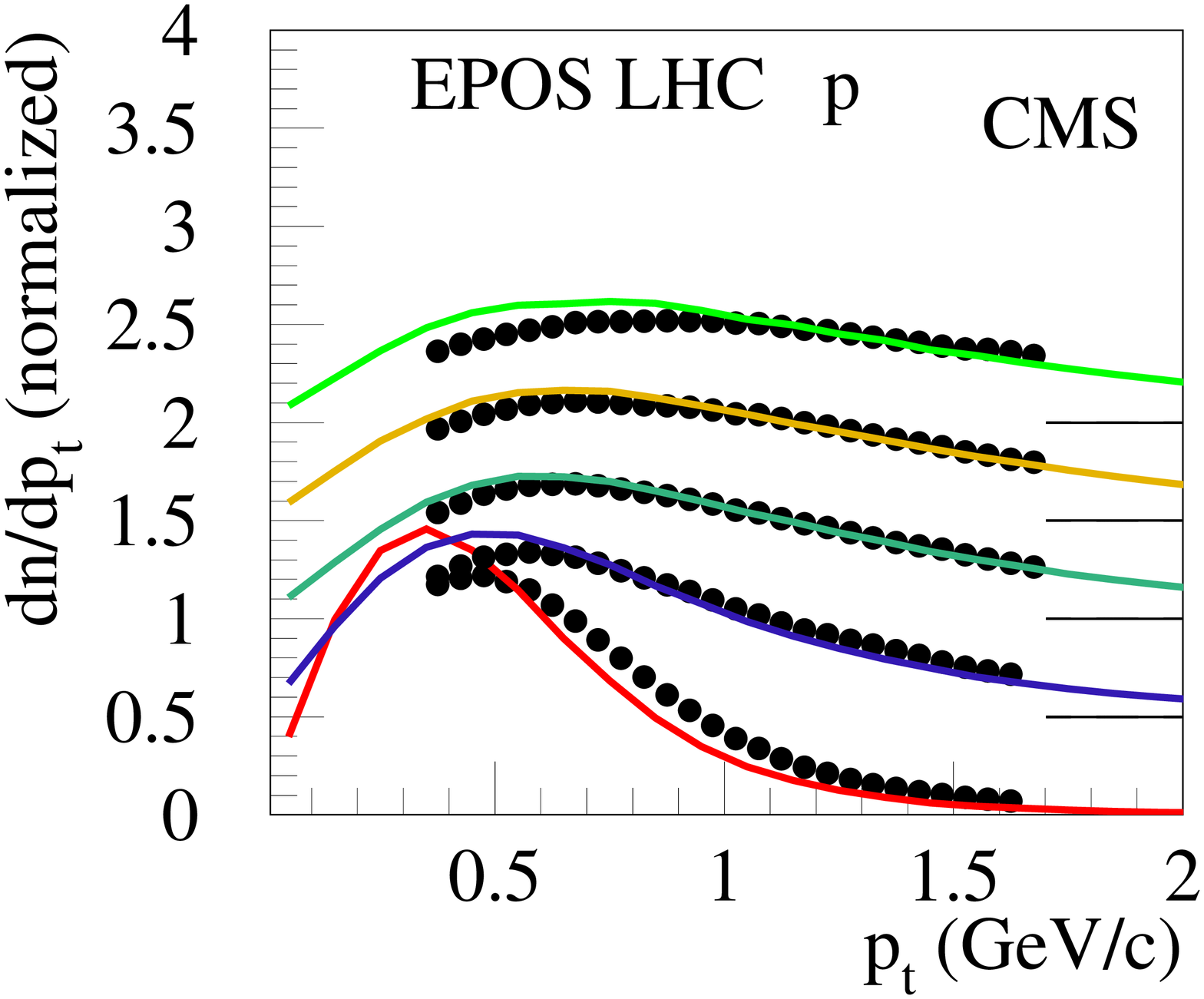}\hspace*{-0.5cm}\includegraphics[angle=270,scale=0.2]{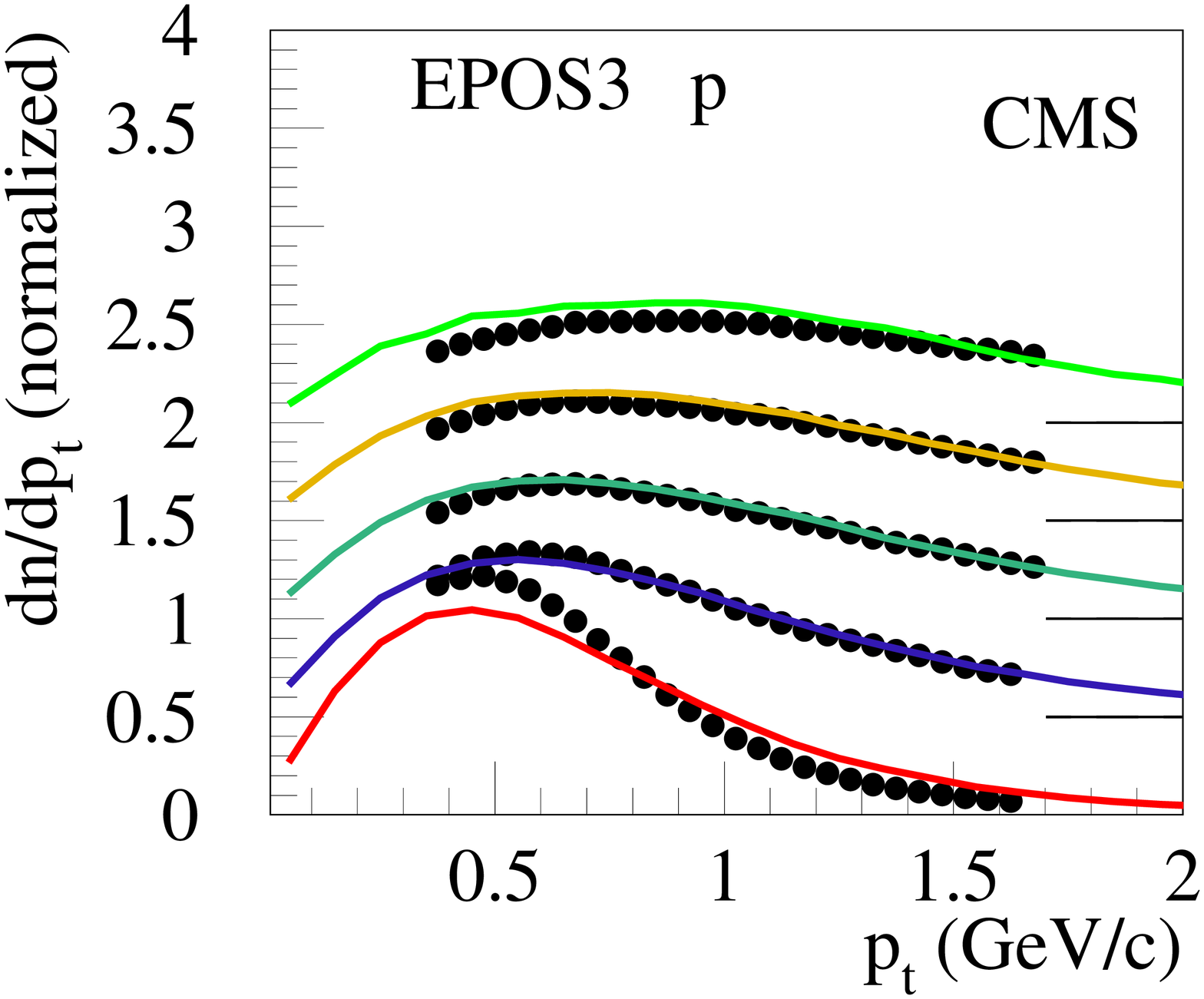}%
\end{minipage}%
\end{minipage}

\noindent \caption{(Color online) Same as fig. \ref{fig:cms1}, but for protons.\label{fig:selpi18}}

\end{figure}
In fig. \ref{fig:selpi10}, we compare experimental data \citet{cmspp2}
for pions (black symbols) with the simulations from QGSJETII (upper
left figure), PYTHIA6 (upper right), EPOS$\,$LHC (lower left), and
EPOS3 (lower right). We use the tune Perugia 2011 (350) of PYTHIA6.4.27
(also in figs.\ref{fig:selpi14} and \ref{fig:selpi18}). The different
curves in each figure refer to different centralities, with mean values
(from bottom to top) of 7, 40, 75, 98, and 131 charged tracks. They
are shifted relative to each other by a constant amount. Concerning
the models, QGSJETII is the easiest to discuss, since here the curves
for the different multiplicities are identical. The data, however,
show a slight centrality dependence: the spectra get somewhat harder
with increasing multiplicity. The other models, PYTHIA, EPOS$\,$LHC,
and EPOS3 are close to the data. In fig. \ref{fig:selpi14}, we compare
experimental data \citet{cmspp2} for kaons (black symbols) with the
simulations. In the data, the shapes of the $p_{t}$ spectra change
considerably with multiplicity: they get much harder with increasing
multiplicity. In QGSJETII, there is again no change, whereas PYTHIA,
EPOS$\,$LHC, and EPOS show the right trend. EPOS3 reproduces better
the high multiplicity curves, PYTHIA and EPOS$\,$LHC the low multiplicity
results. In fig. \ref{fig:selpi18}, we compare experimental data
\citet{cmspp2} for protons (black symbols) with the simulations.
Again, as for kaons, the experimental shapes of the $p_{t}$ spectra
change considerably, getting much harder with increasing multiplicity.
In QGSJETII, the curves for the different multiplicities are identical.
The PYTHIA model shows some change with multiplicity, but the shapes
are not correct. EPOS$\,$LHC and EPOS3 give a reasonable description
of the data. The hardening of the shapes with multiplicity, more and
more pronounced with increasing particle mass, is here due to the
radial flow. \textbf{It seems that hydrodynamical flow again helps
considerably to reproduce the data, even in proton-proton scattering}.

\begin{figure}[tb]
\begin{centering}
\vspace*{-0cm}\includegraphics[angle=270,scale=0.3]{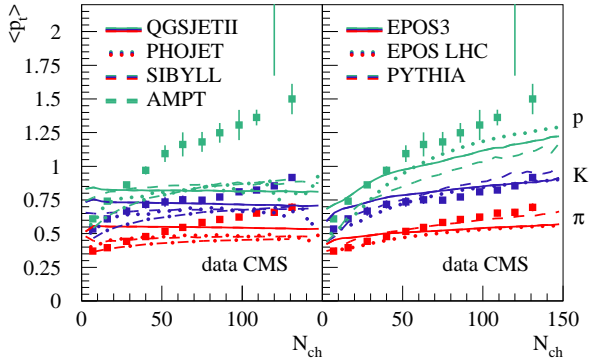}\\

\par\end{centering}

\caption{(Color online) Multiplicity dependence of the average transverse momentum
of protons (green), kaons (blue), and pions (red) in p-p scattering
at 7 TeV. We show data from CMS \citet{cmspp2} (symbols) and simulations
from QGSJETII, PYTHIA6, EPOS$\,$LHC, EPOS3, and in addition PHOJET,
SIBYLL, and AMPT. \label{fig:selpi01}}

\end{figure}
Based on these multiplicity dependent $p_{t}$ spectra, one obtains
the multiplicity dependence of the average transverse momentum $\left\langle \right.p_{t}\left.\right\rangle $,
as shown in fig. \ref{fig:selpi01}, where we plot the multiplicity
dependence of the average transverse momentum of protons (green),
kaons (blue), and pions (red) in p-p scattering at 7 TeV. We show
data from CMS \citet{cmspp2} (symbols) and simulations from QGSJETII,
PYTHIA6, EPOS$\,$LHC, EPOS3, and in addition PHOJET, SIBYLL, and
AMPT. Whereas QGSJETII, PHOJET, SIBYLL, and AMPT shows no or little
multiplicity dependence, PYTHIA, EPOS$\,$LHC and EPOS3 increase with
multiplicity, and this increase is more pronounced for heavier particles.
In EPOS$\,$LHC and EPOS3, this is due to the radial flow, in PYTHIA
due to the so-called color reconnection.

\section{Summary}

We described in detail EPOS3, an event generator based on a 3D+1 viscous
hydrodynamical evolution starting from flux tube initial conditions,
generated in the Gribov-Regge multiple scattering framework. Individual
scatterings are referred to as Pomerons, identified with parton ladders,
eventually showing up as flux tubes (or strings). We discussed that
in p-Pb collisions, the geometry is essentially determined by the
number of Pomerons, being proportional to the number of flux tubes
(and eventually to the multiplicity). A large number of flux tubes
means a high probability to create high density matter which will
evolve hydrodynamically. This explains why in our approach with increasing
multiplicity the hydrodynamical flow becomes more and more important,
being visible in terms of a shift of intermediate $p_{t}$ particles
to higher values. This shift is more and more pronounced with increasing
particle mass. These features seem to be present in recent p-Pb and
even in p-p data. To confirm the {}``flow hypothesis'', we compared
EPOS3 simulations with essentially all available data on $p_{t}$
spectra of identifies particles in pPb scattering at 5.02 TeV and
p-p scattering at 7 TeV, and with all available simulations from other
models. In all cases, hydrodynamical flow improves the situation considerably.
It should be said that this is the first publication concerning EPOS3,
the parameters are far from being optimized (it takes 1 month of simulations
on several hundreds of nodes for one parameter set). 

\begin{acknowledgments}
This research was carried out within the scope of the GDRE (European
Research Group) {}``Heavy ions at ultrarelativistic energies''.
Iu.K acknowledges support by the National Academy of Sciences of Ukraine
(Agreement F4-2013) and by the State Fund for Fundamental Researches
of Ukraine (Agreement F33/24-2013). Iu.K. acknowledges the financial
support by the ExtreMe Matter Institute EMMI and Hessian LOEWE initiative.
\end{acknowledgments}

\end{document}